# Title

# Uniaxial Recovery Perspective of Glassy Polymer Nanoindentation

First author: Prakash Sarkar

PhD student, Department of Metallurgical Engineering and Materials Science, Indian Institute of Technology Bombay, Mumbai- 400076, Maharashtra, India

ORCID: 0000-0002-4404-916X

Corresponding author: Hemant Nanavati

Professor, Department of Chemical Engineering, Indian Institute of Technology Bombay, Mumbai- 400076, Maharashtra, India

*E-mail: hnanavati@iitb.ac.in

ORCID: 0000-0002-5982-6531

# Graphic for Table of Contents (TOC) Only

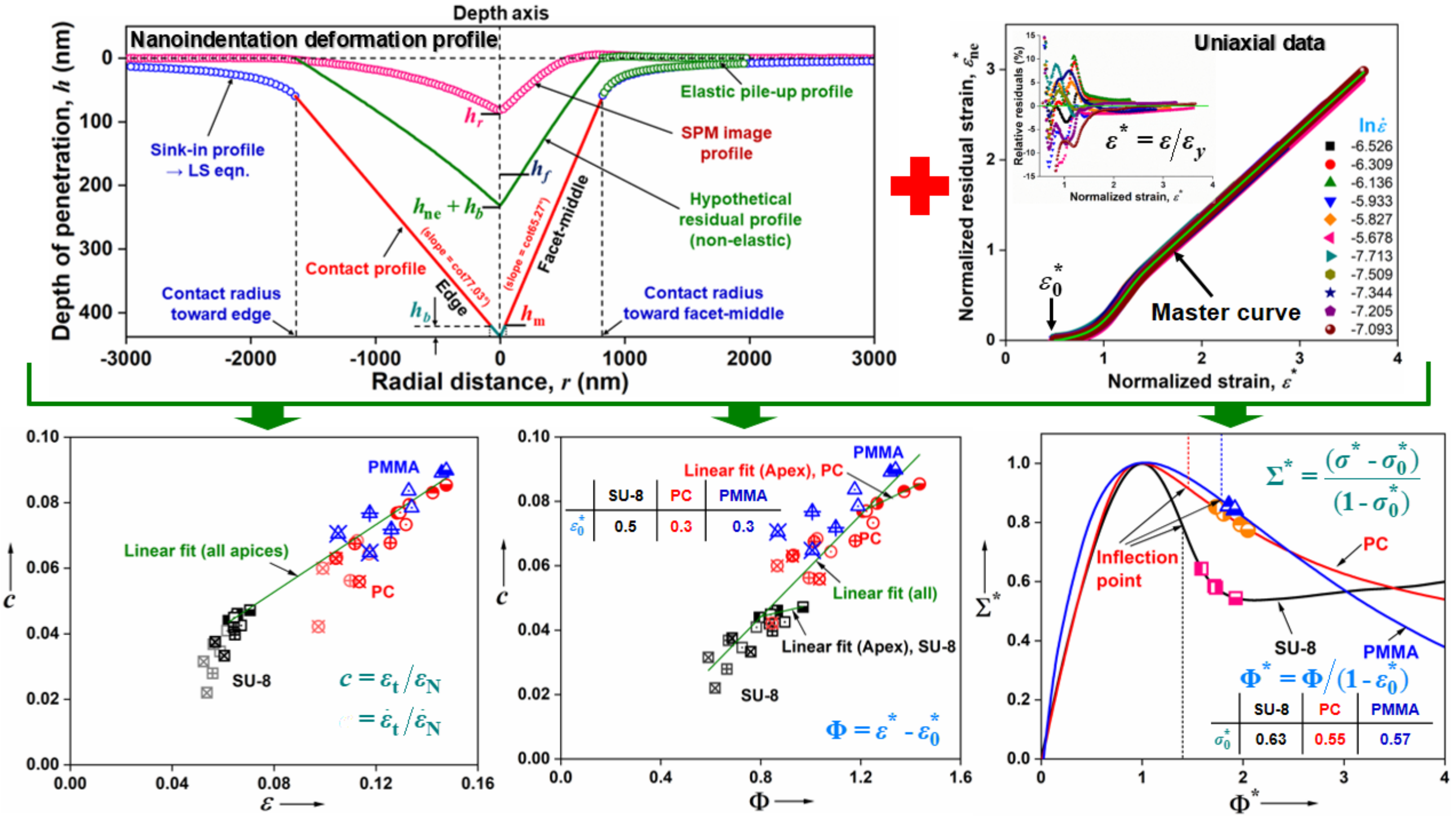

# Abstract

Sharp tip nanoindentation of glassy polymers is a constrained, localized viscoelastoplastic deformation. We interpret this complexity, in terms of the well-understood uniaxial deformation. From the uniaxial compression data in the literature, for PMMA, PC and crosslinked SU-8, we obtain their universal, yield-normalized recovery curves, with $\varepsilon^* = \varepsilon / \varepsilon_y$ , a measure of the corresponding strain states (*CSS*). Nanoindentation recovery is determined from the 2sec constant rate unloading *h-P* data, modelled by a generalized power-law (variable power exponent). Comparing these loading and recovery data-sets, yields the correlation coefficient between the notional nanoindentation strain rate $(\dot{\varepsilon}_{\mathrm{N}} \sim \dot{h}/h)$ and strain and true strain rate and strain, $c = \dot{\varepsilon}_{\mathrm{N}} / \dot{\varepsilon}_{\mathrm{t}} = \varepsilon_{\mathrm{N}} / \varepsilon_{\mathrm{t}}$ . The equivalent strain, $\varepsilon$, and the *c* value for any polymer, are within a narrow range, from the onset of indentation. Combining residual profiles via scanning probe microscopy with mathematical modelling of the indenter tip, provides the strain distribution beneath the tip. *CSS* measures examined here, indicate polymer-specific regions to regions common to glassy polymers, which are reached very early in the nanoindentation.

# 1. Introduction

Nanoindentation is a potentially scale-bridging and a relatively non-invasive technique, to determine mechanical properties, over very small dimensions. The multiaxial, localized and constrained deformation response of any material during nanoindentation, is very complex. This complexity is further manifested, particularly for indentation via sharp tips with sharp edges, such as the Berkovich tip; elastic, viscous and plastic contributions, begin very early during nanoindentation. In addition, the contact regions share part of their deformation with the non-contact region. Finally, the directions of the deformation and the resistance to volume change, are also distributed.

Nanoindentation characterization methods essentially comprise employing the load-displacement (*P*-*h*) data to estimate the hardness, $H = P_{\mathrm{m}}/A_c$ ($P_{\mathrm{m}}$ is the full load, and $A_c$ is the contact area) and the reduced modulus, $E_r = S\sqrt{\pi}/2\sqrt{A_c}$ ($\sim E_{\mathrm{N}} \times \left(1-\nu^2\right)$, for compliant materials), where $\nu$ and $E_{\mathrm{N}}$ are the Poisson's ratio and elastic nanoindentation modulus of the indented sample. The stiffness, $S = dP/dh\big|_{P_{\mathrm{m}}}$, is the unloading onset slope.

Load deformation relationships are best understood in case of uniaxial and simple shear deformations, along initial levels of complexity, such as bending and torsion, which have usually been carried out in the bulk state. Their deformation includes significant and simultaneous visco-plastic (VP) contributions, in addition to elastic contributions. The characteristic uniaxial compression properties are the yield stress, $\sigma_y$ and the uniaxial modulus, *E*. The phenomena they measure, are better understood than those measured by the corresponding properties, $H$ and $E_r$.

As reported previously [1], [2], uniaxial deformation-based understanding can form the basis of extending our analysis of viscoelastoplastic (VEP) nanoindentation deformation, if the elastic, viscous and plastic aspects of the deformation can be separated to the extent possible. We employ uniaxial compression as the basis to identify aspects of nanoindentation deformation of glassy polymers, which exhibit VEP deformation, and are more pliant than other materials such as metals or ceramics. We consider both, thermosets (represented by the cross-linked epoxy, SU-8) and thermoplastics (represented by polycarbonate (PC) and Poly-methyl methacrylate (PMMA)).

This work aims to interpret the complex nanoindentation deformation phenomena, in terms of the uniaxial deformation phenomena, for amorphous glassy polymers. The displacement controlled (DC) uniaxial compression data considered here, are at set strain-rates, and are obtained from the literature. Uniaxial deformation is defined in terms of strain and strain rate. Mayo *et al.* [3]–[5], have phenomenologically defined the nanoindentation strain rate, $\dot{\varepsilon}_{\mathrm{N}} \sim \dot{h}/h$. Lucas *et al.* [6], have phenomenologically obtained $\dot{h}/h = 0.5\,\dot{P}/P$, for load-controlled (LC) nanoindentation. In Section 5, we first explain the necessity of considering LC nanoindentation with DC uniaxial compression, for our framework here. One aspect of expressing nanoindentation deformation in terms of uniaxial deformation, is its interpretation in terms of its equivalent uniaxial strain and strain rate.

The non-uniformly distributed deformation during glassy polymer nanoindentation, would be reflected in the instantaneous equivalent uniaxial strain, which would be greater than the yield strain, $\varepsilon_y$. As first described by Tabor [1], the residual nanoindentation strain post-unload, corresponds to that of the equivalent uniaxial non-elastic strain. This non-elastic strain, could be related to the imposed strain at $P_{\mathrm{m}}$, as shown in Fig. 1. Our analysis applies this idea, ensuring that the total nanoindentation strain is the sum of the objectively defined, recovered strain and the residual strain. Then, if $\varepsilon_y$ is a function of strain rate, we ensure consistency with the equivalent strain rate.

Our methodology involves topographical imaging via Scanning Probe Microscopy (SPM), in order to examine the residual nanoindentation imprint on the material. We thus estimate recovered and residual nanoindentation strain, for mapping with the corresponding uniaxial strains. This interpretation of the equivalent strain is provided in terms of the objectively defined state of deformation, with respect to $\varepsilon_y$. The corresponding states of deformation (as defined by the recovery), suggest universal maps of uniaxial behaviours of glassy polymers, on which the limited regions of nanoindentation can be located; depending on the type of normalization in the corresponding strain states (*CSS*), these regions may be polymer-specific, or universal, across polymers. Our framework provides insights into the actual phenomena during nanoindentation deformation, on the basis of uniaxial compression. These can be employed, for example, to develop more faithful frameworks for predictive modelling or finite element (FE) investigations.

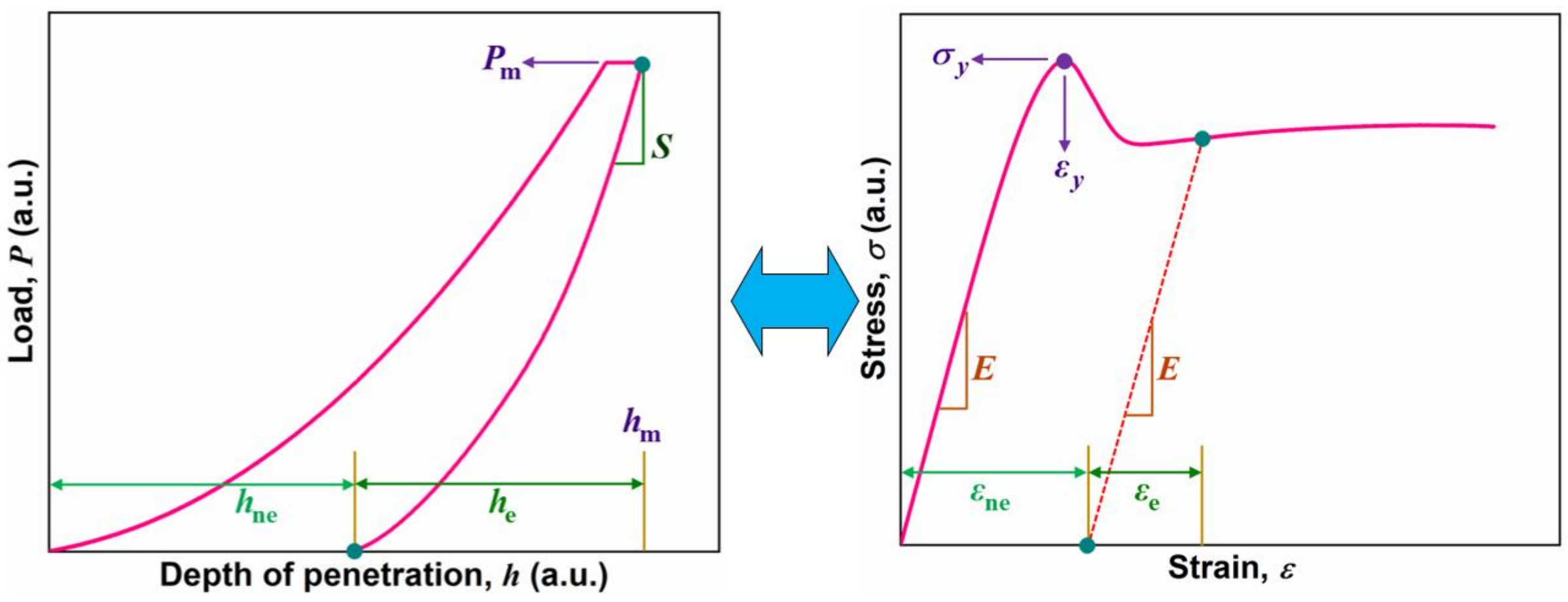

**Figure 1**: LC nanoindentation $P$-$h$ Recovery in terms of DC uniaxial compression $\sigma$-$\varepsilon$ Recovery.

# 2. Background

The first mathematical analyses on indentation were presented by Love [7] and Sneddon [8], [9] (LS) for ideal indenter indentations of ideally elastic materials. Based on these analyses, Oliver and Pharr (OP) [10]–[12]. This framework has been validated for a range of elastoplastic (EP) materials, whose moduli range from 68 GPa (Aluminum) to 440 GPa (sapphire). We will call these materials OP materials. Currently, nanoindentation analyses broadly deal with estimating the nanoindentation properties, $H$ and $E_r = S\sqrt{\pi}/2\sqrt{A_c} = \left[\left(\left(1-\nu^2\right)E_{\mathrm{N}}^{-1}\right)+\left(\left(1-\nu_i^2\right)E_i^{-1}\right)\right]^{-1}$ ($\nu_i$ and $E_i$ are the Poisson's ratio and elastic modulus of the indenter, respectively), via the OP framework.

Based on the $A_c$ obtained by the OP method, Poisl *et al.* [2], have been the first to report the mapping of nanoindentation and uniaxial strain rates, as $\dot{\varepsilon} = c\dot{\varepsilon}_{\mathrm{N}}$, ($c$ is the correlation coefficient). They have examined nanoindentation creep following a step load on selenium (in the Newtonian viscosity strain rate regime), and assuming Tabor's phenomenological relationship, $H = 3\sigma_y$, they have found that $c$ ~ 0.09. This value has been considered for polymers as well [13], [14].

Material dependence of $c$ has been reported [2], [15]–[20], by mapping $\dot{\varepsilon}_{\mathrm{N}}$ to uniaxial creep rates, and recognizing that visco-elastic (VE) deformation and yield, occur simultaneously during nanoindentation [21]. Kermouche and co-workers [22]–[25], have synthesized LC uniaxial deformation curves for various $\dot{P}/P$ values, at $P_{\mathrm{m}}$ = 15 mN, employing published uniaxial modulus and ideal yield strain formulations, along with the G'sell-Jonas model [26], for the flow stress. Subsequently, their nanoindentation data provides the strain-rate sensitivity parameter, enabling estimation of the representative strains and strain rates during LC nanoindentation. Attempts to understand the constrained deformation via Berkovich nanoindentation (including frameworks in terms of uniaxial deformation [27][28]), have been reported previously [29], [30] as well.

Expressing LC nanoindentation deformation in terms of equivalent uniaxial compressive strain, requires analysis of the anelastic nanoindentation deformation component. Anelastic effects are well understood for uniaxial deformation. While there are a several reports [23]–[25], [31]–[37] on the VE effects during nanoindentation of polymers, these nanoindentation analyses consider $A_c$ from OP-type methods, which, as we find in this work, are inappropriate for amorphous glassy polymers.

One common feature in previous nanoindentation vs uniaxial deformation analyses, is that they relate the creep during the hold period of LC nanoindentation, and relate it to the creep during experimental deformation of fabricated micropillars [20] or under simulated uniaxial deformation. Subsequently, the variation of nanoindentation properties with nanoindentation strain rate, has been mapped to mechanical properties as function of uniaxial strain rate.

In this work, uniaxial compression data for amorphous glassy polymers for a range of DC deformation rates, have been sourced from the literature: (i) for cross-linked SU–8, we consider DC micropillar compression [14], and (ii) for PC [38], [39] and for PMMA [40]–[43], we consider the DC bulk compression data. We consider the $P - h$ data of our nanoindentation experiments, and the SPM images of their residual imprints. We obtain the hypothetical pure elastic recovery measure from the unloading curves, fit to a generalized power law (GPL) [44]. The non-elastic strain is then computed systematically as per the schematic (Fig. 1) for each stress-strain set, and is interpreted in terms of the non-elastic residual imprint depth. In previous works, the strain has not been considered. Since nanoindentation begins with a preload, integrating the strain rate over any part of the nanoindentation duration, yields the strain difference, but not the absolute strain. We overcome this difficulty by employing experimentally obtained elastically recovered strain, as our basis. Then, we determine only the effective strain difference between the imposed strain and the recovered strain, to obtain the residual strain.

# 3. Material and sample preparation

The materials we investigate here are amorphous glassy polymers, i.e.,

1. For thermosets, we have considered highly cross-linked SU-8 epoxy. We begin with SU-8 2050 grade resin, to fabricate thin film samples on silicon wafer substrate, via a standard photo-lithography process [45], [46]. The fabrication steps are: (i) RCA cleaning of 2" dia. silicon wafer,

(ii) dehydration baking of wafer at 110°C for 30 min, (iii) spin coating of SU-8 2050 grade at optimized parameters to obtain ~ 22 μm thin film, (iv) pre-exposure baking at 65°C for 10 min and then 95°C for 5 min, (v) UV-exposure for 13 s, (vi) post-exposure baking (PB) at 95°C for 30 min, and (vii) hard baking (HB) at 150°C for 15 min.

2. Commercial PC and PMMA, of unspecified grade, represent thermoplastics, and have been indented as received.

# 4. Nanoindentation experiments

We carry out nanoindentation employing a Hysitron nanoindenter (model: TI Premier, Bruker corporation™) with a Berkovich tip (radius ~ 100 nm), at two maximum load levels, $P_{\mathrm{m}}$ (1000 μN and 9000 μN), and at two relative loading rates, $\dot{P}/P$ = 0.1 /s and 1.0 /s. The loading is followed by hold (70 s for cross-linked SU-8, 100 s for PC, 150 s for PMMA), concluding with 2 s of unloading, for all indentations. We have performed four indents at each load. Immediately after withdrawing the load after each indent, the residual indent impressions have been consecutively imaged 3 times with the same tip via SPM (making the scans ~ 9 min apart), at a frequency of 0.8 Hz.

# 5. Results and discussion

In our framework to employ uniaxial compression to understand nanoindentation deformation:

a) Data from previous DC uniaxial compression experiments on highly cross-linked SU-8 micropillars [14] and uniaxial bulk compression of PC [38], [39] and PMMA [40]–[43], have been re-examined, in terms of residual strains across various strain rates.

b) For each of our LC nanoindentations at *set* $\dot{P}/P$ we have estimated the hypothetical residual indent profile after high rate unloading (post elastic recovery).

c) We have then combined the results of steps ‘a)’ and ‘b)’ above, to relate the indentation depth and the recovery, to the equivalent uniaxial strain state and recovery, respectively. We ensure that all our analyses are based on the *actual* load and displacement and on the corresponding actual rates.

We explain first, the necessity of the apparent dichotomy – of carrying out nanoindentation in LC mode and interpreting it via uniaxial compression performed in DC mode. Nanoindentation experiments need to be via LC mode, primarily because:

1. The LC mode is a direct operation, whereas the DC mode is an indirect operation
2. In DC nanoindentation unloading, the indented material loses contact with the tip at $h_f > 0$.
3. In DC nanoindentation, the unload begins during the hold period, with $dP/dh|_{P_\mathrm{m}} \to \infty$. This makes estimating $S$ difficult. However, hold is necessary to remove viscoelastic effects.

In contrast, uniaxial compression data are required in DC mode because:

1. DC compression provides a clearly discernible $\sigma_y$, in terms of the maxima, the post-yield inflection and the “flat”, constant stress region.
2. LC compression would lead to very high post yield strain rates. This would prevent assignment of a clear strain rate to the uniaxial deformation, for relating to the equivalent nanoindentation strain rate. In this framework, the recovery is measured in terms of strain. Hence, the actual strain in uniaxial compression and the recovery thence, are the relevant parameters.

We begin with an examination, for our glassy amorphous polymers, of the uniaxial compression data available in the literature.

## 5.1. Analysis of existing uniaxial compression data

Detailed uniaxial compression experiments on highly cross-linked SU-8 micropillars after toe-correction [47], over a strain rate range, $8.31\times10^{-4}$ /s to $3.42\times10^{-3}$ /s, have been reported previously [14]. These micropillars were fabricated via materials and methods similar to those employed for the SU-8 nanoindentation samples in this work.

For PC and for PMMA, we have examined bulk uniaxial compression experiments from various sources, over various ranges of strain rates. We recognize here, that the PC and PMMA samples examined in the literature, would be of grades, different from those of our nanoindentation samples.

Of the multiple sources for uniaxial compression data for PC, we consider only the data set [38], [39], encompassing an engineering strain rate range, $10^{-3}$ /s to $10^{-1}$ /s. Similarly, for PMMA, we consider data at various set engineering strain rates (over the combined range, $10^{-4}$ /s to 1 /s) [40]–[43]. There are only limited data, at set engineering strain rates (or displacement rates).

The engineering strain rate basis, makes our analysis consistent with that for cross-linked SU-8. The other reason, is that one can analytically interchange between true strain and engineering strain as well as the corresponding rates, only at a given strain (e.g., $\varepsilon_y$). This is because, true strain, $\varepsilon_{\mathrm{t}} = \ln(1+\varepsilon)$, which means that engineering strain, $\varepsilon = \exp\varepsilon_{\mathrm{t}} - 1$, and engineering strain rate, $\dot{\varepsilon} = \dot{\varepsilon}_{\mathrm{t}} \exp\varepsilon_{\mathrm{t}}$. Both, $\dot{\varepsilon}$ and $\varepsilon$, are simultaneously related to both, $\dot{\varepsilon}_{\mathrm{t}}$ and $\varepsilon_{\mathrm{t}}$. Thus, during mapping (Section 5.4), $\sigma_y$ and $E$ at one type of strain rate (over the entire strain range) cannot be analytically mapped, exclusively, to another type of strain rate.

From the DC uniaxial micro-compression (SU-8) and bulk compression (PC and PMMA) stress-strain data, we generate normalized stress-strain plots (Fig. 2) for the three polymers, i.e., $\sigma^* = \sigma/\sigma_y$ vs. $\varepsilon^* = \varepsilon/\varepsilon_y$ (a type of *CSS* described in section 1), for all rates.

The variation of $\sigma_y$ with $\ln\dot{\varepsilon}$ is a well reported phenomenon; e.g., see [48], [49]. Similarly, a broadly monotonic, direct correlation is also expected between $\ln\dot{\varepsilon}$ and the compressive modulus, $E$. Then, in this work, we retain generality in our framework, by considering a linear relationship between $\ln\dot{\varepsilon}$ and $\varepsilon_y$ (eqn. 1), (Fig. 3).

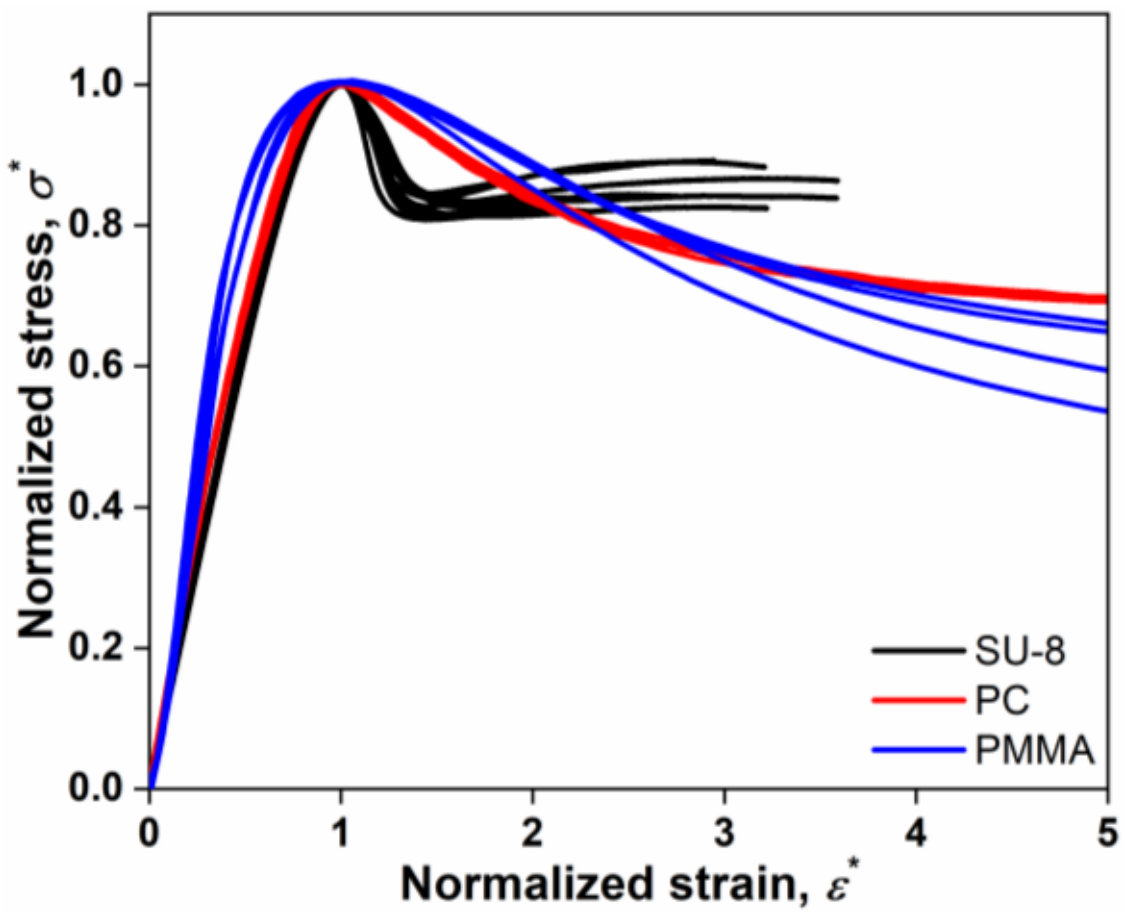

**Figure 2**: *CSS* $\sigma^*$-$\varepsilon^*$ curves for cross-linked SU-8 (black curves), PC (red) and PMMA (blue). For clarity, we have shown the PMMA *CSS* curves, of only the data by Ames *et al*. [42].

$$\varepsilon_y = p\ln\dot{\varepsilon} + q \qquad \text{eqn. 1}$$

The linear fit parameters for the three polymers are listed in Table 1.

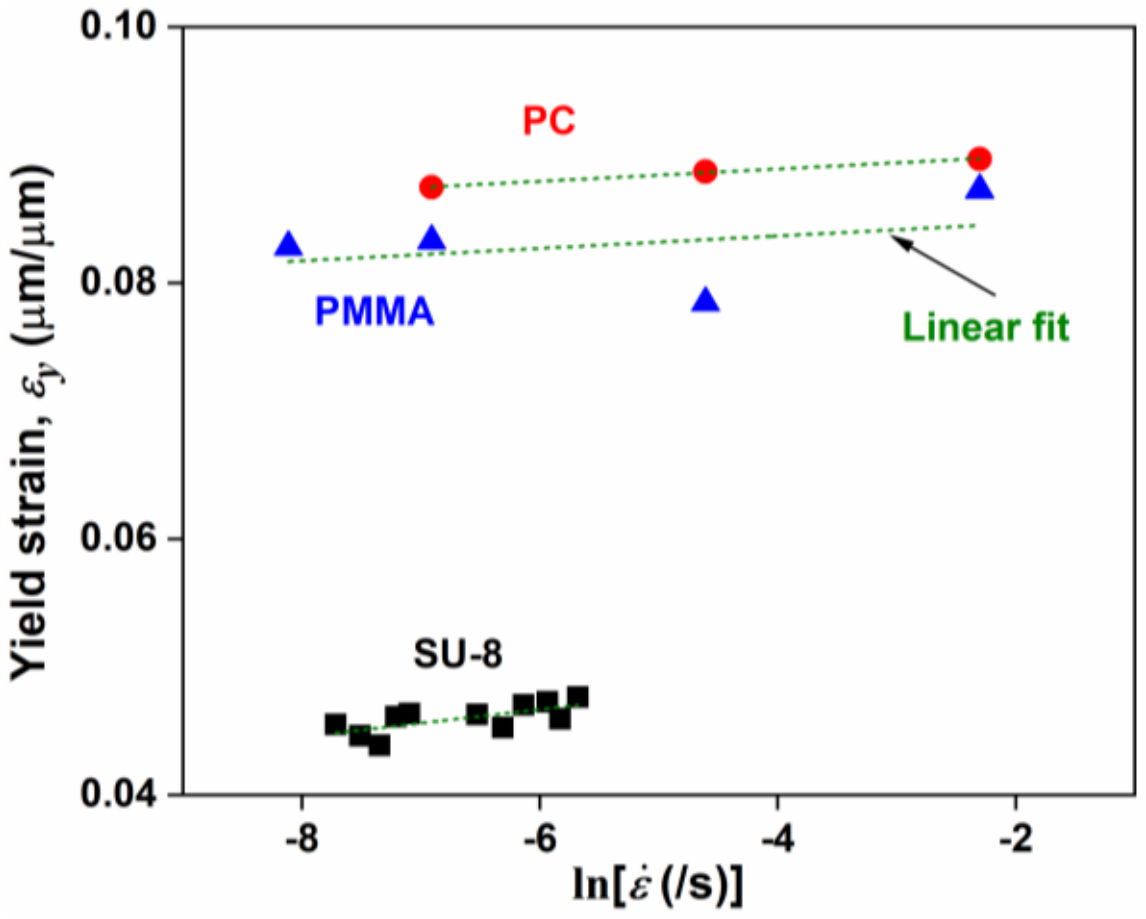

**Figure 3**: Variation of $\varepsilon_y$ with $\ln\dot{\varepsilon}$ for cross-linked SU-8, for PC and for PMMA. For PMMA, we have shown only the data by Ames *et al*. [42].

Based on the *CSS* measure, $\varepsilon^* = \varepsilon(\dot{\varepsilon})/\varepsilon_y(\dot{\varepsilon})$, we determine the non-elastic residual strain, $\varepsilon_{\mathrm{ne}}$ [25], [50], [51], from each of these curves. $\varepsilon_{\mathrm{ne}}$ are obtained for the entire strain-range, by dropping a Hookean line of slope $E(\dot{\varepsilon})$, from the corresponding point on the stress-strain plot, as shown in Fig. 1. Thus, the normalized residual strain is $\varepsilon_{ne}^*$, and $\varepsilon_e^*$ is the normalized elastic strain (eqn. 2).

$$\varepsilon_{ne}^* = \varepsilon^* - \varepsilon_e^* = \varepsilon^* - \frac{\sigma(\dot{\varepsilon})}{E(\dot{\varepsilon})\varepsilon_y(\dot{\varepsilon})} \quad \text{eqn. 2}$$

**Table 1**: Parameters fitted to eqn. 1 for all polymers.

| | $p$ | $q$ | Constant $\varepsilon_y$ |
|---|---|---|---|
| Cross-linked SU-8 | $9.46\times10^{-4}$ | $5.23\times10^{-2}$ | 0.045 |
| PC | $4.81\times10^{-4}$ | $9.08\times10^{-2}$ | 0.089 |
| PMMA [40] | $4.18\times10^{-3}$ | $11.2\times10^{-2}$ | 0.094 |
| PMMA [41] | $-2.52\times10^{-3}$ | $7.44\times10^{-2}$ | 0.092 |
| PMMA [42] | $4.86\times10^{-4}$ | $8.56\times10^{-2}$ | 0.083 |

There is a single source each, for PC and cross-linked SU-8 data. There are multiple sources for PMMA data. These enable examination of the effect of the grade in terms of specific features of the stress-strain plots, on the $\varepsilon_{ne}^*$ vs. $\varepsilon^*$ plots. We combine Fig. 2 and eqn. 2, to estimate $\varepsilon_{ne}^*$ for any strain. For each of the three polymers, over the strain rate range of their respective data, *CSS* behaviour (Fig. 4(a)) contains polymer-specific aspects and aspects common to all polymers.

These aspects include:

1. We find that qualitatively, the PC and PMMA correlations are similar to each other (no inflection exhibited) and different from the correlation for cross-linked SU-8 (inflection at $\varepsilon^* \sim 1.2$).
2. For *CSS* up to the elastic limit, $\varepsilon_0^*$, there is no residual strain, i.e., $\varepsilon_{ne}^* \approx 0$. For any one source, over various strain rates, there is only a slight variation in $\varepsilon_y$, and in $\varepsilon_0^*$ (~ 0.5 for cross-linked SU-8, $\varepsilon_0^* \sim 0.3$ for PC, and for PMMA, $\varepsilon_0^* \sim 0.3$ [41],[42], or $\varepsilon_0^* \sim 0.4$ [40]). Therefore, the corresponding $\varepsilon_{ne}^*$ vs. $\varepsilon^*$ plots agree well with each other. We note from the stress-strain data, that the grades of PMMA samples from different sources [41],[42], differ in $\sigma_y$, $\varepsilon_y$, *E*, and the strain-hardening behaviour. However, these differences, do not affect the correspondence in their $\varepsilon^*_{\rm ne}$ plots. The qualitatively similar *E*, $\varepsilon^*_0$ and $\varepsilon_y$, ensure that the differences in the recoveries, due to the differences in post-yield stress (i.e., hardening), do not significantly affect the unrecovered strain. However, differences in $\varepsilon^*_0$ (PMMA samples of refs. [41],[42] vs. those of ref. [40]), shift the $\varepsilon^*_{\rm ne}$ curve laterally. Still, the plots of the two thermoplastics are broadly similar, in that, they do not exhibit any discernible inflection.

3. For *CSS* beyond the elastic limit, we define $\Phi = \varepsilon^* - \varepsilon^*_0$; Fig. 4(b) indicates that the normalized recovery curves merge at high $\Phi$. However, they are polymer specific at low $\Phi$. Thus, we visually divide the behaviour into two regions. In the first region, from $\varepsilon^*_0$ up to approximately $\varepsilon^*$ = $\varepsilon^*_{\rm L}$~1.5 for cross-linked SU-8 and $\varepsilon^*$ = $\varepsilon^*_{\rm L}$~1.2 for PC and for PMMA, the relationship can be fit to a polynomial in $\Phi$ as per eqn. 3.

$$\varepsilon^*_{\rm ne} = \sum_{i=2}^{6} k_i \Phi^i \qquad \text{eqn. 3}$$

For the region, $\varepsilon^* > \varepsilon^*_{\rm L}$, the relationship is approximately linear; i.e.,

$$\varepsilon^*_{\rm ne} = m_{\rm L}\varepsilon^* + k_{\rm L} \qquad \text{eqn. 4}$$

Details of the analyses are shown in Supporting Information 1. We note here that the location of $\varepsilon^*_{\rm L}$ is approximate, and has been identified visually. The choice of a linear relationship for $\varepsilon^* > \varepsilon^*_{\rm L}$, is to facilitate the numerical analysis in section 5.4. The fitted polynomial parameters (eqn. 3) and the linear fit parameters (eqn. 4) are listed in Table 2. The polynomial region beyond $\varepsilon^*$ = 1, is typically the softening region, where the stress decreases during extension. For PC and PMMA, $\varepsilon^*_{\rm L}$ is effectively the inflection point in Fig. 4. For SU-8, the inflection point is at $\varepsilon^* \sim 1.2$. The possible physical significance of $\varepsilon_L^*$ will be discussed in a sections 5.4 and 5.6.

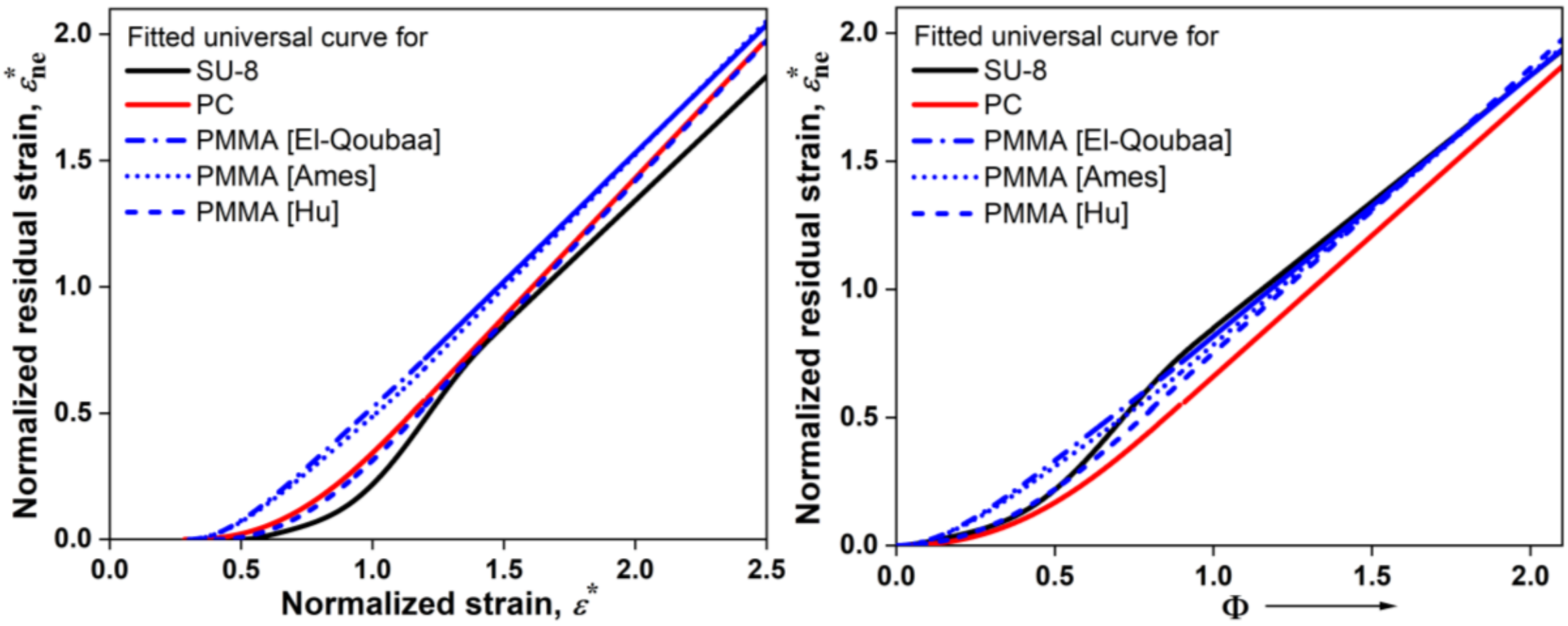

**Figure 4**: Fitted universal curve for *CSS* $\varepsilon_{ne}^*$ for cross-linked SU-8, PC and PMMA. Similar correlation for two PMMA samples by El-Qoubaa [41] and Ames [42] with similar $\varepsilon_0^*$, but different $\varepsilon_y$ and different quantitative correlations in case of different $\varepsilon_0^*$ of Hu [40] and Ames [42]: (a) vs. $\varepsilon^*$. Lateral shifts correspond to effects of $\varepsilon_0^*$ and (b) vs $\Phi = \varepsilon^* - \varepsilon_0^*$. Broadly similar behavior after removal of the lateral shift, with polymer-specific behavior at lower $\Phi$

**Table 2:** Fitted parameters values of universal residual strain curves.

| | $k_6$ | $k_5$ | $k_4$ | $k_3$ | $k_2$ | $k_L$ | $m_L$ |
|---|---|---|---|---|---|---|---|
| Cross-linked SU-8 | 8.57 | -25.58 | 26.21 | -10.67 | 2.32 | -0.65 | 1.00 |
| PC | -- | -- | -0.49 | 0.72 | 0.44 | -0.77 | 1.10 |
| PMMA [40] | -- | -- | -0.407 | 0.337 | 0.812 | -0.804 | 1.11 |
| PMMA [41] | -- | -- | 0.996 | -2.50 | 2.328 | -0.504 | 1.017 |
| PMMA [42] | -- | -- | 1.082 | -2.505 | 2.216 | -0.5885 | 1.056 |

## 5.2. Nanoindentation data analysis

This section comprises two parts. In Part I, we analyse the *P-h* data. In Part II, we examine the SPM imaging data and discuss the “internal” upflow (indicated by these data) and its consequences.

# I. Analysis of Nanoindentation *h-P* data

The $P$-$h$ curves for all indentations at 0.1 /s and 1.0 /s $\dot{P}/P$ for all three polymers, indicate excellent reproducibility (Supporting Information 2). We consider here, the deformation response, in terms of the unloading $P$-$h$ data. For the unloading step, the tip begins from rest, and accelerates to reach the *set* unloading rate at $\sim 0.95\,P_{\mathrm{m}}$. Deceleration begins on reaching $\sim 0.3\,P_{\mathrm{m}}$, and the subsequent decelerating recovery, up to $h_f$, the residual depth at $P$ = 0 µN (obtained from the nanoindentation data), includes significant VE effects (See Supporting Information 3).

To clearly separate out the various contributions to deformation, it is necessary to determine the hypothetical non-elastic residual penetration depth, $h_{\mathrm{ne}}$, obtained after purely elastic recovery, $h_{\mathrm{e}} = h_{\mathrm{m}} - h_{\mathrm{ne}}$ (at the *set* unloading rate). Separation of phenomena requires that the first part of the recovery should be purely elastic (instant recovery). As described previously [44] and summarized here, we consider two options for estimating the elastic recovery:

1. <u>Modelling the unloading *h-P* data via generalized power law (GPL) method</u>:

   We model the unloading *h-P* data, only within the range, from $\sim 0.95\,P_{\mathrm{m}}$, up to $\sim 0.3\,P_{\mathrm{m}}$. Conventional methods [10]–[12] employ the power law (PL) model (written in log-log form as $\ln P = m \times \ln\left(h - h_{\mathrm{ne}}\right) + \ln g$). In our work, $h_{\mathrm{ne}2}$ is the residual non-elastic displacement is based on 2s unloading time. For polymers in this work, we recognize that 1/*m*, the instantaneous slope in the $\ln\left(h - h_{\mathrm{ne}2}\right)$ vs $\ln P$ correlation, varies with $\ln P$, either linearly or quadratically. Therefore, we employ the GPL expression, $h = gP^{\left(k_2 (\ln P)^2 / 3\right) + (k_1 \ln P/2) + k_0} + h_{\mathrm{ne}2}$. The fitted parameters are $k_0$, $k_1$, $k_2$, $g$ and $h_{\mathrm{ne}2}$. For the linear variation of $1/m$ with $\ln P$, $k_2$ = 0.

2. Estimating Data for Instant Unload:

The variation in $1/m$ with $\ln P$, during the unloading over 2s, suggests that the expected constant rate elastic unloading, is compromised by VE effects. We observe these effects explicitly via SPM, in the time-dependent recovery of the indentation profile, as described in Part II. To circumvent these effects, we consider the top 10% of the constant rate unloading curve, i.e., from $\sim 0.95\,P_{\mathrm{m}}$, up to $\sim 0.885\,P_{\mathrm{m}}$. We model the unloading *h-P* data in this range via the PL framework, i.e., $h = gP^{(1/m)} + h_{\mathrm{ne}0}$. $h_{\mathrm{ne}0}$ is the residual non-elastic displacement, based on effectively instant (zero time) unloading. We find $m \sim 1.25$, across all the polymers, as well as across, both $\dot{P}/P$ and $P_{\mathrm{m}}$. This is consistent with the elastic unloading of EP materials considered in the OP framework development [10]–[12]. Hence, unload to $h_{\mathrm{ne}0}$, could be instantaneous elastic recovery.

Figure 5 indicates the two fits of the unloading *h-P* data (from $\sim 0.95\,P_{\mathrm{m}}$, up to $\sim 0.3\,P_{\mathrm{m}}$, and from $\sim 0.95\,P_{\mathrm{m}}$, up to $\sim 0.885\,P_{\mathrm{m}}$), and the estimates for $h_{\mathrm{ne}2}$ and $h_{\mathrm{ne}0}$. The accuracy of these methods is established by the very low relative residuals (< 0.05%), throughout the corresponding fitting ranges. The fitted GPL parameters for 2 s unloading as well as the PL parameters for zero-time unloading, are listed in Supporting Information 4. Table 3 lists the maximum displacement and the hypothetical residual displacements. Both methods yield identical values for the stiffness, *S*.

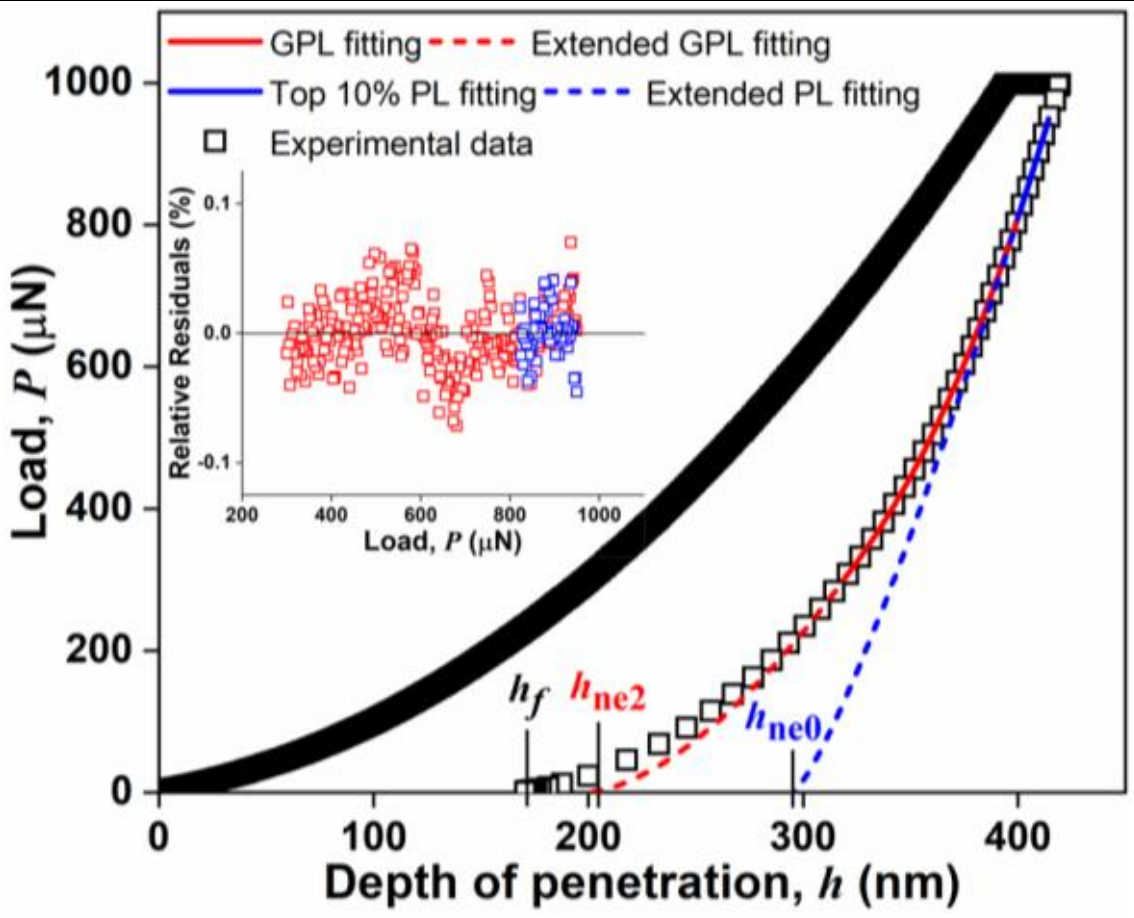

**Figure 5**: *P-h* curve for $P_m = 1000$ μN for cross-linked SU-8. Unload *h-P* data, fits (details in text) extended to the *h*-axis, $h_{ne0}$ and $h_{ne2}$. Inset: relative residuals of both fits.

**Table 3**: The maximum and hypothetical residual nanoindentation displacements for glassy polymers

| | $\dot{P}/P$ (/s) | $P_m$ (μN) | $h_m$ (nm) | $h_{ne2}$ (nm) | $h_{ne0}$ (nm) |
|---|---|---|---|---|---|
| Cross-linked SU-8 | 0.1 | 1000 | 420 ± 1 | 209 ± 4 | 297 ± 1 |
| | | 9000 | 1374 ± 7 | 704 ± 13 | 1023 ± 8 |
| | 1.0 | 1000 | 423 ± 6 | 200 ± 5 | 299 ± 5 |
| | | 9000 | 1366 ± 7 | 536 ± 8 | 1021 ± 7 |
| PC | 0.1 | 1000 | 563 ± 4 | 337 ± 11 | 432 ± 6 |
| | | 9000 | 1860 ± 16 | 1144 ± 15 | 1440 ± 17 |
| | 1.0 | 1000 | 559 ± 3 | 348 ± 17 | 427 ± 7 |
| | | 9000 | 1816 ± 6 | 981 ± 10 | 1393 ± 8 |
| PMMA | 0.1 | 1000 | 478 ± 8 | 331 ± 10 | 372 ± 11 |
| | | 9000 | 1537 ± 10 | 1039 ± 19 | 1195 ± 9 |

The choice of the appropriate estimate of the elastic recovery ($h_{ne0}$ or $h_{ne2}$) vis-à-vis uniaxial deformation, requires a comparison between the nanoindentation modulus, $E_N$, and the uniaxial deformation modulus, *E*. In addition to stiffness estimates, estimating $E_N$ requires an accurate estimate of the actual, load-bearing contact area, $A_c$. This estimate is obtained from the SPM imaging of the residual nanoindentation imprint.

## II. Analysis of SPM Data

Along with the *P*-*h* data, the post-unloading SPM images provide insights into the deformation phenomena. We first examine the top view of the residual imprint after unloading, as shown in Fig. 6 for cross-linked SU-8 (imprints for all three polymers, are available in Supporting Information 5).

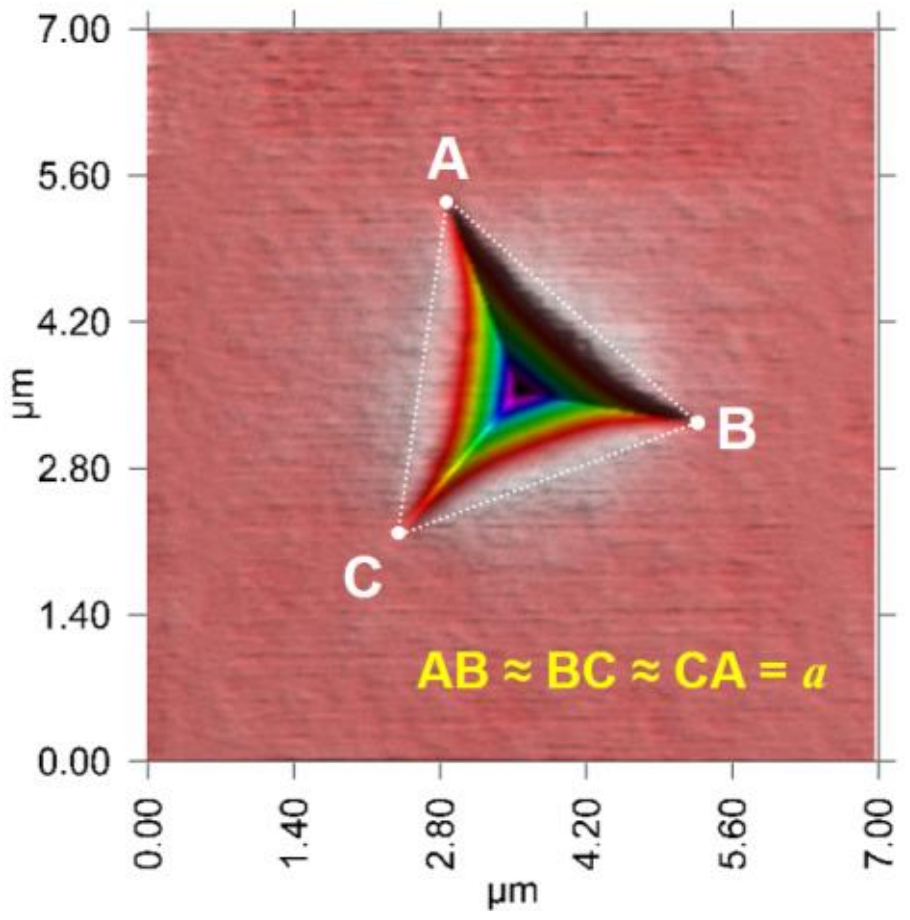

**Figure 6**: Top view of residual imprint for cross-linked SU-8. The ends of the edge imprints at full load form an equilateral ΔABC, of side *a*.

Figure 6 is the image is of a residual imprint, which is due to the residual non-elastic displacement corresponding to VEP downflow beyond the elastic displacement. As described above, during unload and over time at *P*=0, part of this depth is recovered, but part remains as a permanent residual deformation. SPM imaging of the residual imprint of the indent requires ~ 9 min after unloading. Thus, the SPM image profile, corresponds to the indented region ~ 6 min after complete unload. The sharp edges of the indenter cause yield at contact with the sample, giving rise to permanent line imprints. The outermost tips of the permanent imprints, form the corners of the contact area. This principle is consistent with our observation that the $a$ value (corner to corner distance of the ΔABC in Fig. 6) is constant for more than 12 hrs. after indentation [52], [53]. The contours are concave outwards, causing previous researchers to consider the lower penetration but still concave contours, as contact boundaries [54], [55], with the resulting contact area estimate becoming smaller than that of ΔABC; $A_c < \sqrt{3}a^2/4$. It is important to note that the concave contours are of the residual imprint, post-unload, while $A_c$ should correspond to the contact at full load. To estimate this contact area, we first consider the vertical sections of the residual imprint.

Figure 7(a) is a representative image of a vertical SPM section. The maximum residual depth, $h_r$, corresponds to the reduced residual indentation, after additional zero-load recovery. We have plotted (Fig. 7(b)) the vertical SPM sections with the origin at the $z$ = 0 level, at the base of the *depth-axis*, i.e., at the centroid of horizontal tip section. We have indicated $h_{\mathrm{ne}2}$, $h_f$ and $h_r$, as per our experimental observations. Thus, even as the apex depth varies over that duration, $a$, in Fig. 6, remains constant. As indicated by the SPM sections, we make the following observations:

a) The tip of a real indenter is never an ideal point of size zero. There is a blunt depth, $h_b$, which has been estimated by modifying the fitting analysis of the contact area calibration with the quartz standard (Supporting Information 6).

b) The $a$ values (Fig. 6) from the permanent imprints of the sharp edges of the Berkovich tip, yield the geometric contact height, $h_c^{\Delta} = h_c + h_b = a/2\sqrt{3}\tan 65.27°$. Thus, the permanent imprints unambiguously indicate there is sink-in at full load along the edges, and the sink-in depth, $h_s^{\Delta} = h_{\mathrm{m}}^{\Delta} - h_c^{\Delta}$, where $h_{\mathrm{m}}^{\Delta} = h_{\mathrm{m}} + h_b$. The sink-in is non-contact deformation, and is fully recovered at unload. For the remainder of this report, $h_c \Rightarrow h_c^{\Delta}$ [56].

c) There is *pile-up* (PU) along the facet. The normalized maximum pile-up, $\left(h_{\mathrm{m}}^{\mathrm{PU}}/h_{\mathrm{m}}\right)$, is ~ 1% for cross-linked SU-8, ~ 8% for PC and ~ 3% for PMMA. Since this pile-up is imaged post-unload, it occurs when the sunk-in depth has fully recovered. A previous geometry-based analysis [44] conveys that $\left(A_c^{\mathrm{PU}}/A_c^{\Delta}\right) < \left(h_{\mathrm{m}}^{\mathrm{PU}}/h_{\mathrm{m}}\right)$. Since $E_{\mathrm{N}} \propto A_c^{-0.5}$, we neglect pile-up here, since it reduces our $E_{\mathrm{N}}$ estimates by < 4%.

d) SPM data indicates that $h_{\mathrm{m}}^{\mathrm{PU}} > h_{s}^{\Delta}$ for PC, and $h_{\mathrm{m}}^{\mathrm{PU}} < h_{s}^{\Delta}$ for PMMA and for cross-linked SU-8. Thus, for PC, during loading, along with sink-in along the edge, pile-up occurs along the facet. For SU-8 and PMMA, the pile-up is likely to be during unloading only. In either case, the pile-up would correspond to a further reduced sink-in along the facet, vis-à-vis that along the edge.

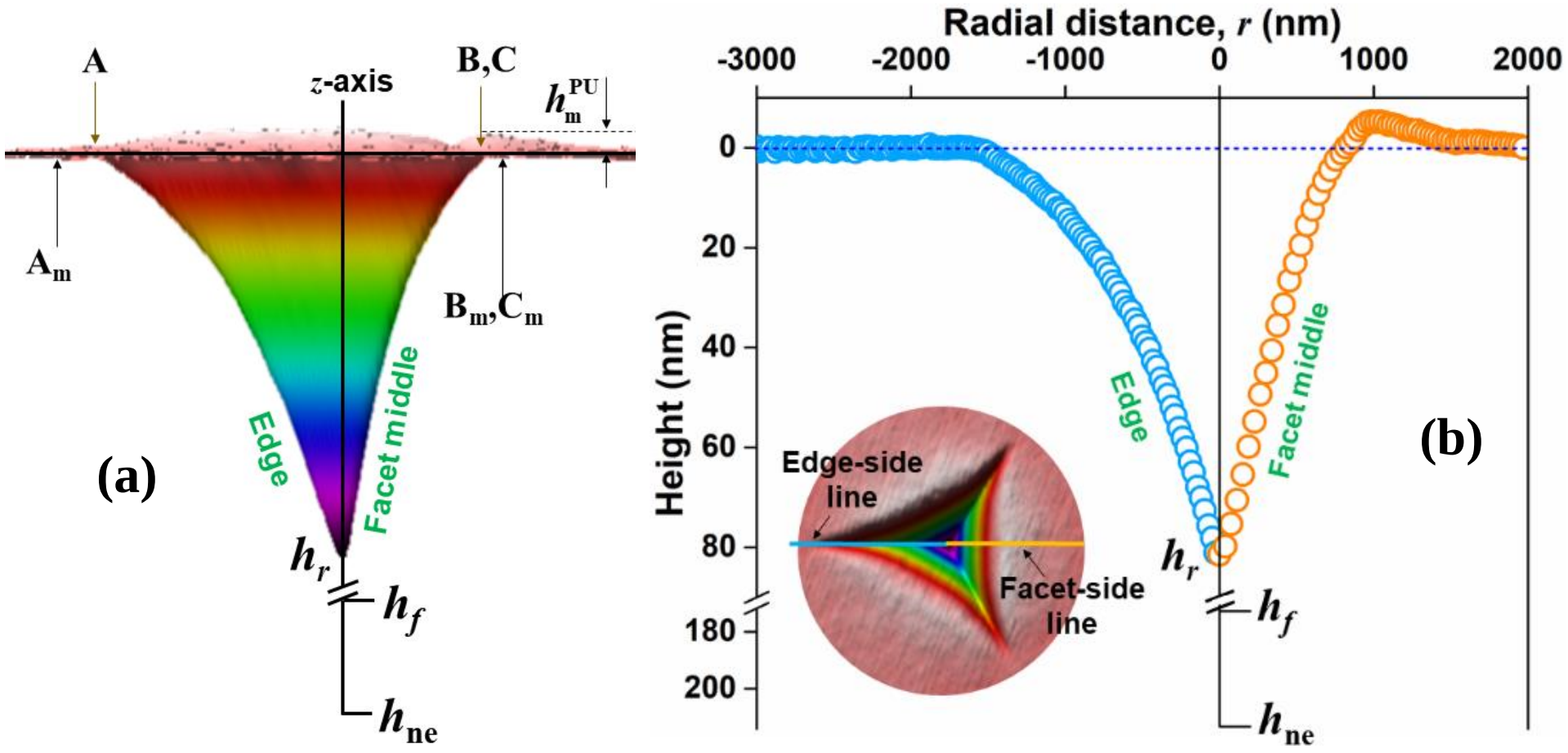

**Figure 7**: Residual indent impression for cross-linked SU-8 at $P_{\mathrm{m}}$ = 1000 µN and $\dot{P}/P$ = 0.1 /s (a) vertical view, (b) scan line which pass through the tip apex depth.

In addition to the contours being convex inward in Fig. 6, the depth profiles in Fig. 7, are convex upward. During nanoindentation, that contact displacement is vertically downward [57], [58], with the consequent recovery being vertically upward. Also, the permanent imprints due to stress concentrations along the edges, correspond to yield. Thus the edge displacement contains a lower recoverable elastic contribution than that away from the edges; i.e., the stress decreases from the edge to the middle of the facet [59]. Similarly, the tip apex deformation contains the greatest stress concentrations, corresponding to the least fractional elastic contribution. The elastic contribution increases with distance from the apex, as the stress concentrations decrease. Then:

(i) The fractional recovery away from the edges would be greater than that at the edges corresponding to the convex inward displacement contours in Fig. 6.

(ii) The fractional recovery increases with distance from the tip apex – in any direction, corresponding to the convex upward displacement profile in Fig. 7.

We summarize here an earlier [44] validation, that on neglecting the slight pile-up, the contact boundary is ΔABC. *In situ* full load optical views of Berkovich [60] and Vickers [61] indentation of transparent PDMS elastomers (from beneath the sample), indicate that the full load contact boundaries of soft elastomers are bowed inwards (greater sink-in along the facet). These are consistent with the *in situ* indentation video frames [62], that indicate that their sink-in along the facets begins further away from the tip surface, than that along the edges; also, there is no discernible post-unload pile-up. Such behavior leads to the concave outward contact boundaries, which were considered previously [54], [55] for glassy polymers. However, when the elastomer surface is coated with a hard Au-Pd layer, the sink-in begins uniformly around the indenter tip. This corresponds to an approximately linear contact boundary between the ends of the edge-imprints. Additionally, a recent [63] FE simulation has also predicted a linear contact boundary for PMMA indentation. Therefore, for glassy polymers, the contact region is ΔABC, with $A_c = \sqrt{3}a^2/4$.

We find that the $h_c/h_{\mathrm{m}}^{\Delta}$ ratios for our polymer materials, are greater than those obtained via the OP-type methods [10]–[12]. This too is consistent with the FE simulation finding [63] mentioned above. Pile-up occurrence means that the sink-in along the facet is further reduced, as compared to that along the edge. We attribute these reductions in sink-in to upflow, which we will discuss later in this section. $\left(h_s^{\Delta}/h_{\mathrm{m}}^{\Delta}\right) = 1 - \left(h_c/h_{\mathrm{m}}^{\Delta}\right)$, is usually considered to be a measure of the elastic contribution to the deformation, and is realized from the loading step. However, when there is upflow, for a given depth, the contact increases and sink-in decreases. Thus, $\left(h_s^{\Delta}/h_{\mathrm{m}}^{\Delta}\right)$ decreases due to reasons other than additional downward deformation under the indenter tip. Hence, for such cases, $h_s$ cannot be considered as the sole measure of the elastic contribution. Then, recovery, if estimated objectively, is a more reliable measure of the elastic contribution.

At the onset of unload (which occurs after an adequate hold period), there is no memory of the VE effects of the upflow during loading, except of the instantaneous effects in terms of the contact area and the state of strain. This means that recovery is governed by this distribution of the deformation under the tip, and by the contact at that load. The strain in the layers beneath the tip, represents the events at the tip apex, such as displacement and recovery, and the effects of upflow, in terms of increased contact and absorbed strain. Hence, the elastically recovered displacement on unloading, $h_{\mathrm{e}} = h_{\mathrm{m}} - h_{\mathrm{ne}}$ ($h_{\mathrm{ne}} = h_{\mathrm{ne}0}$ or $h_{\mathrm{ne}2}$), and the elastic fraction, $h_{\mathrm{e}}/h_{\mathrm{m}}$, are the relevant measures of the elastic aspects of the deformation. We can estimate the likely sink-in, $h_s^{\mathrm{e}} = h_{\mathrm{e}}(\pi - 2)/\pi$, for recovery $h_{\mathrm{e}}$, had there been no upflow. This interpretation suggests that the upflow, $h_{uf} = h_s^{\mathrm{e}} - h_s^{\Delta}$ (illustrated in Fig. 8).

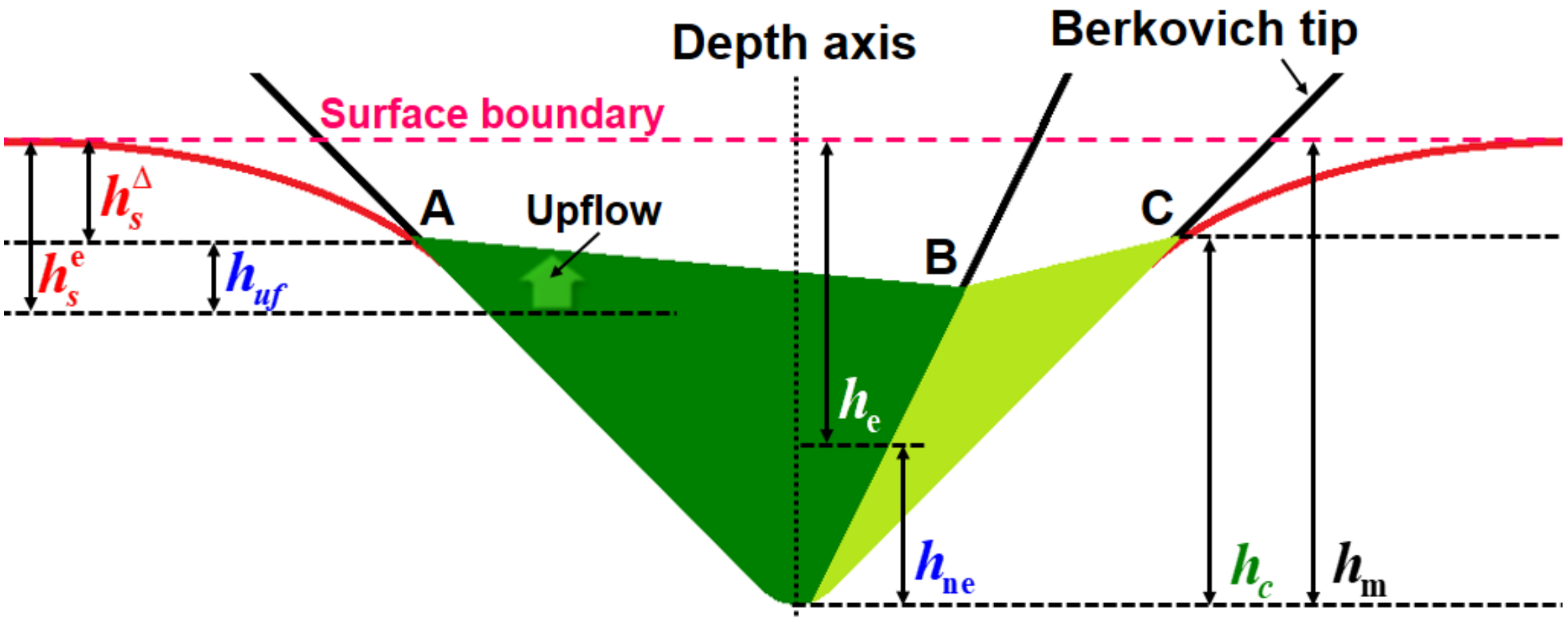

**Figure 8**: Schematic of elastic and anelastic displacements considering existing upflow events.

In our analyses, we employ the explicitly obtained $A_c$ via SPM imaging, as opposed to that via the indirect method of the OP framework [10]–[12]. Table 4 lists the $a$ values, the corresponding $h_c/h_{\mathrm{m}}^{\Delta}$, the estimated $h_{uf}/h_{\mathrm{m}}$ and the resultant elastic fraction, $h_{\mathrm{e}}/h_{\mathrm{m}}$. We find that $h_c/h_{\mathrm{m}}^{\Delta}$ is independent of rate. It depends slightly on $h_{\mathrm{m}}$ for SU-8, but not for PC and PMMA. The significant upflow (not accounted for by the LS and OP developments) which causes a greater $A_c$, is possibly a characteristic of compliant materials such as glassy polymers. $h_{uf}/h_{\mathrm{m}}$ exhibits a mixed correlation with deformation rate (decreases with $\dot{P}/P$ at low $P_{\mathrm{m}}$ increases with $\dot{P}/P$ at high $P_{\mathrm{m}}$) and a decreasing correlation with $P_{\mathrm{m}}$. Hence, the upflow is a result of both, solid-like and fluid-like, physically governed volume-conserving phenomena (illustrated in Fig. 9(a)-(b)). These effects are:

(i) In terms of solid deformation, the upflow corresponds to a form of transferred or secondary Poisson expansion. The primary Poisson deformation at the tip surface (i.e., inner regions), is directed laterally away from the tip. This lateral expansion of the inner regions tends to laterally compress the outer regions. This lateral compression would cause a secondary vertical Poisson extension. Therefore, away from the contact, the lateral Poisson expansion of the inner regions, "turns" upward, toward the unconstrained sample surface above (the path of least resistance), resulting in upflow. Friction-effects, reported in [64], [65], which could reduce the Poisson effect, act only at horizontal contact with the tip (as would be expected in case of a flat punch indenter). In case of a Berkovich indenter, the only horizontal contact, is the tangential point contact at the blunted tip apex. The lateral Poisson expansion displacement is directed away from the apex the tip surface, thus losing contact, and precluding any surface friction there.

(ii) In terms of the flow effects, the layer wetting the tip exhibiting yield flow, is constrained to move downward because of the downward force imposed by the tip. This has been clarified by Chaudhri *et al.* [66], [67], for conical indentation, and would be analogous for Berkovich indentation. Thus, the yielded deforming contact region of the surface, moves axially downwards, along with the tip surface. Then the downward flowing contact layers, are continuous with the layers just beyond the contact layer; the latter oppose the downward motion, so as to undergo volume conserving *laminar-like* upflow towards the $z = 0$ surface. The net result is an upflow, as depicted in Fig. 9(b).

Table 4 indicates that $h_{\mathrm{e}}/h_{\mathrm{m}}$ depends on rate, as well as on $h_{uf}/h_{\mathrm{m}}$. As depicted in Fig. 1(b) and described in section 5.4, increasing $h_{\mathrm{e}}/h_{\mathrm{m}}$ is very well correlated with a decrease in $\varepsilon$. This increase in $h_{\mathrm{e}}/h_{\mathrm{m}}$ due to a combination of the upflow and high rate, can be understood as follows:

(i) The increased upflow, absorbs part of the downward compressive strain in the elements in contact with the tip, which affects the strain distribution in the material.

(ii) During a slow loading step, there is time available for the flow component of the upper contact layers, to absorb part of the elastic deformation of the lower layers. Hence, the increase in load, compresses more, the already yielded and therefore, more pliable contact layers at lower rates (higher deformation gradient with depth), than it does at higher rates (Fig. 10(a)). Rapid deformation compresses more layers (greater pervaded deformed depth), but does so relatively equitably (lower deformation gradient with depth), as shown in Fig. 10(b). Consequently, the strain in the layers below the tip, decreases gradually, over a greater depth. Hence, lesser yield occurs, with more layers that are less strained and elastically compressed, for a similar $h_{\mathrm{m}}$.

**Table 4:** Nanoindentation SPM Analysis – contact *a* values, $h_c/h_{\mathrm{m}}^{\Delta}$, fractional upflow, $h_{uf}/h_{\mathrm{m}}$, and recovery-based elastic contribution estimates ($h_{\mathrm{e}}/h_{\mathrm{m}}$, with $h_{\mathrm{ne}} = h_{\mathrm{ne}\,2}$).

| | $\dot{P}/P$ (/s) | $P_{\mathrm{m}}$ (μN) | $a$ (μm) | $h_c/h_{\mathrm{m}}^{\Delta}$ | $h_{uf}/h_{\mathrm{m}} \times 10^2$ | $h_{\mathrm{e}}/h_{\mathrm{m}}$ |
|---|---|---|---|---|---|---|
| Cross-linked SU-8 | 0.1 | 1000 | 2.84 ± 0.01 | 0.86 | 4.1 ± 0.4 | 0.50 |
| | | 9000 | 9.65 ± 0.03 | 0.92 | 9.9 ± 0.8 | 0.49 |
| | 1.0 | 1000 | 2.83 ± 0.04 | 0.86 | 3.0 ± 0.3 | 0.53 |
| | | 9000 | 9.65 ± 0.02 | 0.93 | 11.2 ± 0.5 | 0.61 |
| PC | 0.1 | 1000 | 4.05 ± 0.03 | 0.93 | 7.5 ± 1.3 | 0.40 |
| | | 9000 | 13.24 ± 0.10 | 0.94 | 8.2 ± 0.4 | 0.39 |
| | 1.0 | 1000 | 3.96 ± 0.03 | 0.91 | 4.4 ± 1.3 | 0.43 |
| | | 9000 | 13.00 ± 0.06 | 0.94 | 9.7 ± 0.8 | 0.46 |
| PMMA | 0.1 | 1000 | 3.46 ± 0.03 | 0.93 | 4.5 ± 0.7 | 0.31 |
| | | 9000 | 10.94 ± 0.03 | 0.94 | 5.6 ± 0.7 | 0.32 |

In the next section, we first determine the appropriate measure of nanoindentation recovery (i.e., whether $h_{\mathrm{ne}\,2}$ or $h_{\mathrm{ne}\,0}$), based on the relationship between the uniaxial compressive modulus $E$ and the corresponding nanoindentation modulus, $E_{\mathrm{N}}$. This will enable development of a framework (Section 5.4) to interpret the nanoindentation strain and recovery, in terms of the equivalent uniaxial compression strain and recovery, respectively.

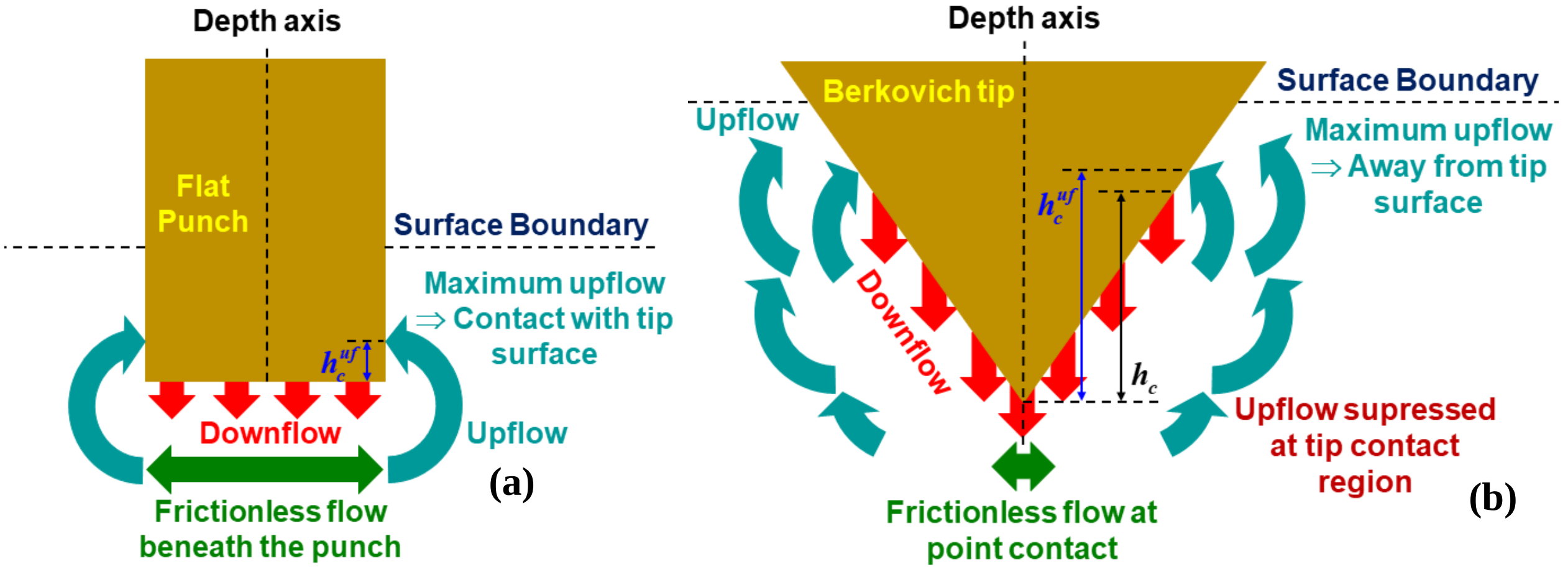

**Figure 9**: Schematic of frictionless deformation and upflow event (a) only due to upflow for flat punch (vertical surface contact), (b) due to both, downflow and upflow for Berkovich tip (inclined surface contact).

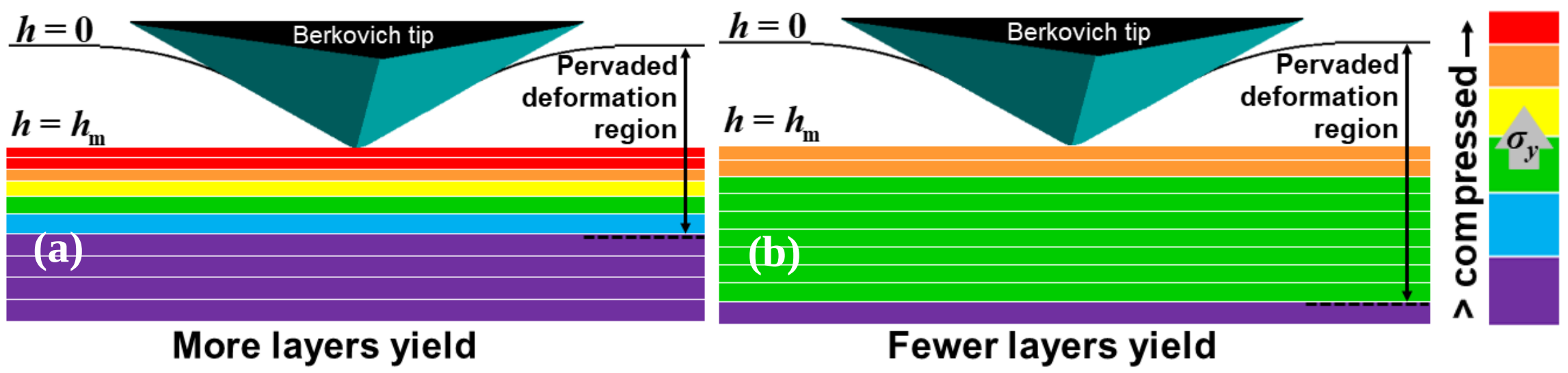

**Figure 10:** Schematic of the compressed layers under the tip apex at $P_{\mathrm{m}} = 9000$ µN, for $\dot{P}/P$ : (a) 0.1 /s: the pervaded deformation region is smaller, with fewer deformed layers, but with greater (yield) fraction, (b) 1.0 /s: the pervaded deformation region is greater, with more layers with recoverable, elastic deformation

## 5.3. $E_{\mathrm{N}}$ Effects on Mapping Nanoindentation Recovery

In this section, we first determine the conventionally defined nanoindentation modulus, $E_{\mathrm{N}}$, as described in Section 1. This involves employing the estimate of $S = dP/dh|_{P_{\mathrm{m}}}$, from the unloading $h$-$P$ data, and $A_c = \left(\sqrt{3}/4\right)a^2$, where $a$ is determined by SPM imaging (Table 4). $S$ estimates are identical, whether from the GPL modelling of the unloading $h$-$P$ data in the range from ~ 0.95 $P_{\mathrm{m}}$, up to ~ 0.3 $P_{\mathrm{m}}$, or from the PL modelling of that data from ~ 0.95 $P_{\mathrm{m}}$, up to ~ 0.885 $P_{\mathrm{m}}$.

Table 5 indicates that $E_{\mathrm{N}}/E$ ~1.1 to 1.9. The nanoindentation recovery measured in terms of indenter displacement, can be translated in terms of strain, via an indirect procedure, involving $\ln h$ and a correlation constant, *c*, as described in the next section. The equivalent uniaxial elastic recovery (residual depth = $h_{\mathrm{ne}0}$), corresponds to a strain recovery (Fig. 1-b) at a greater slope (~ $E_{\mathrm{N}}$ ), than that for the actual uniaxial deformation (recovery slope ~ *E*).

In order to mitigate potential errors due to the unavoidable mapping of unequal moduli, we deliberately add consistently reproducible VE effects to the elastic recovery; $h_{\mathrm{e}} = h_{\mathrm{m}} - h_{\mathrm{ne}2}$, is then the measure of the equivalent elastic recovery, for mapping to uniaxial compression. Also, since $h_{\mathrm{ne}2}$ is obtained by fitting a reasonable data range, it provides a reproducible equivalence correction. It is a balance, between accommodating VE effects, and accounting for $E_{\mathrm{N}}/E > 1$.

We explain this concept in terms of the elastic recovery and the stiffness. For the GPL framework which yields $h_{\mathrm{ne}2}$, the unloading *h-P* data correlation yields $S = dP/dh|_{P_{\mathrm{m}}} = \left(dh/dP|_{P_{\mathrm{m}}}\right)^{-1}$. Considering the PL framework, $P = B\left(h - h_{\mathrm{ne}}\right)^m$, $S = mP_{\mathrm{m}}/\left(h_{\mathrm{m}} - h_{\mathrm{ne}}\right)$. Then keeping $A_c$, $B$ and $m$ fixed, we can estimate the equivalent uniaxial modulus from the nanoindentation data, in terms of $E_{\mathrm{N}0}$ and $E_{\mathrm{N}2}$; $\left(E_{\mathrm{N}0}/E_{\mathrm{N}2}\right) = \left(h_{\mathrm{m}} - h_{\mathrm{ne}2}\right)/\left(h_{\mathrm{m}} - h_{\mathrm{ne}0}\right)$. This enables comparison with $E_{\mathrm{N}}/E$ (Table 5); i.e., $E_{\mathrm{N}0} \equiv E_{\mathrm{N}}$ and $E_{\mathrm{N}2} \equiv E$. See Fig. 11.

We recognize that this comparison ( $E_{\mathrm{N}0}/E_{\mathrm{N}2}$ vs $E_{\mathrm{N}}/E$ ) is coarse over the three polymers; it is a reasonable approximation for the case of SU-8, where, both, the nanoindentation and uniaxial compression experiments have been carried out on the similarly synthesized polymers, employing resins of similar grade. In subsequent analyses, as we correlate the nanoindentation strain and its recovery to the corresponding uniaxial states, we will consider $h_{\mathrm{ne}} = h_{\mathrm{ne}2}$.

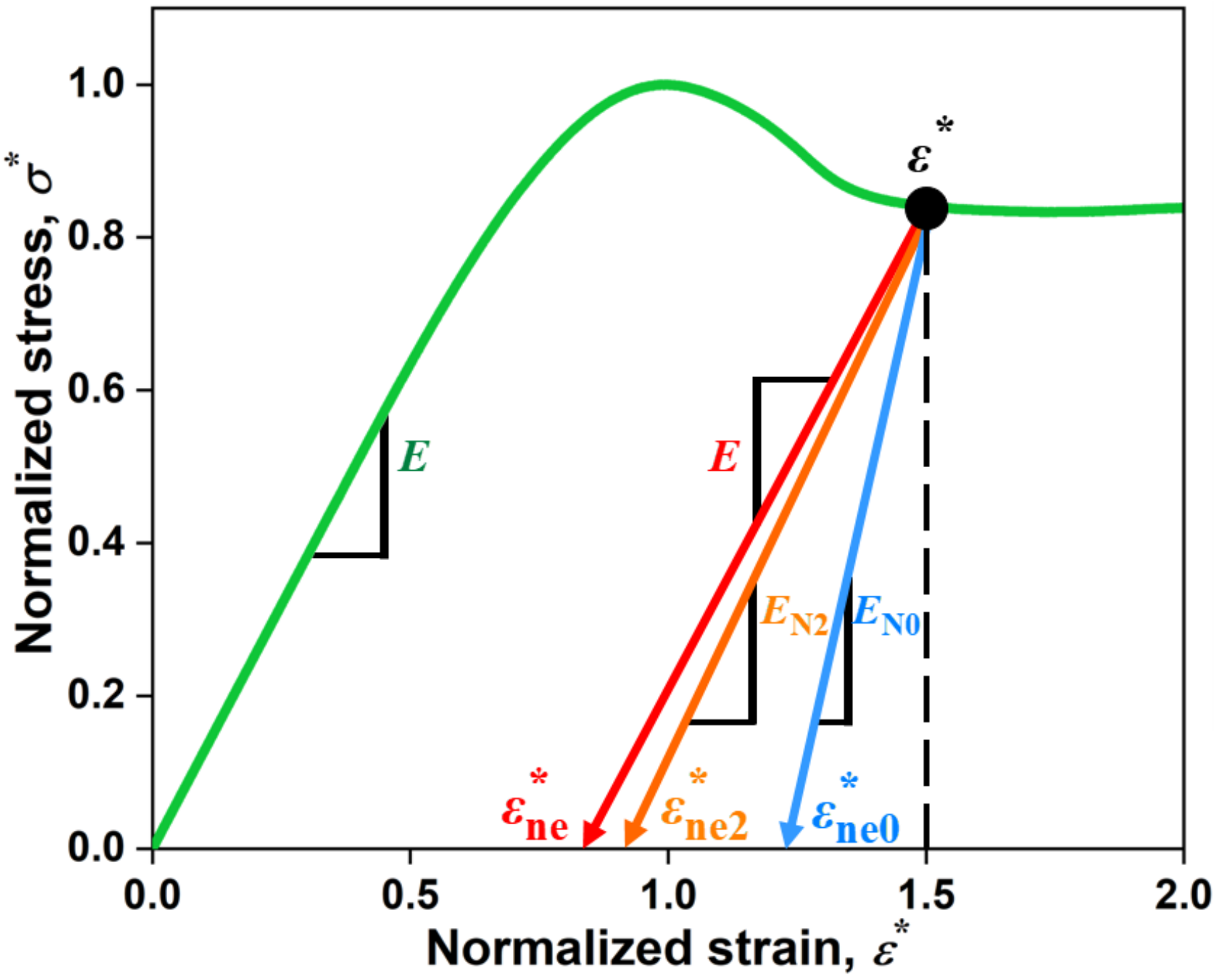

**Figure 11:** The choice $h_{\text{ne}} = h_{\text{ne}2}$, illustrated in terms of the uniaxial recovery slope – which should be the uniaxial modulus, $E \sim E_{\text{N}2}$.

**Table 5**: Effect of choice of recovery measure vis-à-vis $E_{\text{N}}/E$

| | $\dot{P}/P$ (/s) | $P_{\text{m}}$ (μN) | $E_{\text{N}}/E$ | $(E_{\text{N}0}/E_{\text{N}2})$ |
|---|---|---|---|---|
| Cross-linked SU-8 | 0.1 | 1000 | 1.87 ± 0.02 | 1.71 ± 0.02 |
| | | 9000 | 1.74 ± 0.01 | 1.91 ± 0.01 |
| | 1.0 | 1000 | 1.83 ± 0.03 | 1.79 ± 0.00 |
| | | 9000 | 1.81 ± 0.01 | 2.42 ± 0.03 |
| PC | 0.1 | 1000 | 1.22 ± 0.01 | 1.71 ± 0.04 |
| | | 9000 | 1.07 ± 0.01 | 1.74 ± 0.04 |
| | 1.0 | 1000 | 1.21 ± 0.01 | 1.58 ± 0.03 |
| | | 9000 | 1.12 ± 0.01 | 1.83 ± 0.03 |
| PMMA | 0.1 | 1000 | 1.39 ± 0.00 | 1.36 ± 0.00 |
| | | 9000 | 1.20 ± 0.01 | 1.46 ± 0.04 |

## 5.4. Correlation between nanoindentation and compressive strains

In this section, we exploit the separation of VEP nanoindentation loading deformation, into purely elastic recovery, followed by VE recovery and yield residue – to phenomenologically map the events, corresponding to the uniaxial phenomena. In uniaxial deformation, beyond the elastic limit, $\varepsilon_0^*$, recovery from uniaxial deformation is partial, and $\varepsilon_{\text{ne}}^*$ increases monotonically with $\varepsilon$.

In case of DC uniaxial compression, the strain rate, $\dot{\varepsilon}$, is proportional to the fixed displacement rate, $\dot{h}$ [14]. In case of LC nanoindentation, phenomenologically, the notional strain rate, $\dot{\varepsilon}_{\mathrm{N}} \sim \dot{h}/h$ [3]–[5], which can be correlated with $\dot{P}/P$ [6]. We also note that the nanoindentation strains, and residual as well as recovered strains, phenomenologically correspond to the true strain scale, because over time $t_1$ to $t_2$, $\Delta\varepsilon_{\mathrm{N}} \sim \ln(h_1/h_2)$; thus, $\dot{\varepsilon}_{\mathrm{N}} \sim \Delta\varepsilon_{\mathrm{N}}/\Delta t \sim \ln(h_1/h_2)/(t_1 - t_2)$. In addition, the nanoindentation stress is considered in terms of the instantaneous total projected contact area, which is analogous to the case of true stress. Therefore, the uniaxial data and property correlations, which are in terms of engineering stress and strain, need to be converted to true stress and strain terms.

During nanoindentation loading, $\dot{h}/h$ decays asymptotically from a very high value at $h$ ~ 0 (shown in Appendix A for cross-linked SU-8). However, from $h \sim h_{\mathrm{e}}$ $(= h_{\mathrm{m}} - h_{\mathrm{ne}})$ onwards, up to $h = h_{\mathrm{m}}$, $\dot{h}/h$ decreases by only ~ 2%. Nanoindentation beyond $h_{\mathrm{e}}$, would to correspond to the *CSS* beyond $\varepsilon_{\mathrm{e}}^*$. The corresponding full unload residual strain (at $h_{\mathrm{ne}}$), would correspond to the *CSS* at unload, $\varepsilon_{\mathrm{ne}}^*$. Thus, for any $h$ > $h_{\mathrm{e}}$, $\varepsilon_{\mathrm{N}} = \varepsilon_{\mathrm{N,e}} + \varepsilon_{\mathrm{N,ne}} = \varepsilon_{\mathrm{N,e}} + \ln(h/h_{\mathrm{e}})$.

The normalized elastic limit up to $\varepsilon_0^*$, is effectively independent of uniaxial strain rate (Fig. 2). The subsequent deformation until yield, also exhibits very slight dependence on rate. We then assume that this rate-independence extends to the elastically recovered *CSS*, $\varepsilon_{\mathrm{e}}^*$, which would be less than the normalized yield strain, $\varepsilon_y^* = 1$. $\varepsilon_{\mathrm{e}}^*$ corresponds to $\varepsilon_{\mathrm{N,e}}$, which is attained over the time period, during which, there is a significant variation in nanoindentation strain rate. Since this strain rate variation is inconsequential, $\varepsilon_{\mathrm{N,e}}$ will be considered rate-independent in the subsequent analysis.

As a first approximation, we consider that the equivalent uniaxial compressive true strain, is proportional to the nanoindentation strain; i.e., $\varepsilon_{\mathrm{t}} = c\varepsilon_{\mathrm{N}}$. This requires that, for the corresponding strain rates, $\dot{\varepsilon}_{\mathrm{t}} = c\dot{\varepsilon}_{\mathrm{N}}$ [2], [15]–[20]. Then, $\varepsilon_{\mathrm{t}} = \varepsilon_{\mathrm{t,e}} + \varepsilon_{\mathrm{t,ne}}$, means that

$$\ln \frac{h}{h_{\mathrm{e}}} = \varepsilon_{\mathrm{N}} - \frac{\varepsilon_{\mathrm{t,e}}}{c} = \varepsilon_{\mathrm{N}} - \frac{\varepsilon_{\mathrm{t}} - \varepsilon_{\mathrm{t,ne}}}{c} \qquad \text{eqn. 5}$$

As described in section 5.1, in terms of true strain, eqn. 1 becomes, $\varepsilon_y = p\left(\varepsilon_{\mathrm{t}} + \ln \dot{\varepsilon}_{\mathrm{t}}\right) + q$, which gives $\varepsilon^* = \left(\exp \varepsilon_{\mathrm{t}} - 1\right) / \left(p\left(\varepsilon_{\mathrm{t}} + \ln \dot{\varepsilon}_{\mathrm{t}}\right) + q\right)$.

At $P_{\mathrm{m}}$, the nanoindentation strain rate from the reference depth, $h_{\mathrm{e}}$,

$$\dot{\varepsilon}_{\mathrm{N}} = \frac{\ln\left(h_{\mathrm{m}}/h_{\mathrm{e}}\right)}{t_{\mathrm{m}} - t_{\mathrm{e}}} = \frac{\dot{\varepsilon}_{\mathrm{t}}}{c} \qquad \text{eqn. 6}$$

Hence, the nanoindentation strain,

$$\varepsilon_{\mathrm{N}} = \frac{\varepsilon_{\mathrm{t}}}{c} = \ln \frac{h_{\mathrm{m}}}{h_{\mathrm{e}}} + \frac{\varepsilon_{\mathrm{t}} - \varepsilon_{\mathrm{t,ne}}}{c} \qquad \text{eqn. 7}$$

Next, we correlate the residual nanoindentation strain (corresponding to $h_{\mathrm{ne}}$) with the residual uniaxial compression strain, $\varepsilon_{\mathrm{ne}}$ (see Appendix B and Supporting Information 7). From eqn. 7,

$$\left(\frac{h_{\mathrm{ne}}}{h_{\mathrm{e}}}\right)^c = \frac{1 + \varepsilon_y \varepsilon_{\mathrm{ne}}^*}{1 + \varepsilon_y \left(\varepsilon^* - \varepsilon_{\mathrm{ne}}^*\right)} \qquad \text{eqn. 8}$$

When $\varepsilon_0^* < \varepsilon^* < \varepsilon_{\mathrm{L}}^*$ (combining eqn. 3 with eqn. 8),

$$\left(\frac{h_{\mathrm{ne}}}{h_{\mathrm{e}}}\right)^{c}=\frac{1+\left(p\left(\varepsilon_{\mathrm{t}}+\ln\dot{\varepsilon}_{\mathrm{t}}\right)+q\right)\left[\sum_{i=2}^{6}k_i\left(\frac{\exp\varepsilon_{\mathrm{t}}-1}{p\left(\varepsilon_{\mathrm{t}}+\ln\dot{\varepsilon}_{\mathrm{t}}\right)+q}-\varepsilon_0^{*}\right)^{i}\right]}{1+\left(p\left(\varepsilon_{\mathrm{t}}+\ln\dot{\varepsilon}_{\mathrm{t}}\right)+q\right)\left[\frac{\exp\varepsilon_{\mathrm{t}}-1}{p\left(\varepsilon_{\mathrm{t}}+\ln\dot{\varepsilon}_{\mathrm{t}}\right)+q}-\sum_{i=2}^{6}k_i\left(\frac{\exp\varepsilon_{\mathrm{t}}-1}{p\left(\varepsilon_{\mathrm{t}}+\ln\dot{\varepsilon}_{\mathrm{t}}\right)+q}-\varepsilon_0^{*}\right)^{i}\right]} \qquad \text{eqn. 9}$$

Again, if $\varepsilon_y$ is considered to be invariant with $\ln\dot{\varepsilon}$, eqn. 9 simplifies to

$$\left(\frac{h_{\mathrm{ne}}}{h_{\mathrm{e}}}\right)^{c}=\frac{1+\varepsilon_y\left[\sum_{i=2}^{6}k_i\left(\frac{\exp\varepsilon_{\mathrm{t}}-1}{\varepsilon_y}-\varepsilon_0^{*}\right)^{i}\right]}{1+\varepsilon_y\left[\frac{\exp\varepsilon_{\mathrm{t}}-1}{\varepsilon_y}-\sum_{i=2}^{6}k_i\left(\frac{\exp\varepsilon_{\mathrm{t}}-1}{\varepsilon_y}-\varepsilon_0^{*}\right)^{i}\right]} \qquad \text{eqn. 9a}$$

If when $\varepsilon^{*} > \varepsilon_{\mathrm{L}}^{*}$ (combining eqn. 4 with eqn. 8),

$$\left(\frac{h_{\mathrm{ne}}}{h_{\mathrm{e}}}\right)^{c}=\frac{1+\left[p\left(\varepsilon_{\mathrm{t}}+\ln\dot{\varepsilon}_{\mathrm{t}}\right)+q\right]\left(m_{\mathrm{L}}\frac{\exp\varepsilon_{\mathrm{t}}-1}{p\left(\varepsilon_{\mathrm{t}}+\ln\dot{\varepsilon}_{\mathrm{t}}\right)+q}+k_{\mathrm{L}}\right)}{1+\left[p\left(\varepsilon_{\mathrm{t}}+\ln\dot{\varepsilon}_{\mathrm{t}}\right)+q\right]\left(\frac{\exp\varepsilon_{\mathrm{t}}-1}{p\left(\varepsilon_{\mathrm{t}}+\ln\dot{\varepsilon}_{\mathrm{t}}\right)+q}-\left(m_{\mathrm{L}}\frac{\exp\varepsilon_{\mathrm{t}}-1}{p\left(\varepsilon_{\mathrm{t}}+\ln\dot{\varepsilon}_{\mathrm{t}}\right)+q}+k_{\mathrm{L}}\right)\right)} \qquad \text{eqn. 10}$$

If $\varepsilon_y$ is considered as invariant with $\ln\dot{\varepsilon}$, eqn. 10 simplifies to

$$\left(\frac{h_{\mathrm{ne}}}{h_{\mathrm{e}}}\right)^{c}=\frac{1+\varepsilon_y\left(m_{\mathrm{L}}\frac{\exp\varepsilon_{\mathrm{t}}-1}{\varepsilon_y}+k_{\mathrm{L}}\right)}{1+\varepsilon_y\left(\frac{\exp\varepsilon_{\mathrm{t}}-1}{\varepsilon_y}-\left(m_{\mathrm{L}}\frac{\exp\varepsilon_{\mathrm{t}}-1}{\varepsilon_y}+k_{\mathrm{L}}\right)\right)} \qquad \text{eqn. 10a}$$

In eqns. 9, 9a, 10, 10a, $\varepsilon_y$ and $\varepsilon_0^{*}$ are constant for a given polymer. The objective is to identify $c$, $\dot{\varepsilon}_{\mathrm{t}}$ and $\varepsilon_{\mathrm{t}}$ for a given nanoindentation. Eqns. 6 and 7, relate $\dot{\varepsilon}_{\mathrm{t}}$ and $\varepsilon_{\mathrm{t}}$ to the nanoindentation data, $h_{\mathrm{ne}}$, $h_{\mathrm{m}}$ and $h_{\mathrm{e}}$, through $c$. Equation 7 is incorporated into eqns. 9 (or 9a) and 10 (or 10a). Thus, these equations contain only nanoindentation data, $h_{\text{ne}}$, $h_{\text{m}}$ and $h_{\text{e}}$, and are implicit in $c$, the only unknown. The regime that the nanoindentation strain belongs to (whether $\varepsilon^* < \varepsilon_{\text{L}}^*$ or $\varepsilon^* > \varepsilon_{\text{L}}^*$), is still unknown. Solving eqns. 6 and 9 (or eqn. 9a only, if $\varepsilon_y$ is a constant) is simultaneously, yields a value set for $\varepsilon_{\text{t}}$ and $c$; eqns. 6 and 10 (or eqn. 10a only, if $\varepsilon_y$ is a constant), provide another value set for $\varepsilon_{\text{t}}$ and $c$. The value of $c$ for that nanoindentation, depends on which regime contains the resultant $\varepsilon^*$ (corresponding to the resultant $\varepsilon_{\text{t}}$ values).

Table 6 lists the $\varepsilon$, $\dot{\varepsilon}$ and $c$ values, for the various experimental conditions for the three polymers. There is negligible pile-up in cross-linked SU-8; for PC and PMMA, the contact pile-up ~ $0.015\,h_{\text{m}}^{\Delta}$. However, we find that despite any pile-up, there is negligible variation between edge-side and facet side estimates of $\varepsilon$, $\dot{\varepsilon}$ and $c$ values. In addition, we find that the uncertainty in the relationships between $\varepsilon_y$ vs $\ln\dot{\varepsilon}$, corresponds to a relative uncertainty of ~ 2.5%, in the $c$ values (listed in Supporting Information 8). A similar uncertainty exists while considering $\varepsilon_y$ vs $\ln\dot{\varepsilon}$ to be constant.

**Table 6:** $\varepsilon$, $\dot{\varepsilon}$ and $c$ values with respect to tip apex displacement.

| | $\dot{P}/P$ (/s) | $P_{\text{m}}$ (μN) | $\dot{\varepsilon} \times 10^2$ (/s) | $\varepsilon \times 10^2$ | $c$ |
|---|---|---|---|---|---|
| Cross-linked SU-8 | 0.1 | 1000 | 0.28 ± 0.00 | 6.21 ± 0.08 | 0.045 ± 0.000 |
| | | 9000 | 0.28 ± 0.00 | 6.44 ± 0.08 | 0.047 ± 0.000 |
| | 1.0 | 1000 | 3.18 ± 0.29 | 6.31 ± 0.02 | 0.046 ± 0.000 |
| | | 9000 | 2.78 ± 0.03 | 5.86 ± 0.03 | 0.044 ± 0.000 |
| PC | 0.1 | 1000 | 0.53 ± 0.01 | 12.79 ± 0.38 | 0.083 ± 0.002 |
| | | 9000 | 0.46 ± 0.04 | 13.19 ± 0.32 | 0.085 ± 0.002 |
| | 1.0 | 1000 | 5.01 ± 0.17 | 12.16 ± 0.58 | 0.079 ± 0.003 |
| | | 9000 | 4.50 ± 0.02 | 11.44 ± 0.07 | 0.077 ± 0.000 |
| PMMA [40] | 0.1 | 1000 | 0.67 ± 0.02 | 16.98 ± 0.23 | 0.110 ± 0.001 |
| | | 9000 | 0.71 ± 0.01 | 17.21 ± 0.14 | 0.111 ± 0.000 |
| PMMA [41] | 0.1 | 1000 | 0.57 ± 0.03 | 14.16 ± 0.22 | 0.094 ± 0.001 |
| | | 9000 | 0.61 ± 0.01 | 14.34 ± 0.15 | 0.095 ± 0.000 |
| PMMA [42] | 0.1 | 1000 | 0.62 ± 0.01 | 14.57 ± 0.28 | 0.089 ± 0.001 |
| | | 9000 | 0.66 ± 0.02 | 14.76 ± 0.17 | 0.090 ± 0.000 |

We first consider across polymers, the inter-related effects of $\dot{P}/P$ and $P_{\mathrm{m}}$, on the equivalent uniaxial $\varepsilon$ and $\dot{\varepsilon}$. For a given $\dot{P}/P$, we find that $\dot{\varepsilon}$ decreases marginally, when $P_{\mathrm{m}}$ increases by a factor of 9. This is because, for VE materials, the strain rate leads the stress, and asymptotically decreases [14] towards the Hertzian limit, $\dot{h}/h \sim 0.5\,\dot{P}/P$ (shown in Appendix A). $\dot{\varepsilon}$ decreases with increase in $P_{\mathrm{m}}$, as the higher load provides additional time for $\dot{\varepsilon}_{\mathrm{N}}$, to decrease asymptotically. For a given $P_{\mathrm{m}}$, when $\dot{P}/P$ is increased by factor of 10, from 0.1 /s to 1.0 /s, $\varepsilon$ decreases, although the $h_{\mathrm{m}}$ values are comparable. Decrease in $\varepsilon$ is consistent with the increase in $h_{\mathrm{e}}/h_{\mathrm{m}}$ (relative elastic contribution) with strain rate, in Table 4. $\dot{\varepsilon}$ increases with increasing $\dot{P}/P$, as expected.

For the polymers in Table 6, $c$ values ~ 0.044 to 0.047 for cross-linked SU-8, ~ 0.077 to 0.085 for PC, and ~ 0.09 to ~ 0.11 for PMMA (depending on the uniaxial compression data source). When $P_{\mathrm{m}}$ is increased by a factor of 9 from 1000 μN to 9000 μN, the variation in $c$ is insignificant for all polymers. At $P_{\mathrm{m}}$ =9000 μN, $c$ decreases by 7%-10%, when $\dot{P}/P$ is increased from 0.1 /s to 1.0 /s.

Thus, we find that for a given polymer, the $c$ value increases with $\varepsilon$. Since they both do not vary with $P_{\mathrm{m}}$, they are reached very early the during nanoindentation. For all three polymers, the same $P_{\mathrm{m}}$ values and $\dot{P}/P$ (i.e., $\dot{\varepsilon}_{\mathrm{N}}$) values, result in different values of $\varepsilon$ and of $\ln\dot{\varepsilon}$. We find that across polymers also, $c$ increases with $\varepsilon$. Since $\varepsilon_y$ also increases across polymers, corresponding to the increase in $\varepsilon$ and $c$, we examine them next, in terms of the nature of the strain (*CSS* metrics).

In order to examine the variations in nanoindentation deformation between polymers, at this time, combined with the *CSS*, $\Phi$, defined in section 5.1, we define $\Phi^* = \Phi/\left(1-\varepsilon_0^*\right)$. $\Phi$ corresponds to the extent of the strain beyond the elastic limit, and $\Phi^* = \Phi/\Phi_y$.

For the polymers, we closely examine first, the plots of $\varepsilon_{\mathrm{ne}}^{*}$ $(=\varepsilon^{*}-\varepsilon_{\mathrm{e}}^{*})$ vs $\Phi$ and $\Phi^{*}$ (Fig. 12(a) and (b), respectively), on which we mark the equivalent *CSS* of our nanoindentation experiments. For the three polymers, the nanoindentation deformation regimes are, $(\Phi)_{\mathrm{PMMA}},(\Phi)_{\mathrm{PC}}>(\Phi)_{\mathrm{SU-8}}$, where for PMMA, $(\Phi)_{\varepsilon_0^*=0.3}<(\Phi)_{\varepsilon_0^*=0.4}$.

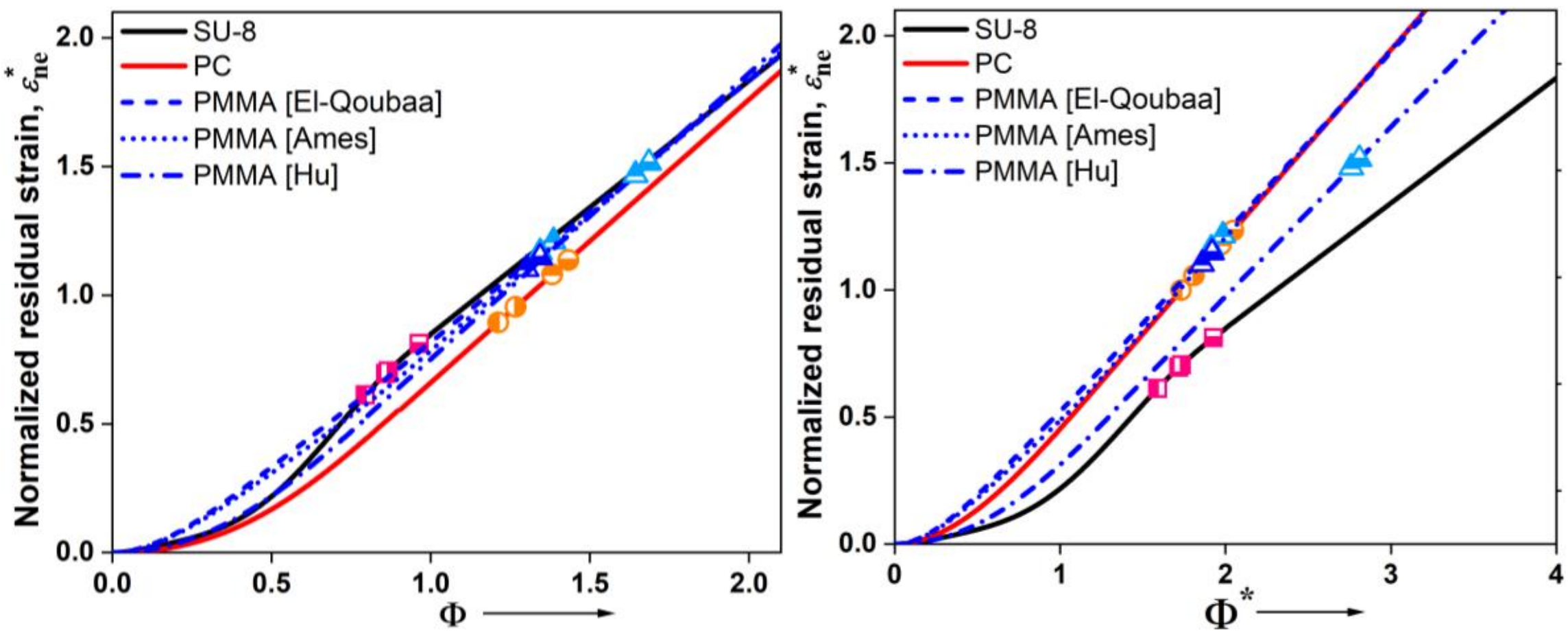

(a) (b)

**Figure 12:** Correlations for *CSS* metrics $\varepsilon_{\mathrm{ne}}^{*}$. The equivalent *CSS* of our nanoindentation experiments are marked; ■, ◒, ▲, △ represent $P_{\mathrm{m}}$ = 1000 μN and $\dot{P}/P$ = 0.1 /s, ■, ◒, ▲, △ represent $P_{\mathrm{m}}$ = 9000 μN and $\dot{P}/P$ = 0.1 /s, ◨, ◑ represent $P_{\mathrm{m}}$ = 1000 μN and $\dot{P}/P$ = 1.0 /s, and ◨, ◑ represent $P_{\mathrm{m}}$ = 9000 μN and $\dot{P}/P$ = 1.0 /s. (a) Polymer-specific zones can be determined with respect to $\Phi$ (b) There appears to be a common region for all polymers with respect to $\Phi^{*}$. For PMMA, if $\varepsilon_0^{*}$~0.4, nanoindentation regions are not common with other polymers.

The nanoindentation regimes of the equivalent *CSS*, $\varepsilon^{*}$ and $\Phi^{*}$, are located past the post-yield inflection. For cross-linked SU-8 the nanoindentation points are past the inflection, in the convex upward region. However, they are in the $\varepsilon^{*}<\varepsilon_{\mathrm{L}}^{*}$ region (section 5.1). The data for PC and PMMA are in the $\varepsilon^{*}>\varepsilon_{\mathrm{L}}^{*}$ region. We recognize that, in Fig 4, for the thermoplastics, the region where the second derivative is zero, is reached only at $\varepsilon^{*}\sim\varepsilon_{\mathrm{L}}^{*}$. Thus, the thermoplastic polymer nanoindentation region is also past the post-yield inflection, as seen in Figs. 4 and 12 (discussed further in section 5.6). Figure 12(a) suggests that there may be polymer-specific nanoindentation regimes on the $\Phi$ scale. For normalization on the $\Phi^*$ scale (Fig. 12(b)), there could be a common regime for all polymers if for PMMA, $\varepsilon_0^*$~0.3. This elastic limit is observed in a significant majority of the literature data for PMMA, and will consider so, in subsequent analyses.

For a given polymer, the *CSS* ranges, $\Phi$ and $\Phi^*$, are small, for a 9-fold range of $P_{\mathrm{m}}$, and for a decade-wide range of $\dot{P}/P$. Hence, it is likely that these *CSS* values are reached very early during nanoindentation, for any amorphous glassy polymer. In order to understand the deformation phenomena further, we consider next, the equivalent uniaxial compression during nanoindentation, as function of relative radial distance, $r/r_{\mathrm{m}}$.

## 5.5. Deformation distribution under the indenter tip

This section consists of two parts. We first estimate the nanoindentation profiles at full load, and after the hypothetical elastic recovery. We then estimate the equivalent strains, based on the radially distributed displacements and recoveries, employing the method described in Section 5.4.

### I. Estimate of state-specific topological profiles

During indentation, the surface of the region away from the tip, i.e., the non-contact region, undergoes radial deformation inwards, as well as axial deformation downwards. The LS analyses [7]–[9], [57], [58], describe the response to indentation by a perfect conical indenter, of a purely elastic material, which recovers completely post-indentation. Based on the *P-h* data, topological imaging and upflow analysis, we estimate next, the topological profiles at specific deformation states for glassy polymers.

### I.1. Profile as per SPM imaging

This profile is shown in Fig. 7(b). We call this profile the $h_r$ profile, $h_r^p(r)$. As indicated in the figure, this profile can be fit to a power-law expression (details in Appendix C).

### I.2. Pure elastic nanoindentation profile where $h_m = h_e$, the elastic depth

We estimate the profile for an ideal tip $(h_b = 0)$ Berkovich nanoindentation, assuming that the $h_e$ value obtained above, corresponds to the maximum displacement for purely elastic nanoindentation deformation. The corresponding sink-in depth would be $h_s^e$, as described above in Section 5.2 II.

The maximum depth profile corresponds to maximum sample contact. Hence, the profile closely follows the indenter geometry. There is no fundamental development for the non-contact surface profile for ideal elastic Berkovich nanoindentation. Therefore, we propose a functional approximation based on existing expressions for conical indentation [7]–[9], [57], [58], which could approximate the profile. The objective is to map the real non-contact profile for VEP materials.

Therefore, as opposed to a single angle describing a cone, we approximate a Berkovich tip indenter, as a family of cones, with semi-angles spanning the range between the two included angles, of the indenter. While traversing any facet, the inclination angle (to the vertical) decreases from the edge angle to the mid-facet angle, before symmetrically increasing back to the edge angle at the opposite edge. These two limiting angles for the Berkovich tip are the mid-facet semi-apex angle of 65.27° and the edge angle, 142.30° - 65.27° = 77.03° (the total included angle =142.30°).

Previous studies have provided optical video stills [61] of the full load vertical sections of transparent elastomer nanoindentation, as well as simulations [60] of edge-side and facet side profiles, which match qualitatively, the non-contact profiles for the conical indentation with these two corresponding cone half-angles. This qualitatively validates our cone-family approximation.

In this case of ideal elastic indentation, the profile from $h_s^{\mathrm{e}}$ to $h_{\mathrm{e}}$, is in contact with the indenter tip. The profile from $z = 0$ to $z = h_s^{\mathrm{e}}$, corresponds to the LS equation [7]–[9], [57], [58], for a conical indenter (eqn. 11).

$$h_s^{ep}(r) = \cot\theta\left[r_{\mathrm{e}}\sin^{-1}\left(\frac{r_{\mathrm{e}}}{r}\right) + \sqrt{r^2 - r_{\mathrm{e}}^2} - r\right] \qquad \text{eqn. 11}$$

Based on the two limiting values for cone half-angles, $r_{\mathrm{e},ed} = 2\sqrt{3}h_{\mathrm{e}}\tan 65.27° / \sqrt{3} = 2h_{\mathrm{e}}\tan 65.27°$, the contact radius in the edge direction and along the facet-middle direction, $r_{\mathrm{e},fa} = h_{\mathrm{e}}\tan 65.27°$.

### I.3. Estimate profile at $P_{\mathrm{m}}$

The contact profile at $P_{\mathrm{m}}$ is up to $h_{\mathrm{m}}$. At $h_{\mathrm{m}}^{\Delta}$, a further depth $h_b$ below $h_{\mathrm{m}}$, the edge and facet lines meet. For VEP deformation, the contact profile would follow the ideal Berkovich tip from $h_s^{\Delta}$ to $h_{\mathrm{m}}^{\Delta}$, as in the case of $h_s^{\mathrm{e}}$ to $h_{\mathrm{e}}$, above (for purely elastic deformation). However, for the real Berkovich tip for VEP deformation, the contact profile is truncated at $h_{\mathrm{m}}$.

The profile from $z = 0$ to $z = h_s^{\Delta}$, we consider the sink-in $z$-$r$ relationships (eqn. 11) for our experiment, to correspond to radially outward-shifted $z$-$r$ relationships (i.e., by shifting the $z$-axes) for the cones of depth $h_{\mathrm{e}}^{\Delta}$ $\left(h_s^{\Delta}/h_{\mathrm{e}}^{\Delta} = (\pi - 2)/\pi\right)$, such that the shifted $h_{\mathrm{e}}^{\Delta}$ cones share the surface lines with the real Berkovich tip. $h_{\mathrm{e}}^{\Delta}$ corresponds to the maximum ideal elastic displacement corresponding to $h_s^{\Delta}$, the geometric sink-in estimate via SPM imaging. This axis shift-based contact profile, ensures that the displacement and its derivatives with respect to radial distance from the apex, are continuous. We note here that the entire deformation of the sunk-in region is recovered, is provided here for completeness and to provide a visual understanding of the deformation. It does not participate further in our analyses.

### I.4. *P* = 0 profile post hypothetical pure elastic recovery

As described in Section 5.2 I, our unloading $P$-$h$ analysis indicates that a hypothetical purely elastic recovery, would correspond to a residual apex depth, $h_{\mathrm{ne}2}$. We estimate the topological profile for this hypothetical depth, based on the $P_{\mathrm{m}}$ profile, determined above, as well as on the actual contact parameters and geometrical analysis of the SPM profiles from two consecutive scans. There is an excellent correlation between the consecutive residual SPM profiles ($h_{r1}^{p}(r)$, $h_{r2}^{p}(r)$ and $h_{r3}^{p}(r)$, see Appendix D) and the corresponding tip apex residual displacements ($h_{r1}$, $h_{r2}$, $h_{r3}$) (eqn. 12).

$$\frac{(h_c - h_{r1})}{(h_c - h_{r2})} = \frac{\left(h_c^p(r) - h_{r1}^p(r)\right)}{\left(h_c^p(r) - h_{r2}^p(r)\right)}; \qquad \frac{(h_c - h_{r1})}{(h_c - h_{r3})} = \frac{\left(h_c^p(r) - h_{r1}^p(r)\right)}{\left(h_c^p(r) - h_{r3}^p(r)\right)} \qquad \text{eqn. 12}$$

We employ this proportionality, which is based on the geometrically obtained $h_c$, to estimate the profile for the hypothetical residual imprint, $h_{\mathrm{ne}}^{p}(r)$ (eqn. 13); $h_r^p(r)$ is obtained by the power law fitting of the SPM section boundary (Fig. 7(b), Appendix C).

$$\frac{(h_c - h_{\mathrm{ne}})}{(h_c - h_r)} = \frac{\left(h_c^p(r) - h_{\mathrm{ne}}^p(r)\right)}{\left(h_c^p(r) - h_r^p(r)\right)} \qquad \text{eqn. 13}$$

These profiles are plotted in Fig. 13.

We devise an alternative method to estimate the $h_{\mathrm{ne}}$ profile, without employing SPM imaging. The $h_{\mathrm{e}}$ estimates are for tip apex displacements. $h_{\mathrm{e}}/h_{\mathrm{m}}^{\Delta}\left(P_{\mathrm{m}} = 9000\mu\mathrm{N}\right) < h_{\mathrm{e}}/h_{\mathrm{m}}^{\Delta}\left(P_{\mathrm{m}} = 1000\mu\mathrm{N}\right)$. Hence, we interpolate the $h_{\mathrm{e}}$ values between $P_{\mathrm{m}}$ = 1000 μN and $P_{\mathrm{m}}$ = 9000 μN, via the power law equation, $h_e = B_e h_m^{m_e}$, to estimate the $h_e$ value for the displacement at any $r/r_m$ position at $P_m$ = 9000 μN. The fitted $B_e$ and $m_e$ values, are listed in Appendix E. We find that at various radial positions, these power-law $h_{ne}$ values agree well with those obtained via eqn. 13.

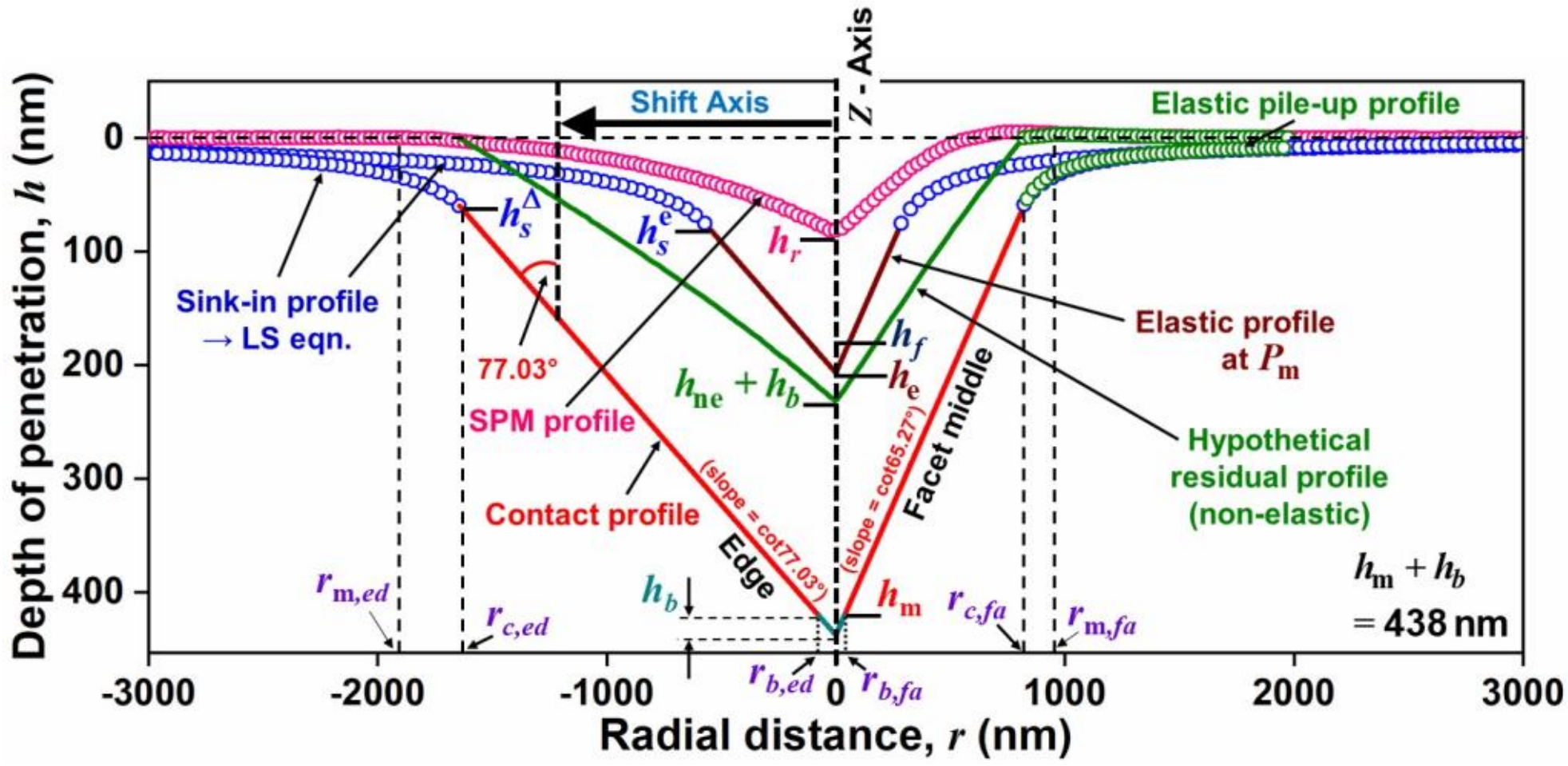

**Figure 13:** Deformation profiles at $P_m$, after elastic recovery ($P$ = 0), hypothetical pure elastic nanoindentation, and as captured by SPM imaging for cross-linked SU-8. The indicated $h_e$ is the recovered displacement, $(h_m - h_{ne})$, transferred toward the surface to indicate elastic nanoindentation. The hypothetical non-elastic residual profile exhibits an intermediate curvature (appears flat at current magnification) between that of the acquired $h_r$ profile and the linear contact profile at full load. The shifted axis intersects the full load deformation profile at depth $h_e^{\Delta} = h_s^{\Delta}\pi/(\pi-2)$, such that the shifted $h_e^{\Delta}$ cones share the surface lines with the real Berkovich tip.

## II. Equivalent strain variation with radial distance from tip apex

At $P_m$, the maximum radial distance along the indenter tip, $r_m$ ($r_{m,ed}$ and $r_{m,fa}$), is obtained by extending the contact profile lines from the depth-axis at $h_m^{\Delta}$, to $z$ = 0 (shown in Fig. 13). In the contact region, the maximum load depths at various $r/r_m$ values, are equal to $h_m^{\Delta}/(1-(r/r_m))$. We, compute the strains and strain rates via the framework in the previous sections, (eqns. 1-10), employing the zero-load $h_{ne}$ profiles determined above, at various $r/r_m$. This enables estimation of the $c$ values, at these radial positions. The latter are plotted in Fig. 14, and compared with the $c$ values at the tip apex for both loads. See Supporting Information 9 for the combined plot of $c$ values, at various $r/r_m$, at both $P_m$ and $\dot{P}/P$.

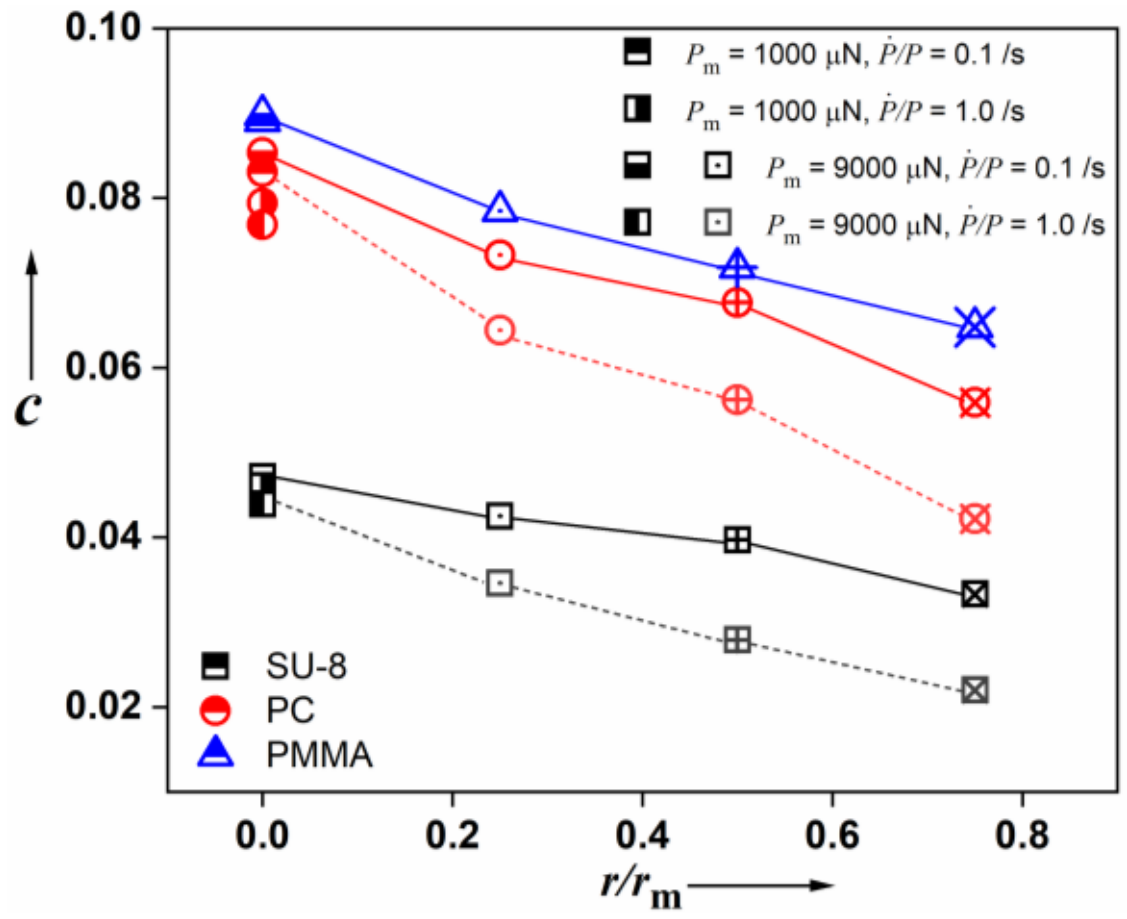

**Figure 14**: The $c$ values for various sections of $r/r_m$ for cross-linked SU-8, PC and PMMA. The lighter red and grey markers, correspond to SU-8 and PC data at $\dot{P}/P$ = 0.1 /s.

The $c$ values decrease with $r/r_m$ (e.g., for cross-linked SU-8: from 0.046 to 0.033 for $\dot{P}/P$ = 0.1 /s and from 0.045 to 0.022 for $\dot{P}/P$ = 1.0 /s). The decrease in $c$ due to a 10-fold increase in applied strain rate, is less than the decrease due to radial distance (where the effective strain rate increase, up to $r/r_m$ ~0.75, is only 4-fold). Thus, the decrease in $c$ and $\varepsilon$ with $r/r_m$, is likely to be due to the increase in the increase in the elastic deformation fraction and source of the imposed stress. At the apex, the stress concentration and the degree of yield, are the greatest. These decrease with increase in radial distance. The deformation also becomes more uniformly distributed over the depth below the tip, as radial distance increases. This idea is consistent with various FE studies of sharp-indenter nanoindentation on polymers (e.g., see [68], [69]). Hence, as the nanoindentation displacement decreases with radial distance, the equivalent $\varepsilon$ decreases by a progressively greater extent.

Considering the variation of $c$ with radial distance, Fig. 15(a)-(b) depicts the variations of $c$ with $\varepsilon$ and $\Phi$ values, which bridge the gap between the apex regimes of all the polymers. Over all polymers, there is a clear increasing trend of the apex $c$ values with $\varepsilon$. However, the overall linear trend for all values of $c$, is clearer with $\Phi$, than with $\varepsilon$. The relationships of $c$ vs $\varepsilon^*$ and $\Phi^*$, are localized to individual polymers, and are significantly different across polymers (not shown).

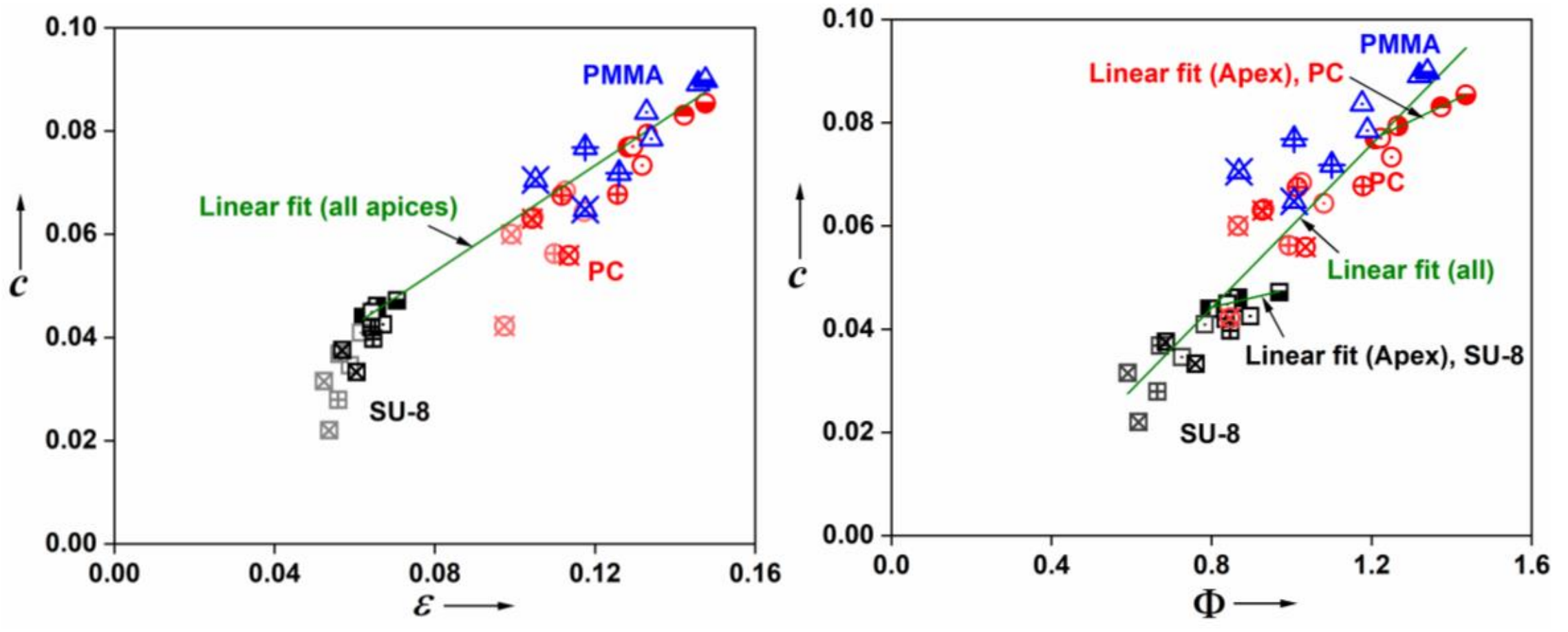

(a) (b)

**Figure 15**: Variation of $c$ values for cross-linked SU-8, PC and PMMA with (a) $\varepsilon$, and (b) $\Phi$. For $\varepsilon$, there is an excellent common fit for the tip apex only, for all polymers. For $\Phi$, the *CSS* measures at the tip apices correlate well in separate local regimes, and are indicated by individual local fits. Considering the effect of radial position, bridges the *CSS* regimes of all polymers, enabling a global linear correlation.

The foregoing discussion has considered the nanoindentation deformation through the prism of uniaxial deformation – in identifying the equivalent deformation ranges, via rationally defined strain state metrics. The fundamental difference between the current work with respect to earlier reports [2], [15]–[20], [22]–[25], is that we employ uniaxial compression strain and recovery, to gain insights into nanoindentation deformation. Our approach incorporates rate effects, (i) by considering two nanoindentation rates, and (ii) by directly determining the unrecovered uniaxial strain for every strain over a range of strain rates. In addition, our model-free analyses interpret the phenomena in terms of the state of the deformation. We explore next, the possible criteria, governing the material specific *CSS* ranges, on dimensionless uniaxial stress-strain plots.

## 5.6. Nanoindentation domains vs equivalent uniaxial *CSS*

In this section, we identify the nanoindentation domains ($\varepsilon$ range from in Table 6), in the corresponding plots of stress-state vs strain-state. Figure 16(a) contains the plots of $\sigma^*$ vs $\varepsilon^*$. These plots are constructed via LOESS smoothening of Fig. 2. As mentioned above, the. The stress states at the elastic limit, $\sigma_0^*$ = 0.63 for cross-linked SU-8, $\sigma_0^*$ = 0.55 for PC, and $\sigma_0^*$ = 0.57 for PMMA.

In order to account for the different elastic limits for the two polymers, similar to those for strain, we define additional stress state metrics, $\Sigma = \sigma^* - \sigma_0^*$ and $\Sigma^* = \Sigma / (1 - \sigma_0^*)$. We plot $\Sigma^*$ vs $\Phi^*$ in Fig. 16(b). The corresponding plots of $\Sigma$ vs $\Phi$, are available in Supporting Information 10.

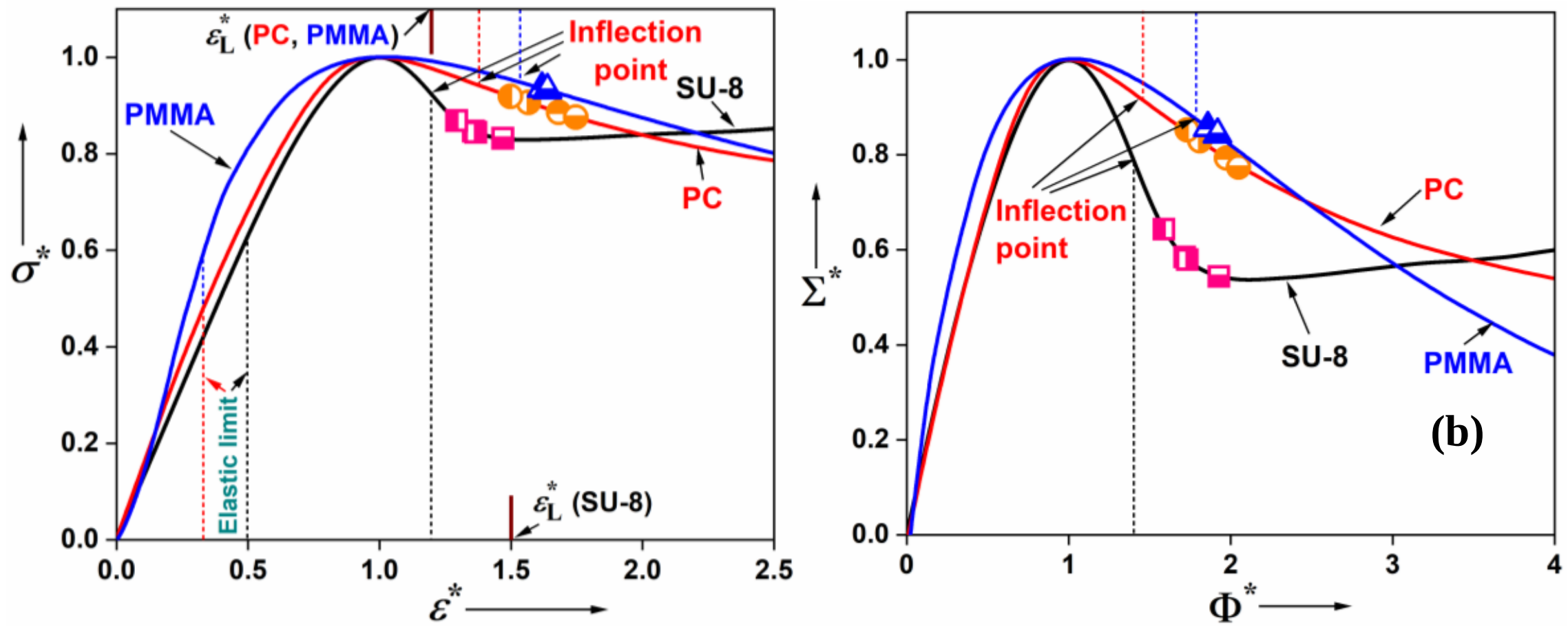

(a) (b)

**Figure 16:** LOESS smoothened curves for cross-linked SU-8, PC and PMMA [42]. The equivalent strains of the nanoindentation experiments are beyond the post-yield inflection. ■, ●, ▲ represent $P_m$ = 1000 µN and $\dot{P}/P$ = 0.1 /s, ■, ●, ▲ represent $P_m$ = 9000 µN and $\dot{P}/P$ = 0.1 /s, ■, ● represent $P_m$ = 1000 µN and $\dot{P}/P$ = 1.0 /s and ■, ● represent $P_m$ = 9000 µN and $\dot{P}/P$ = 1.0 /s. (a) $\sigma^*$ vs $\varepsilon^*$ (b) $\Sigma^*$ vs $\Phi^*$ $\Phi^*$ (nanoindentation range from $\Phi^*$ ~1.7 to $\Phi^*$ ~2.0, common for all polymers).

The nanoindentation data are expectedly beyond the yield with respect to strain and various *CSS* strain metrics. In addition, we find that they are also beyond the post-yield inflection points. These post-yield inflections on the $\sigma^*$ vs $\varepsilon^*$ plots and $\Sigma^*$ vs $\Phi^*$ plots, agree approximately with the inflections on the corresponding $\varepsilon^*_{\mathrm{ne}}$ vs $\varepsilon^*$ (section 5.1) and $\varepsilon^*_{\mathrm{ne}}$ vs $\Phi^*$ plots (section 5.4). We recall from section 5.1 that $\varepsilon^*_{\mathrm{L}}$ are estimated visually. For PC and PMMA, there could be a correlation between these $\varepsilon^*_{\mathrm{L}}$ values and the inflection points. Thus, despite the recovery correlation curves being polymer-specific (in terms of $\varepsilon^*$ and $\Phi$), Thus, despite the recovery correlation curves being polymer-specific (in terms of $\varepsilon^*$ and $\Phi$), we find possible common ranges in terms of $\Phi^*$ (~1.7 to ~2.0), which are reached very early, during the nanoindentation. Our findings in this work are based on thermoplastic as well as thermosetting amorphous glassy polymers. Although we have not nanoindented the same thermoplastic sample that has undergone uniaxial compression testing, our findings describe representative behaviour for such materials. In the next section, we summarize our findings, and provide our conclusions.

# 6. Conclusions

In this investigation, we have employed the well understood uniaxial compression phenomena (data available in the literature) to interpret the complex material response during nanoindentation of glassy polymers. We have considered highly cross-linked SU-8 to represent amorphous glassy thermosets and PC and PMMA, to represent amorphous glassy thermoplastics.

Representing the uniaxial stress-strain data over a range of engineering strain rates, $\dot{\varepsilon}$, in terms of normalized variables, $\sigma^* = \sigma/\sigma_y$ vs. $\varepsilon^* = \varepsilon/\varepsilon_y$, has provided for each polymer, a universal correlation of residual the (non-elastic) strain state, $\varepsilon^*_{\mathrm{ne}}$, vs. strain state, $\varepsilon^*$. $\varepsilon^*$ can be considered to be a measure of corresponding strain state, *CSS*.

We have combined nanoindentation experiments with SPM imaging of residual imprints, to obtain the maximum depth, $h_{\mathrm{m}}$ and the projected contact triangle side, $a$. The ratio of the SPM-obtained contact depth to the corresponding maximum depth, $h_c^{\Delta}/h_{\mathrm{m}}^{\Delta} \sim 0.9$, and is broadly independent of $\dot{P}/P$ and $P_{\mathrm{m}}$. Volume conserving material upflow gives rise to this high value. We have considered a generalized power-law (GPL) model for the constant rate portion (2s unloading) of the unloading data to estimate the corresponding hypothetical non-elastic depth $h_{\mathrm{ne}2}$, and the stiffness, *S*. We note that the recovery over 2s contains viscoelastic (VE) components. From this enabled estimation of the conventionally defined nanoindentation modulus, $E_{\mathrm{N}}$, we find $E_{\mathrm{N}}/E$ ~1.1 to 1.9 (*E* is uniaxial compressive modulus). Here, we consider $\left(h_{\mathrm{m}} - h_{\mathrm{ne}2}\right)$ to be the measure of instant elastic recovery, for $E_{\mathrm{N}}$, which can be reliably and reproducibly mapped to *E*. This further enables rational comparison between nanoindentation elastic recovery and uniaxial elastic recovery.

We have then combined the uniaxial data as function of engineering strain rate, $\dot{\varepsilon}$, and nanoindentation recovery data at the two $\dot{P}/P$ values (which correlate with the nanoindentation strain rates $\dot{\varepsilon}_{\mathrm{N}} \sim \dot{h}/h$). The recovery mapping is in terms of true strain $\varepsilon_{\mathrm{t}} = c\varepsilon_{\mathrm{N}}$ and the non-elastic, residual strain, $\varepsilon_{\mathrm{ne}}$. We consider $\dot{\varepsilon}_{\mathrm{t}} = c\dot{\varepsilon}_{\mathrm{N}}$, if $\varepsilon_y$ is a function of strain rate. We obtain the correlation coefficient, *c*; $c \sim 0.044$ to 0.047 for cross-linked SU-8, and $c \sim 0.077$ to 0.085 for PC, where the slight variation is with $\dot{P}/P$. For PMMA $c \sim 0.09$ when compared with uniaxial data having normalized elastic limit, $\varepsilon_0^*$~0.3 (which we consider to be representative for the polymer).

The elastic fraction of the nanoindentation, $h_{\mathrm{e}}/h_{\mathrm{m}}$ is very well correlated with decreasing strain, $\varepsilon$, which is well correlated with *c* values. At lower rates, there are a fewer deformed layers closer to the apex, but with a greater fraction having yielded, in contrast to a greater number of less strained, elastically deformed layers at greater deformation rates.

We have considered the Berkovich indenter as a family of cones with semi-angle range corresponding to the Berkovich geometry. We have modified the LS analyses and to obtain the profile, $h_{\mathrm{m}}^{p}(r)$ at $P_{\mathrm{m}}$. Proportional scaling analyses of consecutive SPM profiles to account for the VE effects, yield the likely non-elastic residual profiles, $h_{\mathrm{ne}}^{p}(r)$. Thus, we have estimated the $c$ values for various radial positions ($r/r_{\mathrm{m}}$ = 0.25, 0.50, 0.75) from the tip apex (where $r/r_{\mathrm{m}}$ = 0). The decrease of the $c$ values with radial positions is indicative of the deformation distribution beneath the tip; i.e., in regions radially away, the reduced stress concentrations result in a greater contribution of recoverable elastic deformation, leading to a lower $\varepsilon$.

We identify additional metrics for *CSS*, $\Phi = \varepsilon^{*} - \varepsilon_{0}^{*}$, $\Phi^{*} = \Phi \big/ \left(1 - \varepsilon_{0}^{*}\right)$. The radial variation in the $c$ values, bridge the $\Phi$ domains of the tip-apex equivalent strains, of the three polymers. While the tip apex $c$ values, vary linearly with $\varepsilon$, we find a common linear relationship between $c$ and $\Phi$, across polymers and radial locations.

We have extended the development of dimensionless stress-strain relationships beyond those described above, to various stress states metrics, as functions of the *CSS*. The equivalent nanoindentation strain and the *CSS* are always beyond the yield, as well as the post-yield inflection. We thus locate polymer specific zones (in terms of $\varepsilon^{*}$ and $\Phi$) and possible universal zones (in terms of $\Phi^{*}$) of equivalent nanoindentation strain on normalized uniaxial stress-strain curves. These universal zones and the polymer-specific $c$ and $\varepsilon$ ranges, are reached very early during the indentation, and are therefore independent .

# Acknowledgments

We would like to thank the Centre of Excellence in Nanoelectronics (CEN), IIT Bombay, for providing SU-8 epoxy and facilities for the SU-8 sample fabrication. The authors would like to acknowledge the Nanoindenter lab, Central facility, IIT Bombay for the nanoindentation experiments. The authors also acknowledge participation in preliminary discussions by Prof. Prita Pant, Department of Metallurgical Engineering and Materials Science, IIT-Bombay.

# Funding

Student funding for this project was topped up by Prof. Prita Pant's Research Development Fund at IIT-Bombay, RI/0106-10000760.

# Appendix A:Nanoindentation strain rate variation

During nanoindentation loading, $\dot{\varepsilon}_{\mathrm{N}} \sim \dot{h}/h$, decays asymptotically from a very high value. Hence, we can fit $\ln P$ vs $t$ as $\ln P = b_1 t + b_2\left[1-\exp\left(-b_3 t\right)\right] + b_4$, to obtain

$$\frac{1}{P}\frac{dP}{dt} = \frac{\dot{P}}{P} = b_1 + b_2 b_3 \exp\left(-b_3 t\right) \qquad \text{eqn. A.1}$$

$b_1$, $b_2$, $b_3$ and $b_4$ are the fitting constants. Similar fitting (eqn. A.1) is carried out between $\ln h$ and $t$, to obtain $\dot{h}/h$, see Fig. A.1(a)-(b) and Fig. A.2(a)-(b).

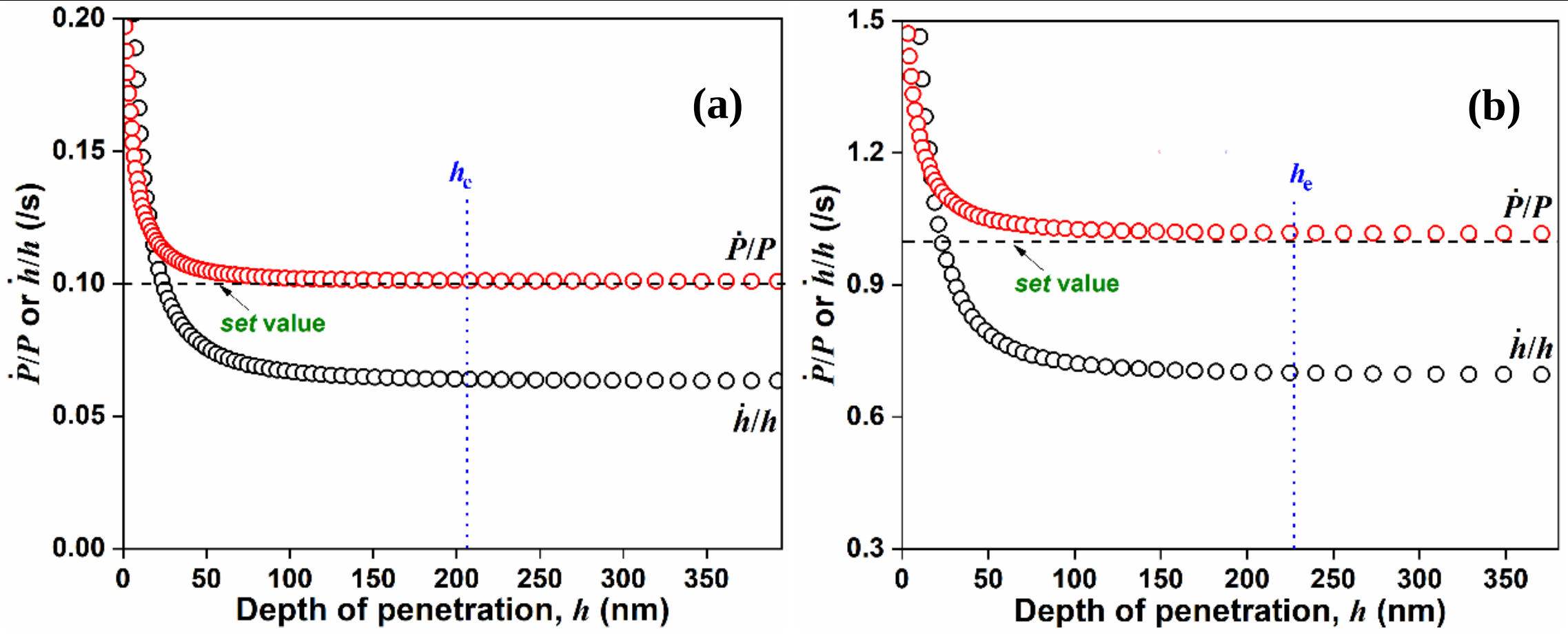

Figure A.1: Variation of normalized rates vs penetration depth for cross-linked SU-8 for $P_m$ = 1000 μN and *set* $\dot{P}/P$ is (a) 0.1 /s, (b) 1.0 /s. The elastic recovery ($h_e$) positions, assumed to be rate independent, are indicated.

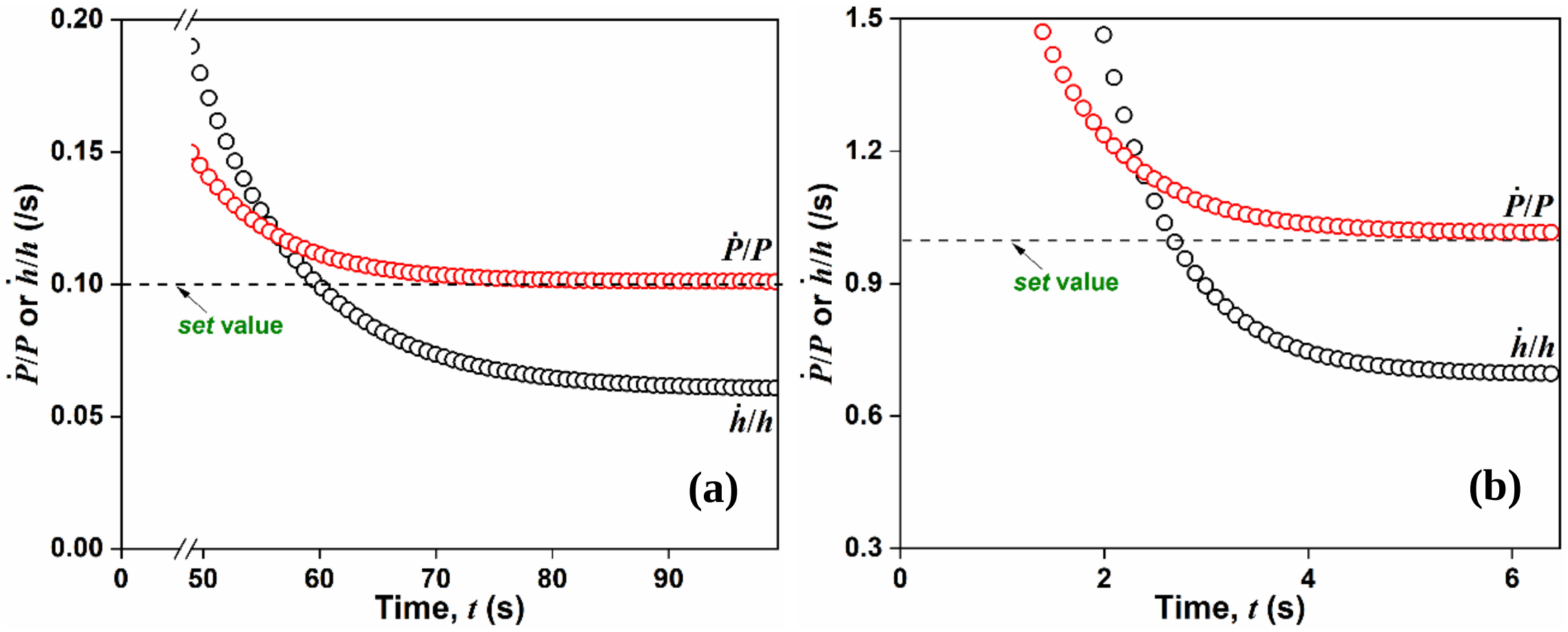

Figure A.2: Variation of normalized rates vs loading time for cross-linked SU-8 for $P_m$ = 1000 μN and *set* $\dot{P}/P$ is (a) 0.1 /s, (b) 1.0 /s.

# Appendix B: True strain correlation with nanoindentation data

We know that, $\varepsilon = \exp\varepsilon_t - 1$ and $\dot{\varepsilon} = \dot{\varepsilon}_t \exp\varepsilon_t$. The premise of this work is:

Fractional residual uniaxial strain = fractional residual nanoindentation strain, $\Rightarrow \frac{\varepsilon_{t,ne}}{\varepsilon_t} = \frac{\varepsilon_{N,ne}}{\varepsilon_N}$.

Nanoindentation residual strain = $\varepsilon_{N,ne} = \ln\frac{h_{ne}}{h_e} + \frac{\varepsilon_{t,e}}{c}$

$$\Rightarrow \varepsilon_{\text{t,ne}} = c\ln\frac{h_{\text{ne}}}{h_{\text{e}}} + \varepsilon_{\text{t,e}},\ \Rightarrow c\ln\frac{h_{\text{ne}}}{h_{\text{e}}} = \varepsilon_{\text{t,ne}} - \varepsilon_{\text{t,e}},\ \Rightarrow c\ln\frac{h_{\text{ne}}}{h_{\text{e}}} = \ln\left(1+\varepsilon_{\text{ne}}\right) - \ln\left(1+\varepsilon_{\text{e}}\right)$$

$$\Rightarrow c\ln\frac{h_{\text{ne}}}{h_{\text{e}}} = \ln\left(1+\varepsilon_y\varepsilon_{\text{ne}}^*\right) - \ln\left(1+\varepsilon_y\varepsilon_{\text{e}}^*\right),\ \Rightarrow \left(\frac{h_{\text{ne}}}{h_{\text{e}}}\right)^c = \frac{1+\varepsilon_y\varepsilon_{\text{ne}}^*}{1+\varepsilon_y\varepsilon_{\text{e}}^*}.$$

Therefore,

$$\left(\frac{h_{\text{ne}}}{h_{\text{e}}}\right)^c = \frac{1+\varepsilon_y\varepsilon_{\text{ne}}^*}{1+\varepsilon_y\left(\varepsilon^* - \varepsilon_{\text{ne}}^*\right)} \qquad \text{eqn. B.1}$$

# Appendix C:Power law fitting of residual image scan profile

The SPM scan section corresponding to the $h_r$ profile (as shown in Fig. C.1), includes the edge corner, the tip apex position and the middle of the facet. The profiles can be expressed as power-law relationships with radial distance, $r$, with the origins shifted to (i) the position A at the $z$ = 0 level for the edge side and (ii) the maximum pile-up point, for the facet side. Thus, the resultant expressions for the power law relationships are (eqn. C.1 and eqn. C.2):

$$z = \beta_{ed}\left(r_{c,ed} - r\right)^{m_{ed}}, \text{ for the edge side} \qquad \text{eqn. C.1}$$

$$z + h_{\text{m}}^{\text{PU}} = \beta_{fa}\left(r_{\text{m}}^{\text{PU}} - r\right)^{m_{fa}}, \text{ for the facet side} \qquad \text{eqn. C.2}$$

Here, $\beta_{ed}$, $\beta_{fa}$, $m_{ed}$, $m_{fa}$ are the fitting constants (tabulated in Table B.1), $h_{\text{m}}^{\text{PU}}$ is the maximum pile-up height and $r_{\text{m}}^{\text{PU}}$ is the corresponding radial distance of maximum pile-up, $r_{c,ed} = a/\sqrt{3}$ and $r_{c,fa} = a/2\sqrt{3}$.

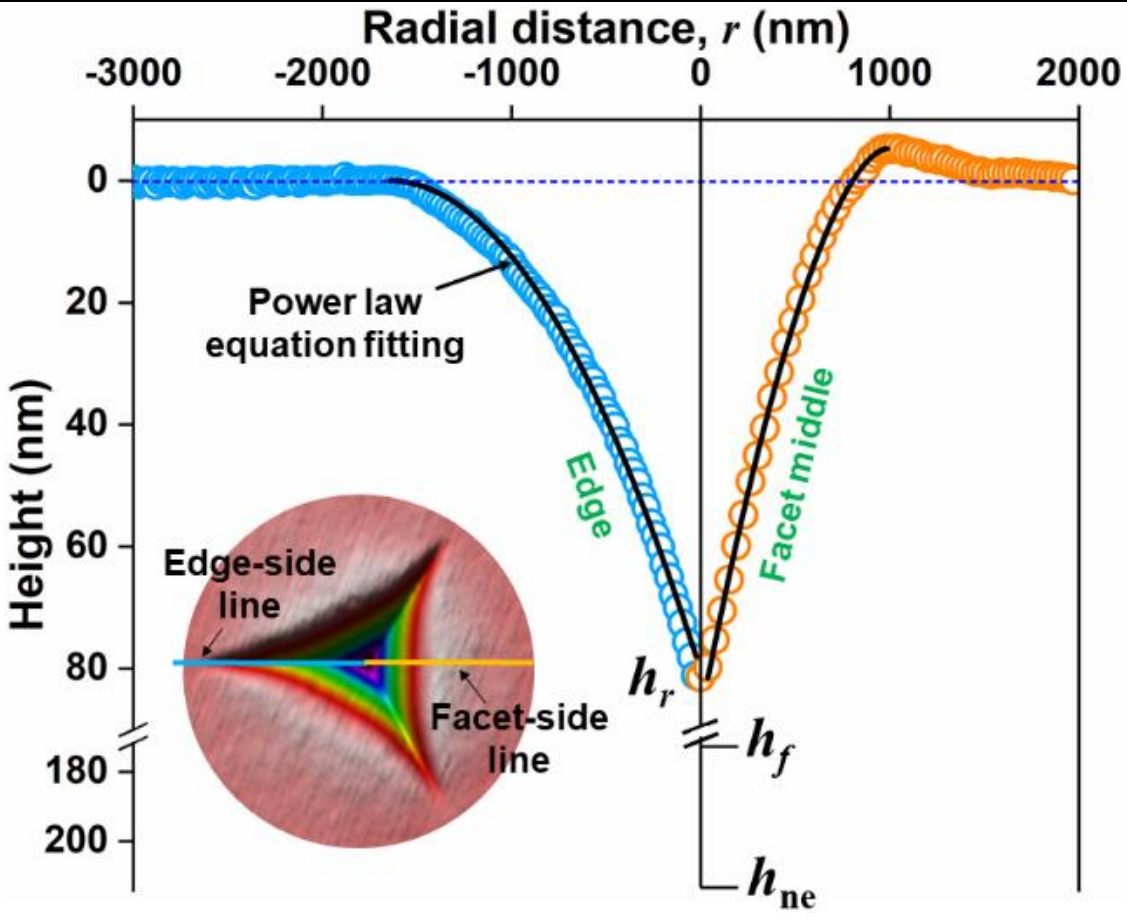

Figure C.1: Residual indent impression scan line with power law equation fitting for cross-linked SU-8.

**Table C.1:** Power law fitting parameters of the scan profile obtained from residual imprints for cross-linked SU-8, PC and PMMA.

| | $\dot{P}/P$ (/s) | $P_m$ (μN) | $\beta_{ed}$ | $\beta_{fa}$ | $m_{ed}$ | $m_{fa}$ |
|---|---|---|---|---|---|---|
| Cross-linked SU-8 | 0.1 | 1000 | $3.83 \times 10^{-5}$ | $5.82 \times 10^{-5}$ | 1.93 | 2.13 |
| | | 9000 | $7.99 \times 10^{-5}$ | $2.18 \times 10^{-5}$ | 1.68 | 2.17 |
| | 1.0 | 1000 | $4.73 \times 10^{-5}$ | $1.30 \times 10^{-2}$ | 1.95 | 1.38 |
| | | 9000 | $8.97 \times 10^{-6}$ | $9.78 \times 10^{-7}$ | 1.97 | 2.55 |
| PC | 0.1 | 1000 | $1.99 \times 10^{-4}$ | $6.13 \times 10^{-3}$ | 1.80 | 1.50 |
| | | 9000 | $1.63 \times 10^{-6}$ | $2.15 \times 10^{-7}$ | 2.08 | 2.98 |
| | 1.0 | 1000 | $6.72 \times 10^{-4}$ | $4.94 \times 10^{-4}$ | 1.59 | 1.98 |
| | | 9000 | $7.73 \times 10^{-4}$ | $4.94 \times 10^{-5}$ | 1.47 | 2.18 |
| PMMA | 0.1 | 1000 | $1.24 \times 10^{-4}$ | $2.35 \times 10^{-4}$ | 1.82 | 1.48 |
| | | 9000 | $6.71 \times 10^{-6}$ | $1.04 \times 10^{-6}$ | 1.24 | 1.86 |

# Appendix D: SPM image profiles vs. corresponding proportional profiles

We find an excellent correlation between the consecutive residual profiles and the corresponding residual displacements of the tip apex (eqns. 12 and 13) as shown in Fig. D.1.

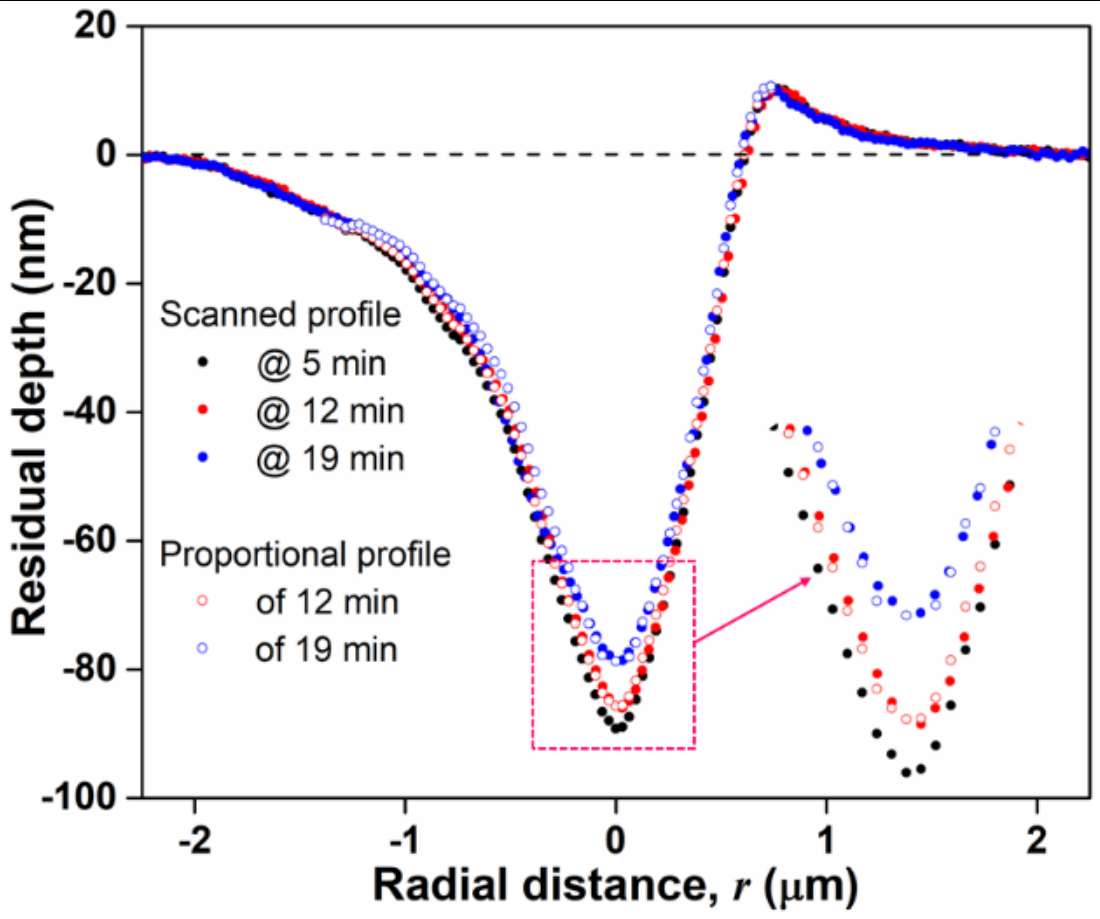

Figure D.1: SPM image profile for cross-linked SU-8 at 5 min, 12 min and 19 min and corresponding proportional profile for $P_m$ = 1000 μN and $\dot{P}/P$ = 1.0 /s. The proportional profile (open circles) agrees very well with the corresponding scanned profile (solid circles).

# Appendix E: Power Law for Elastic Displacement

The fitted $B_e$ and $m_e$ values (Section 5.5 I.4) are listed in Table E.1.

**Table E.1:** Power law fitting parameters between $h_e$ and $h_m$ for cross-linked SU-8, PC and PMMA.

| | Parameters | $\dot{P}/P$ (/s) | |
|---|---|---|---|
| | | 0.1 | 1.0 |
| Cross-linked SU-8 | $B_e$ | 0.70 | 0.47 |
| | $m_e$ | 0.94 | 1.02 |
| PC | $B_e$ | 0.45 | 0.46 |
| | $m_e$ | 0.98 | 0.99 |
| PMMA | $B_e$ | 0.36 | -- |
| | $m_e$ | 0.99 | -- |

# Uniaxial Recovery Perspective of Glassy Polymer Nanoindentation

**Prakash Sarkar[1], Hemant Nanavati[2*]**

[1]Department of Metallurgical Engineering and Materials Science, Indian Institute of Technology Bombay, Mumbai- 400076, Maharashtra, India

[2]Department of Chemical Engineering, Indian Institute of Technology Bombay, Mumbai- 400076, Maharashtra, India

*E-mail: hnanavati@iitb.ac.in

# Supporting Information 1

## Universal curves for residual uniaxial strain

The non-elastic residual strains in uniaxial compression, $\varepsilon_{\mathrm{ne}}$, are obtained for the entire strain-range, by drawing a Hookean line of slope $E(\dot{\varepsilon})$, from the corresponding point on the stress-strain plot, as shown in Fig. 1(a). We define, $\varepsilon^{*}=\varepsilon(\dot{\varepsilon})/\varepsilon_{y}(\dot{\varepsilon})$ and then determine $\varepsilon_{\mathrm{ne}}$ from each $\sigma$-$\varepsilon$ curve, which are then normalized to $\varepsilon_{\mathrm{ne}}^{*}=\varepsilon(\dot{\varepsilon})/\varepsilon_{y}(\dot{\varepsilon})$. The $\varepsilon_{\mathrm{ne}}^{*}$ vs $\varepsilon^{*}$ computations are plotted in Fig. S.1.1 for cross-linked SU-8, in Fig. S.1.2 for PC and in Fig. S.1.3-5 for PMMA. We find an approximately universal corresponding strain state (*CSS*) behaviour, across the strain rates examined here.

### 1. Universal curve for cross-linked SU-8

We have fit a polynomial equation (eqn. S.1.1) for the data range of 0.5 < $\varepsilon^{*}$ < 1.5.

$$\varepsilon_{\mathrm{ne}}^{*}=k_6\left(\varepsilon^{*}-0.5\right)^6+k_5\left(\varepsilon^{*}-0.5\right)^5+k_4\left(\varepsilon^{*}-0.5\right)^4+k_3\left(\varepsilon^{*}-0.5\right)^3+k_2\left(\varepsilon^{*}-0.5\right)^2 \qquad \text{eqn. S.1.1}$$

Hence,

$$\frac{d\varepsilon_{\mathrm{ne}}^{*}}{d\varepsilon^{*}}=6k_6\left(\varepsilon^{*}-0.5\right)^5+5k_5\left(\varepsilon^{*}-0.5\right)^4+4k_4\left(\varepsilon^{*}-0.5\right)^3+3k_3\left(\varepsilon^{*}-0.5\right)^2+2k_2\left(\varepsilon^{*}-0.5\right) \qquad \text{eqn. S.1.2}$$

At $\varepsilon^{*}$ ~ 1.5, $d\varepsilon_{\mathrm{ne}}^{*}/d\varepsilon^{*}$ ~ 1. Therefore, from eqn. S.1.2,

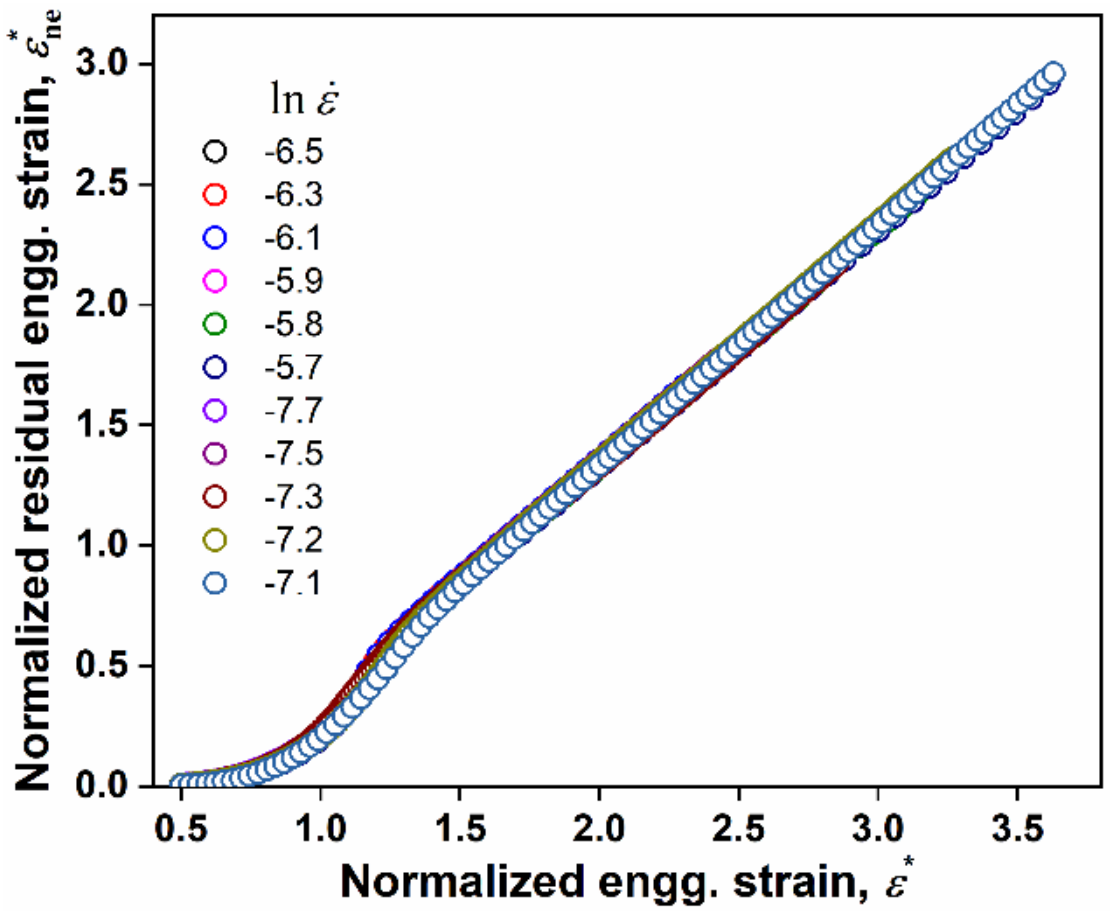

Figure S.1.1: ***CSS*** $\varepsilon^*_{\mathrm{ne}}$ **vs** $\varepsilon^*$ **for cross-linked SU-8 over all strain rates.**

$6k_6\left(\varepsilon^*-0.5\right)^5+5k_5\left(\varepsilon^*-0.5\right)^4+4k_4\left(\varepsilon^*-0.5\right)^3+3k_3\left(\varepsilon^*-0.5\right)^2+2k_2\left(\varepsilon^*-0.5\right)=1$. Hence,

$$k_2=0.5-\left(3k_6+2.5k_5+2k_4+1.5k_3\right) \qquad \text{eqn. S.1.3}$$

Incorporating the eqn. S.1.3 into eqn. S.1.1,

$$\begin{aligned}\varepsilon^*_{\mathrm{ne}}&=k_6\left(\varepsilon^*-0.5\right)^6+k_5\left(\varepsilon^*-0.5\right)^5+k_4\left(\varepsilon^*-0.5\right)^4+k_3\left(\varepsilon^*-0.5\right)^3\\&\quad+\left(0.5-\left(3k_6+2.5k_5+2k_4+1.5k_3\right)\right)\left(\varepsilon^*-0.5\right)^2\end{aligned} \qquad \text{eqn. S.1.4}$$

The resultant fitting constant values are: $k_6$ = 8.57, $k_5$ = -25.58, $k_4$ = 26.21, $k_3$ = -10.67 and from eqn. S.1.4, $k_2$ = 2.32. Therefore, from eqn. S.1.1,

$$\begin{aligned}\varepsilon^*_{\mathrm{ne}}&=8.57\left(\varepsilon^*-0.5\right)^6-25.58\left(\varepsilon^*-0.5\right)^5+26.21\left(\varepsilon^*-0.5\right)^4-10.67\left(\varepsilon^*-0.5\right)^3\\&\quad+2.32\left(\varepsilon^*-0.5\right)^2\end{aligned} \qquad \text{eqn. S.1.5}$$

We have fit a linear equation for the data range of $\varepsilon^*$ > 1.5, $\varepsilon^*_{\mathrm{ne}}=m_{\mathrm{L}}\varepsilon^*+k_{\mathrm{L}}$, $\Rightarrow k_{\mathrm{L}}=\varepsilon^*_{\mathrm{ne}}-m_{\mathrm{L}}\varepsilon^*$.

At $\varepsilon^*$ = 1.5 in eqn. S.1.5, we obtain $\varepsilon^*_{\mathrm{ne}}$ = 0.85 and $m_{\mathrm{L}}$ = 1. Then, $k_{\mathrm{L}}$ = 0.85-1.5 = -0.65. Thus,

$$\varepsilon_{\mathrm{ne}}^{*} = \varepsilon^{*} - 0.65 \qquad \text{eqn. S.1.6}$$

## 2. Universal curve for PC

In the data range of 0.3 < $\varepsilon^{*}$ < 1.2,

$$\varepsilon_{\mathrm{ne}}^{*} = k_4\left(\varepsilon^{*} - 0.3\right)^4 + k_3\left(\varepsilon^{*} - 0.3\right)^3 + k_2\left(\varepsilon^{*} - 0.3\right)^2 \qquad \text{eqn. S.1.7}$$

Hence,

$$\frac{d\varepsilon_{\mathrm{ne}}^{*}}{d\varepsilon^{*}} = 4k_4\left(\varepsilon^{*} - 0.3\right)^3 + 3k_3\left(\varepsilon^{*} - 0.3\right)^2 + 2k_2\left(\varepsilon^{*} - 0.3\right) \qquad \text{eqn. S.1.8}$$

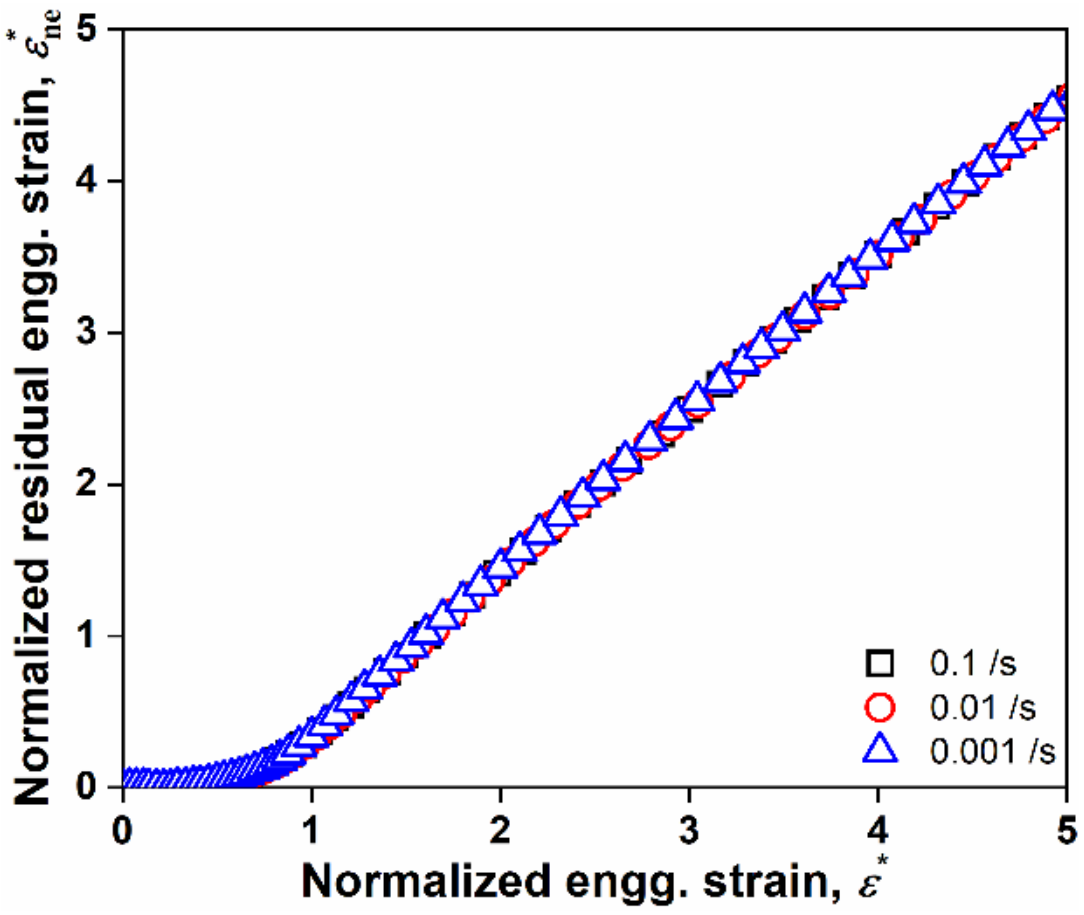

Figure S.1.2: ***CSS*** $\varepsilon_{\mathrm{ne}}^{*}$ **vs** $\varepsilon^{*}$ **for PC over all strain rates.**

At $\varepsilon^{*} \sim 1.2$, $d\varepsilon_{\mathrm{ne}}^{*}/d\varepsilon^{*} \sim 1.10$. Therefore, from eqn. S.1.8,

$4k_4(1.2-0.3)^3 + 3k_3(1.2-0.3)^2 + 2k_2(1.2-0.3) = 1.10$. Therefore,

$$k_2 = 0.61 - \left(1.62k_4 + 1.35k_3\right) \qquad \text{eqn. S.1.9}$$

Combining eqn. S.1.7 and eqn. S.1.9,

$$\varepsilon_{\mathrm{ne}}^{*}=k_4\left(\varepsilon^{*}-0.3\right)^4+k_3\left(\varepsilon^{*}-0.3\right)^3+\left(0.61-\left(1.62k_4+1.35k_3\right)\right)\left(\varepsilon^{*}-0.3\right)^2 \quad \text{eqn. S.1.10}$$

When, $\varepsilon^{*}$ > 1.2,

$$\varepsilon_{\mathrm{ne}}^{*}=1.10\varepsilon^{*}-0.77=0.55 \quad \text{eqn. S.1.11}$$

Thus, eqn. S.1.10 becomes,

$$k_4\left(\varepsilon^{*}-0.3\right)^4+k_3\left(\varepsilon^{*}-0.3\right)^3+\left(0.61-\left(1.62k_4+1.35k_3\right)\right)\left(\varepsilon^{*}-0.3\right)^2=0.55$$

$\Rightarrow k_4\left(1.2-0.3\right)^4+k_3\left(1.2-0.3\right)^3+\left(0.61-\left(1.62k_4+1.35k_3\right)\right)\left(1.2-0.3\right)^2=0.55$. Hence,

$$k_3=-1.8k_4-0.16 \quad \text{eqn. S.1.12}$$

From eqn. S.1.9 and eqn. S.1.12,

$$k_2=0.61-\left(1.62k_4+1.35\left(-1.8k_4-0.16\right)\right),\ \Rightarrow k_2=0.83+0.81k_4$$

Incorporating the above values in eqn. S.1.10,

$$\varepsilon_{\mathrm{ne}}^{*}=k_4\left(\varepsilon^{*}-0.3\right)^4+\left(-1.8k_4-0.16\right)\left(\varepsilon^{*}-0.3\right)^3+\left(0.83+0.81k_4\right)\left(\varepsilon^{*}-0.3\right)^2 \quad \text{eqn. S.1.13}$$

From fitting, $k_4$ = -0.49. Thus, $k_3$ = 0.72, $k_2$ = 0.44. Therefore,

$$\varepsilon_{\mathrm{ne}}^{*}=-0.49\left(\varepsilon^{*}-0.3\right)^4+0.72\left(\varepsilon^{*}-0.3\right)^3+0.44\left(\varepsilon^{*}-0.3\right)^2 \quad \text{eqn. S.1.14}$$

## 3.A. Universal curve for PMMA [28]

In the data range of 0.4 < $\varepsilon^{*}$ < 1.2,

$$\varepsilon_{\text{ne}}^{*}=k_4\left(\varepsilon^{*}-0.4\right)^4+k_3\left(\varepsilon^{*}-0.4\right)^3+k_2\left(\varepsilon^{*}-0.4\right)^2 \qquad \text{eqn. S.1.15}$$

Hence,

$$\frac{d\varepsilon_{\text{ne}}^{*}}{d\varepsilon^{*}}=4k_4\left(\varepsilon^{*}-0.4\right)^3+3k_3\left(\varepsilon^{*}-0.4\right)^2+2k_2\left(\varepsilon^{*}-0.4\right) \qquad \text{eqn. S.1.16}$$

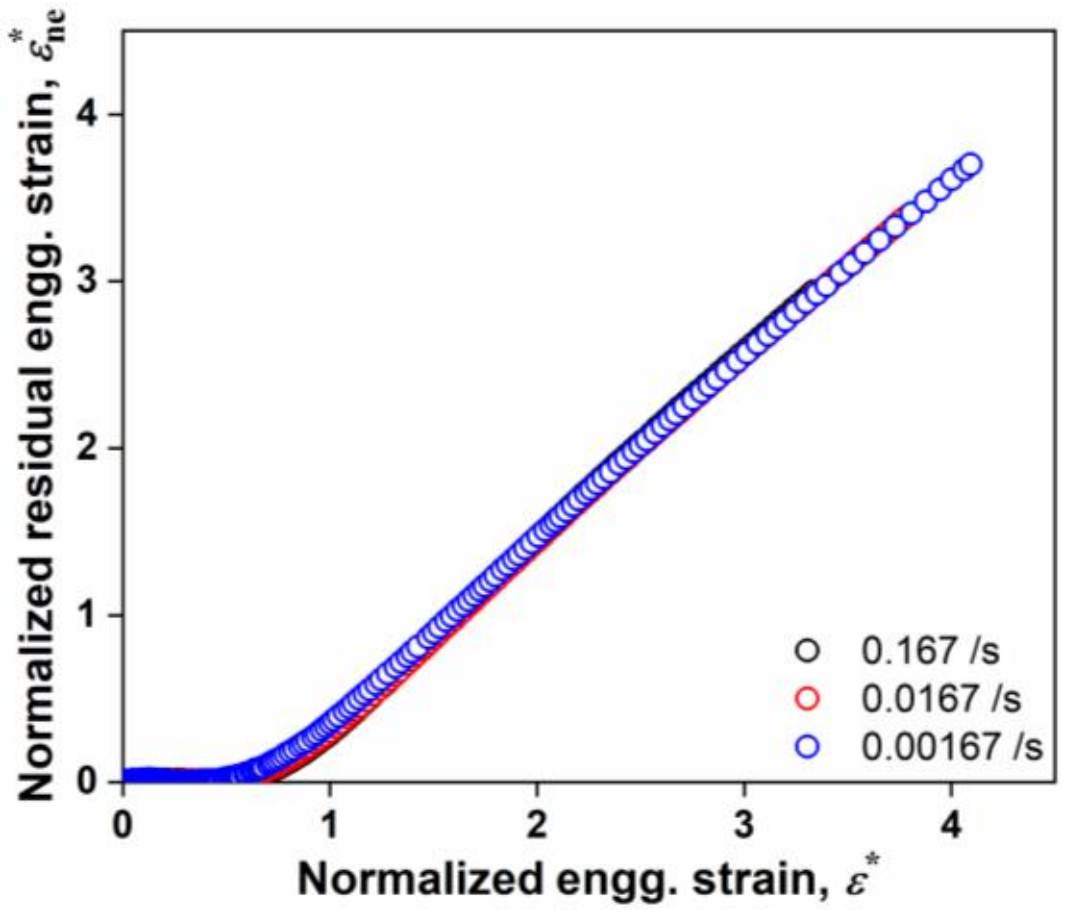

Figure S.1.3: ***CSS*** $\varepsilon_{\text{ne}}^{*}$ **vs** $\varepsilon^{*}$ **for PMMA [28] over all strain rates.**

At $\varepsilon^{*}$ ~ 1.2, $d\varepsilon_{\text{ne}}^{*}/d\varepsilon^{*}$ ~ 1.11. Therefore, from eqn. S.1.16,

$4k_4(1.2-0.4)^3+3k_3(1.2-0.4)^2+2k_2(1.2-0.4)=1.11$. Therefore,

$$2.05k_4+1.92k_3+1.6k_2=1.11$$

$$k_2=0.695-\left(1.28k_4+1.2k_3\right) \qquad \text{eqn. S.1.17}$$

Combining eqn. S.1.15 and eqn. S.1.17,

$$\varepsilon_{\text{ne}}^{*}=k_4\left(\varepsilon^{*}-0.4\right)^4+k_3\left(\varepsilon^{*}-0.4\right)^3+\left(0.695-\left(1.28k_4+1.2k_3\right)\right)\left(\varepsilon^{*}-0.4\right)^2 \qquad \text{eqn. S.1.18}$$

The resultant fitting constant values are: $k_4$ = -0.407, $k_3$ = 0.337 and from eqn. S.1.17, $k_2$ = 0.812.

Therefore, from eqn. S.1.15,

$$\varepsilon_{\mathrm{ne}}^{*} = -0.407\left(\varepsilon^{*} - 0.4\right)^{4} + 0.337\left(\varepsilon^{*} - 0.4\right)^{3} + 0.812\left(\varepsilon^{*} - 0.4\right)^{2} \qquad \text{eqn. S.1.19}$$

We have fit a linear equation for the data range of $\varepsilon^{*}$ > 1.2, $\varepsilon_{\mathrm{ne}}^{*} = m_{\mathrm{L}}\varepsilon^{*} + k_{\mathrm{L}}$, $\Rightarrow k_{\mathrm{L}} = \varepsilon_{\mathrm{ne}}^{*} - m_{\mathrm{L}}\varepsilon^{*}$.

At $\varepsilon^{*}$ = 1.2 in eqn. S.1.19, we obtain $\varepsilon_{\mathrm{ne}}^{*}$ = 0.53 and $m_{\mathrm{L}}$ = 1.11. Then, $k_{\mathrm{L}}$ = 0.53-1.33 = -0.804. Thus,

$$\varepsilon_{\mathrm{ne}}^{*} = 1.11\varepsilon^{*} - 0.804 \qquad \text{eqn. S.1.20}$$

## 3.B. Universal curve for PMMA [29]

In the data range of 0.3 < $\varepsilon^{*}$ < 1.2,

$$\varepsilon_{\mathrm{ne}}^{*} = k_4\left(\varepsilon^{*} - 0.3\right)^{4} + k_3\left(\varepsilon^{*} - 0.3\right)^{3} + k_2\left(\varepsilon^{*} - 0.3\right)^{2} \qquad \text{eqn. S.1.21}$$

Hence,

$$\frac{d\varepsilon_{\mathrm{ne}}^{*}}{d\varepsilon^{*}} = 4k_4\left(\varepsilon^{*} - 0.3\right)^{3} + 3k_3\left(\varepsilon^{*} - 0.3\right)^{2} + 2k_2\left(\varepsilon^{*} - 0.3\right) \qquad \text{eqn. S.1.22}$$

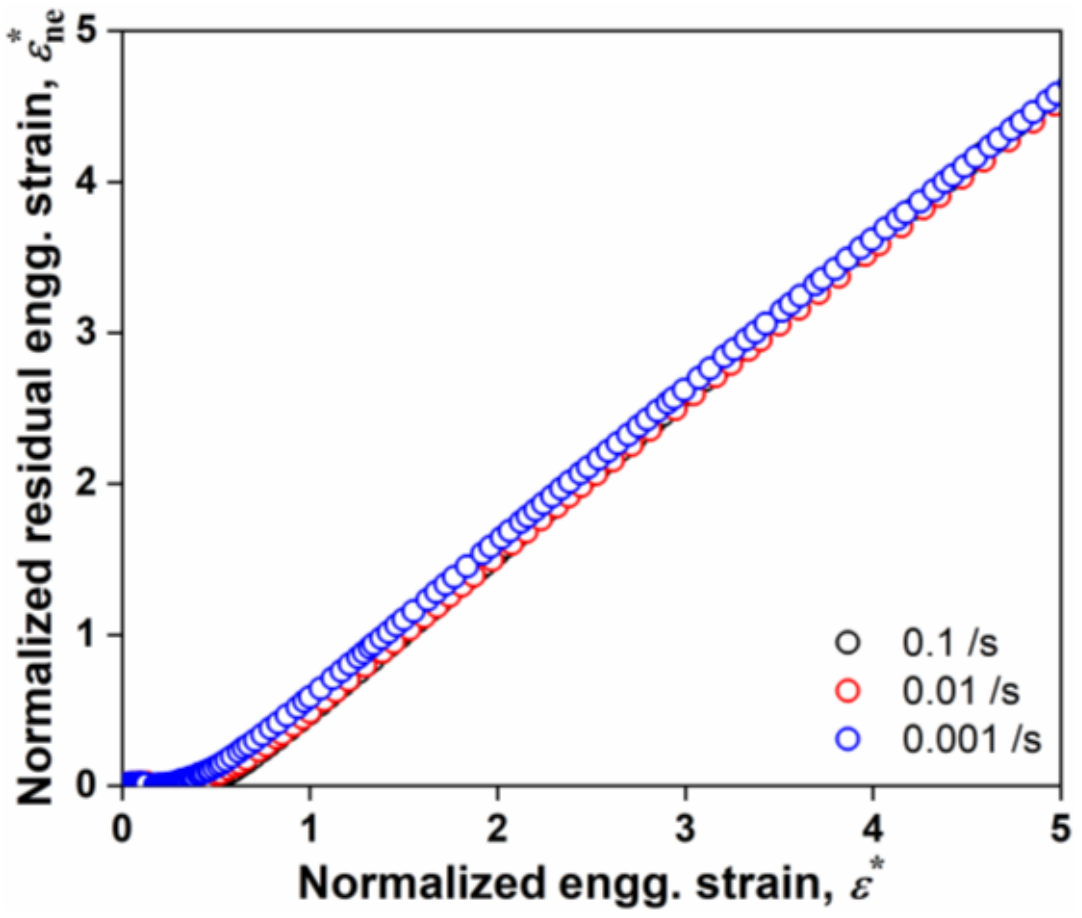

Figure S.1.4: ***CSS*** $\varepsilon_{\mathrm{ne}}^{*}$ **vs** $\varepsilon^{*}$ **for PMMA [29] over all strain rates.**

At $\varepsilon^{*}$ ~ 1.2, $d\varepsilon_{\mathrm{ne}}^{*}/d\varepsilon^{*}$ ~ 1.02. Therefore, from eqn. S.1.22,

$4k_4(1.2-0.3)^3+3k_3(1.2-0.3)^2+2k_2(1.2-0.3)=1.02$. Therefore,

$2.92k_4+2.43k_3+1.8k_2=1.02$

$$k_2=0.565-(1.62k_4+1.35k_3) \qquad \text{eqn. S.1.23}$$

Combining eqn. S.1.21 and eqn. S.1.23,

$$\varepsilon_{\mathrm{ne}}^{*}=k_4\left(\varepsilon^{*}-0.3\right)^4+k_3\left(\varepsilon^{*}-0.3\right)^3+\left(0.565-\left(1.62k_4+1.35k_3\right)\right)\left(\varepsilon^{*}-0.3\right)^2 \qquad \text{eqn. S.1.24}$$

The resultant fitting constant values are: $k_4$ = 0.996, $k_3$ = -2.50 and from eqn. S.1.23, $k_2$ = 2.33. Therefore, from eqn. S.1.21,

$$\varepsilon_{\mathrm{ne}}^{*}=0.996\left(\varepsilon^{*}-0.3\right)^4-2.50\left(\varepsilon^{*}-0.3\right)^3+2.33\left(\varepsilon^{*}-0.3\right)^2 \qquad \text{eqn. S.1.25}$$

We have fit a linear equation for the data range of $\varepsilon^{*}$ > 1.2, $\varepsilon_{\mathrm{ne}}^{*}=m_{\mathrm{L}}\varepsilon^{*}+k_{\mathrm{L}}$, $\Rightarrow k_{\mathrm{L}}=\varepsilon_{\mathrm{ne}}^{*}-m_{\mathrm{L}}\varepsilon^{*}$.

At $\varepsilon^*$ = 1.2 in eqn. S.1.25, we obtain $\varepsilon^*_{\mathrm{ne}}$ = 0.716 and $m_{\mathrm{L}}$ = 1.02. Then, $k_{\mathrm{L}}$ = 0.716-1.22 = -0.504.

Thus,

$$\varepsilon^*_{\mathrm{ne}} = 1.02\varepsilon^* - 0.504 \qquad \text{eqn. S.1.26}$$

## 3.C. Universal curve for PMMA [30]

In the data range of 0.3 < $\varepsilon^*$ < 1.2,

$$\varepsilon^*_{\mathrm{ne}} = k_4\left(\varepsilon^* - 0.3\right)^4 + k_3\left(\varepsilon^* - 0.3\right)^3 + k_2\left(\varepsilon^* - 0.3\right)^2 \qquad \text{eqn. S.1.27}$$

Hence,

$$\frac{d\varepsilon^*_{\mathrm{ne}}}{d\varepsilon^*} = 4k_4\left(\varepsilon^* - 0.3\right)^3 + 3k_3\left(\varepsilon^* - 0.3\right)^2 + 2k_2\left(\varepsilon^* - 0.3\right) \qquad \text{eqn. S.1.28}$$

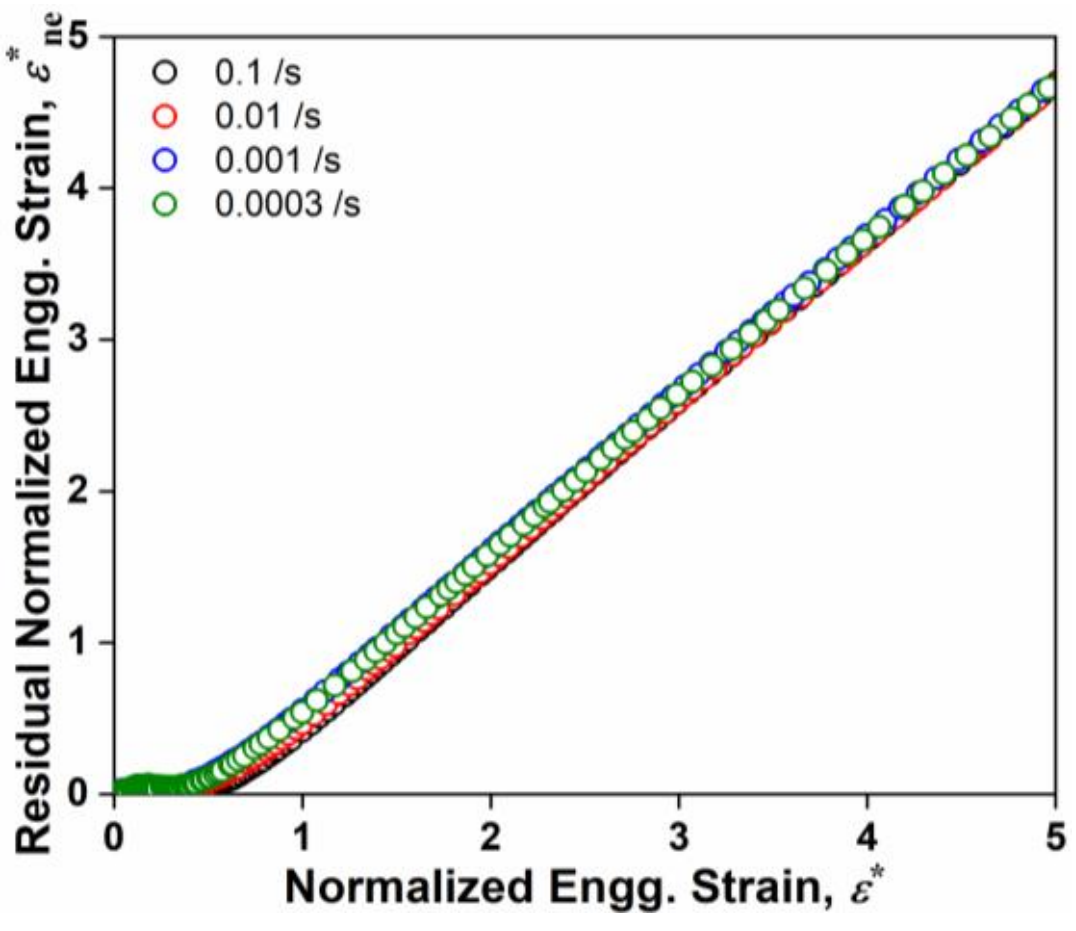

Figure S.1.5: ***CSS*** **$\varepsilon^*_{\mathrm{ne}}$ vs $\varepsilon^*$ for PMMA [30] over all strain rates.**

At $\varepsilon^*$ ~ 1.2, $d\varepsilon^*_{\mathrm{ne}}/d\varepsilon^*$ ~ 1.06. Therefore, from eqn. S.1.28,

$4k_4\left(1.2-0.3\right)^3 + 3k_3\left(1.2-0.3\right)^2 + 2k_2\left(1.2-0.3\right) = 1.06$. Therefore,

$$2.92k_4 + 2.43k_3 + 1.8k_2 = 1.06$$

$$k_2 = 0.587 - \left(1.62k_4 + 1.35k_3\right) \qquad \text{eqn. S.1.29}$$

Combining eqn. S.1.27 and eqn. S.1.29,

$$\varepsilon_{\mathrm{ne}}^{*} = k_4\left(\varepsilon^{*} - 0.3\right)^4 + k_3\left(\varepsilon^{*} - 0.3\right)^3 + \left(0.587 - \left(1.62k_4 + 1.35k_3\right)\right)\left(\varepsilon^{*} - 0.3\right)^2 \qquad \text{eqn. S.1.30}$$

The resultant fitting constant values are: $k_4$ = 1.08, $k_3$ = -2.51 and from eqn. S.1.29, $k_2$ = 2.22. Therefore, from eqn. S.1.27,

$$\varepsilon_{\mathrm{ne}}^{*} = 1.08\left(\varepsilon^{*} - 0.3\right)^4 - 2.51\left(\varepsilon^{*} - 0.3\right)^3 + 2.22\left(\varepsilon^{*} - 0.3\right)^2 \qquad \text{eqn. S.1.31}$$

We have fit a linear equation for the data range of $\varepsilon^{*}$ > 1.2, $\varepsilon_{\mathrm{ne}}^{*} = m_{\mathrm{L}}\varepsilon^{*} + k_{\mathrm{L}}$, $\Rightarrow k_{\mathrm{L}} = \varepsilon_{\mathrm{ne}}^{*} - m_{\mathrm{L}}\varepsilon^{*}$.

At $\varepsilon^{*}$ = 1.2 in eqn. S.1.31, we obtain $\varepsilon_{\mathrm{ne}}^{*}$ = 0.679 and $m_{\mathrm{L}}$ = 1.06. Then, $k_{\mathrm{L}}$ = 0.679-1.27 = -0.589. Thus,

$$\varepsilon_{\mathrm{ne}}^{*} = 1.06\varepsilon^{*} - 0.589 \qquad \text{eqn. S.1.32}$$

# Supporting Information 2

## S.2.1. Load-displacement curves

The $P$-$h$ curves for cross-linked SU-8, PC and PMMA are shown in Fig. S.2.1(a)-(b), Fig. S.2.2(a)-(b) and Fig. S.2.3(a)-(b), respectively, for two levels of maximum load, $P_{\mathrm{m}}$, i.e., 1000 μN and 9000 μN.

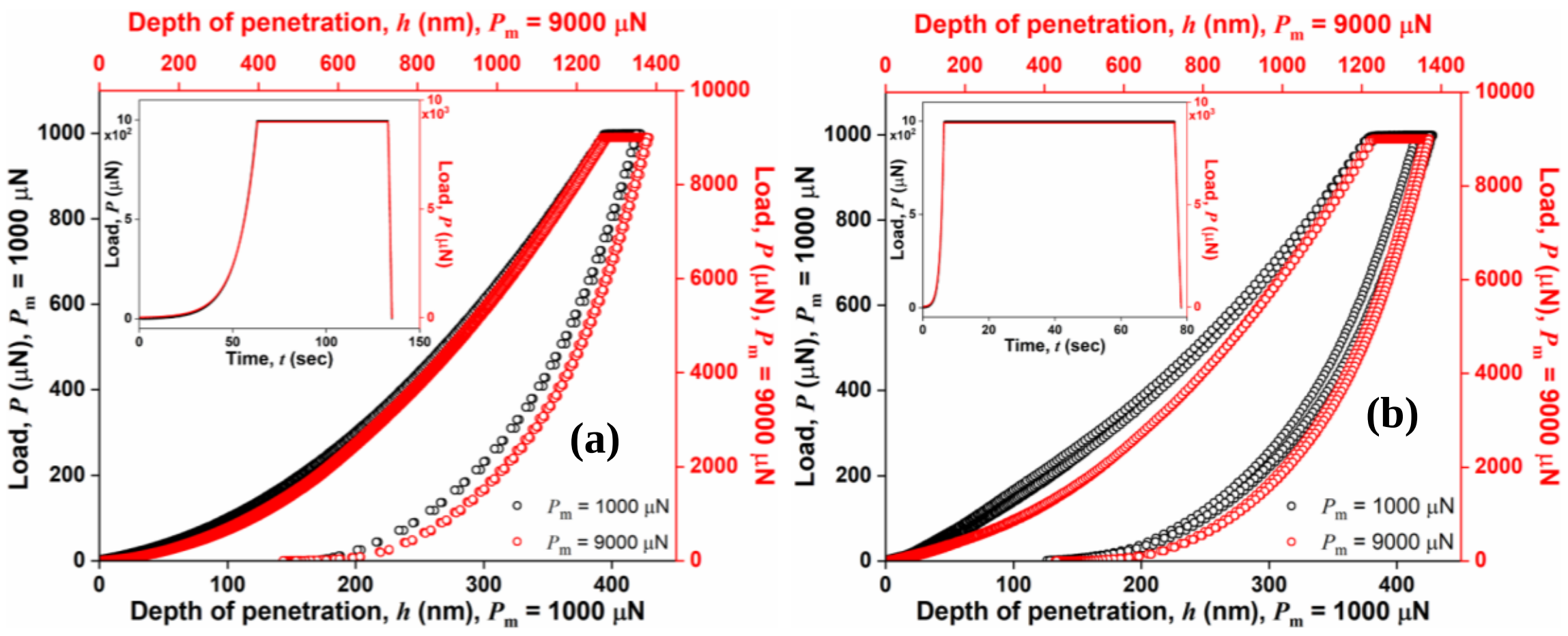

Figure S.2.1: $P$-$h$ **curves for SU-8 for** $P_{\mathrm{m}}$ **= 1000 μN and 9000 μN and** $\dot{P}/P$ **= (a) 0.1 /s, (b) 1.0 /s.**

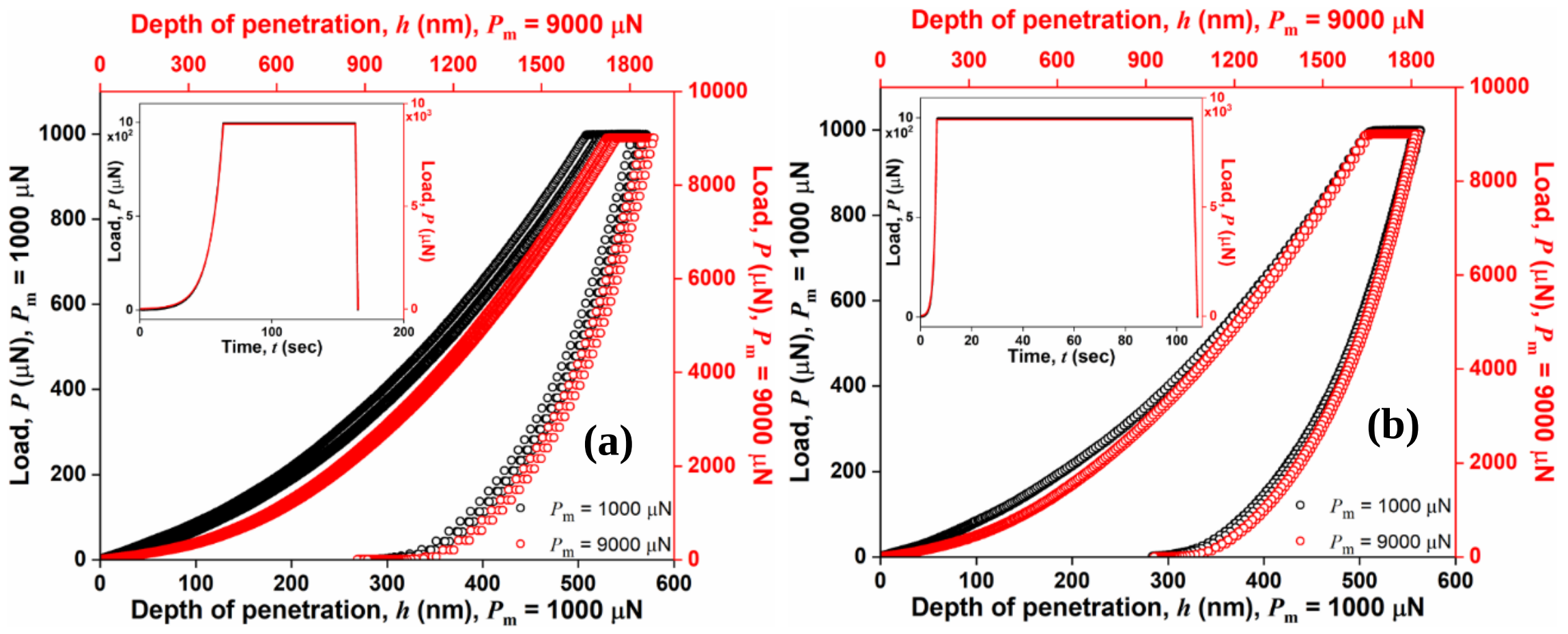

Figure S.2.2: $P$-$h$ **curves for PC for** $P_{\mathrm{m}}$ **= 1000 μN and 9000 μN and** $\dot{P}/P$ **= (a) 0.1 /s, (b) 1.0 /s.**

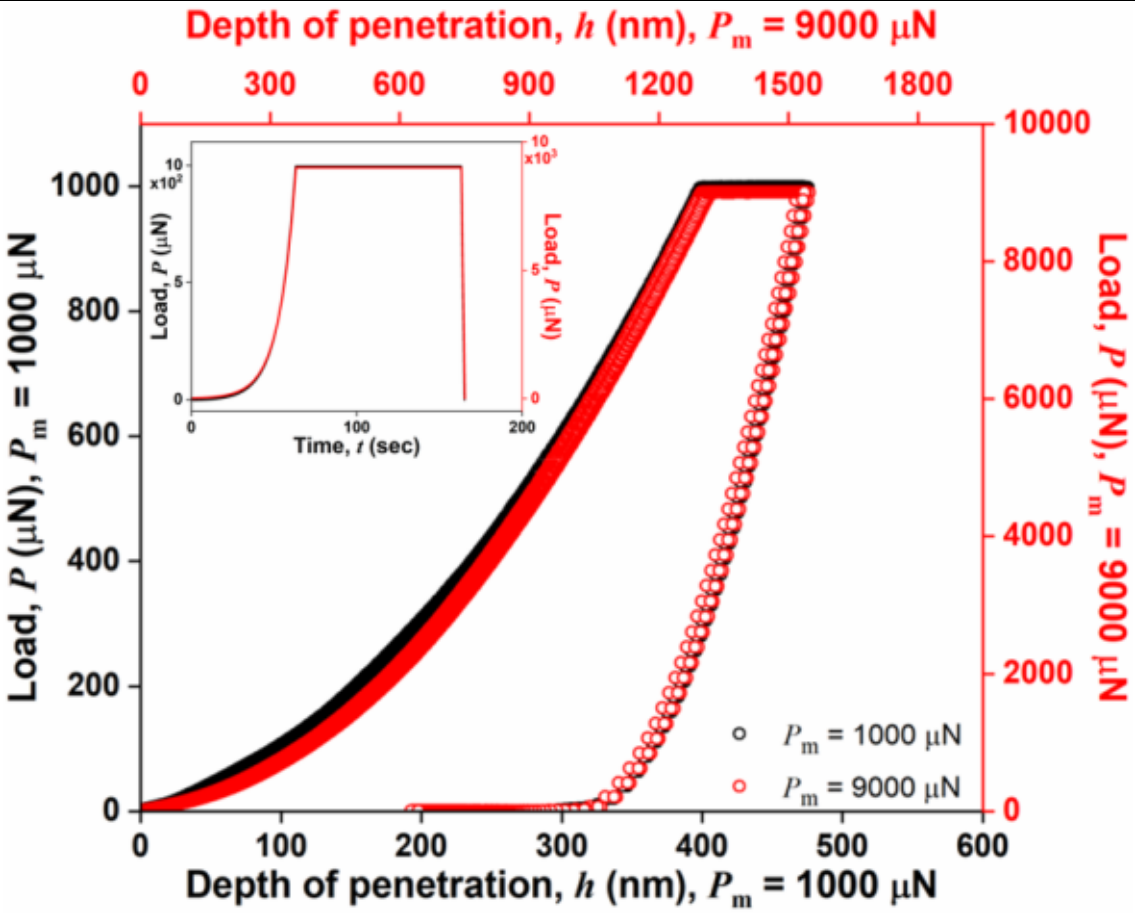

Figure S.2.3: $P$-$h$ **curves for PMMA for** $P_{\mathrm{m}}$ **= 1000 µN and 9000 µN and** $\dot{P}/P$ **= 0.1 /s.**

# Supporting Information 3

## Unloading rate variation during unloading

The unloading step duration is fixed at 2 s. During unloading from Pm, the indenter tip begins from rest, accelerates to reach the set rate at 0.95 $P_m$, unloads at this rate up to 0.3 $P_m$, and then decelerates to zero rate at zero load. This is true for all materials (including all the polymers examined here, and for quartz), and for both maximum loads, $P_m$ = 1000 µN and $P_m$ = 9000 µN. The scaled unloading rate is plotted vs the scaled instantaneous load in Fig. S.4.1.

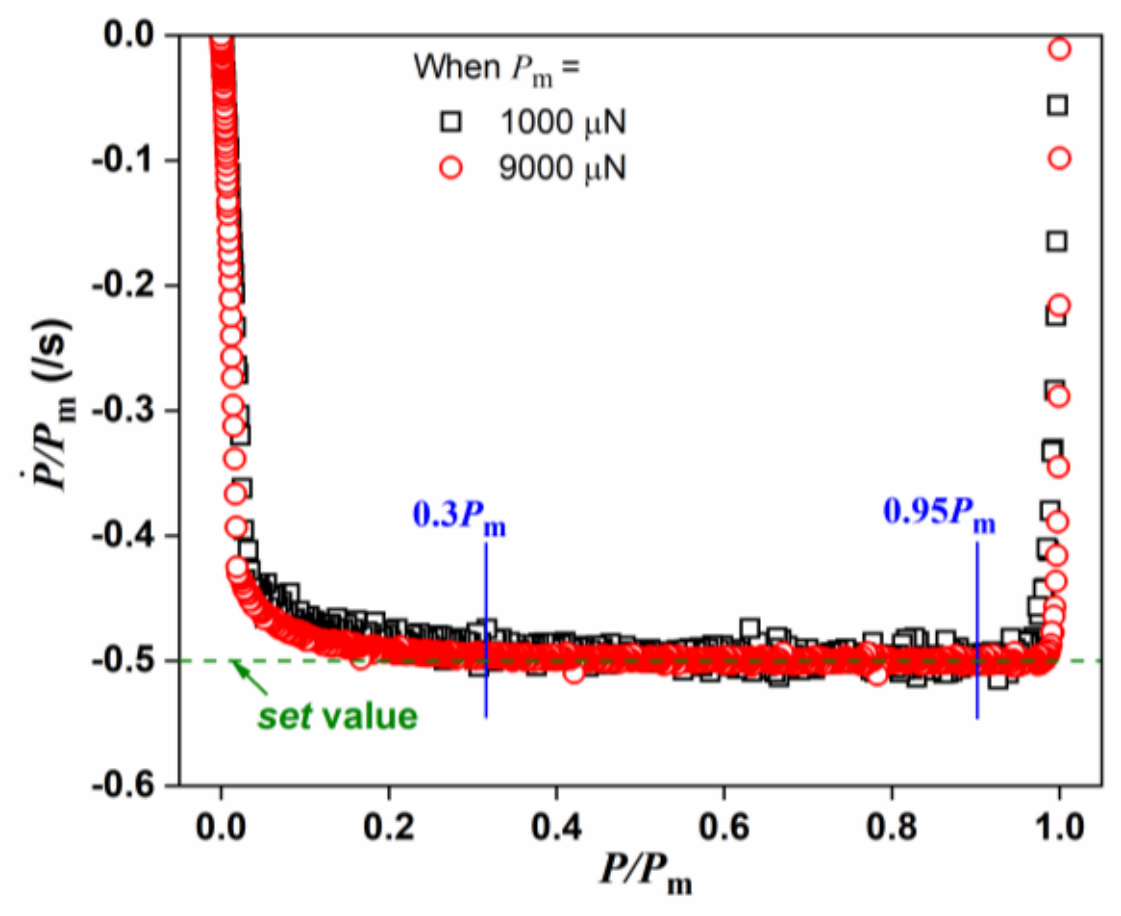

Figure S.3.1: **Scaled unloading rate, $\dot{P}/P_m$ vs scaled load, $P/P_m$, for cross-linked SU-8 for $P_m$ = 1000 µN and $P_m$ = 9000 µN. The acceleration region and deceleration region, are indication by their boundaries at 0.95 $P_m$ and 0.3 $P_m$, respectively.**

# Supporting Information 4

The generalized power law (GPL) fitting has been described in Section 5.2 I. The corresponding fitted values are listed in Table S.4.1.

**Table S.4.1:** GPL fitted constants and their ranges.

| Sample | $\dot{P}/P$ (/s) | $P_{\mathrm{m}}$ (μN) | $k_0$ | $k_1$ | $k_2$ | $1/m$ (Value or range) | |
|---|---|---|---|---|---|---|---|
| | | | | | | GPL | PL |
| Cross-linked SU-8 | 0.1 | 1000 | 7.94 ± 0.76 | -2.20 ± 0.23 | 0.16 ± 0.02 | 0.48 ± 0.00 to 0.62 ± 0.01 | 0.33 |
| | | 9000 | 9.55 ± 0.65 | -1.96 ± 0.15 | 0.11 ± 0.01 | 0.40 ± 0.00 to 0.56 ± 0.01 | 0.34 |
| | 1.0 | 1000 | 9.36 ± 0.80 | -2.59 ± 0.24 | 0.19 ± 0.02 | 0.44 ± 0.01 to 0.62 ± 0.02 | 0.34 |
| | | 9000 | 5.89 ± 0.24 | -1.20 ± 0.06 | 0.06 ± 0.00 | 0.32 ± 0.00 to 0.41 ± 0.00 | 0.36 |
| PC | 0.1 | 1000 | 5.12 ± 0.74 | -1.28 ± 0.22 | 0.09 ± 0.02 | 0.46 ± 0.02 to 0.62 ± 0.03 | 0.35 |
| | | 9000 | 8.74 ± 0.67 | -1.80 ± 0.14 | 0.10 ± 0.01 | 0.45 ± 0.01 to 0.59 ± 0.02 | 0.36 |
| | 1.0 | 1000 | 9.93 ± 2.86 | -2.78 ± 0.81 | 0.21 ± 0.06 | 0.49 ± 0.05 to 0.65 ± 0.11 | 0.38 |
| | | 9000 | 9.52 ± 2.25 | -2.03 ± 0.48 | 0.11 ± 0.03 | 0.41 ± 0.04 to 0.50 ± 0.07 | 0.43 |
| PMMA | 0.1 | 1000 | 1.51 ± 0.01 | -0.13 ± 0.01 | -- | 0.65 ± 0.00 to 0.79 ± 0.01 | 0.52 |
| | | 9000 | 0.89 ± 0.01 | -0.03 ± 0.01 | -- | 0.57 ± 0.00 to 0.60 ± 0.00 | 0.54 |

# Supporting Information 5

## S.5.1. Scanning probe microscopy images

The SPM top view image are presented

on cross-linked SU-8

at $\dot{P}/P$ = 0.1 /s for $P_m$ = 1000 μN (Fig. S.5.1(a)), for $P_m$ = 9000 μN (Fig. S.5.1(b)),

at $\dot{P}/P$ = 0.1 /s for $P_m$ = 1000 μN (Fig. S.5.2(a)), for $P_m$ = 9000 μN (Fig. S.5.2(b)),

on PC

at $\dot{P}/P$ = 0.1 /s for $P_m$ = 1000 μN (Fig. S.5.3(a)), for $P_m$ = 9000 μN (Fig. S.5.3(b)) and

at $\dot{P}/P$ = 1.0 /s for $P_m$ = 1000 μN (Fig. S.5.4(a)), for $P_m$ = 9000 μN (Fig. S.5.4(b)),

and on PMMA

at $\dot{P}/P$ = 0.1 /s for $P_m$ = 1000 μN (Fig. S.5.5(a)), for $P_m$ = 9000 μN (Fig. S.5.5(b)).

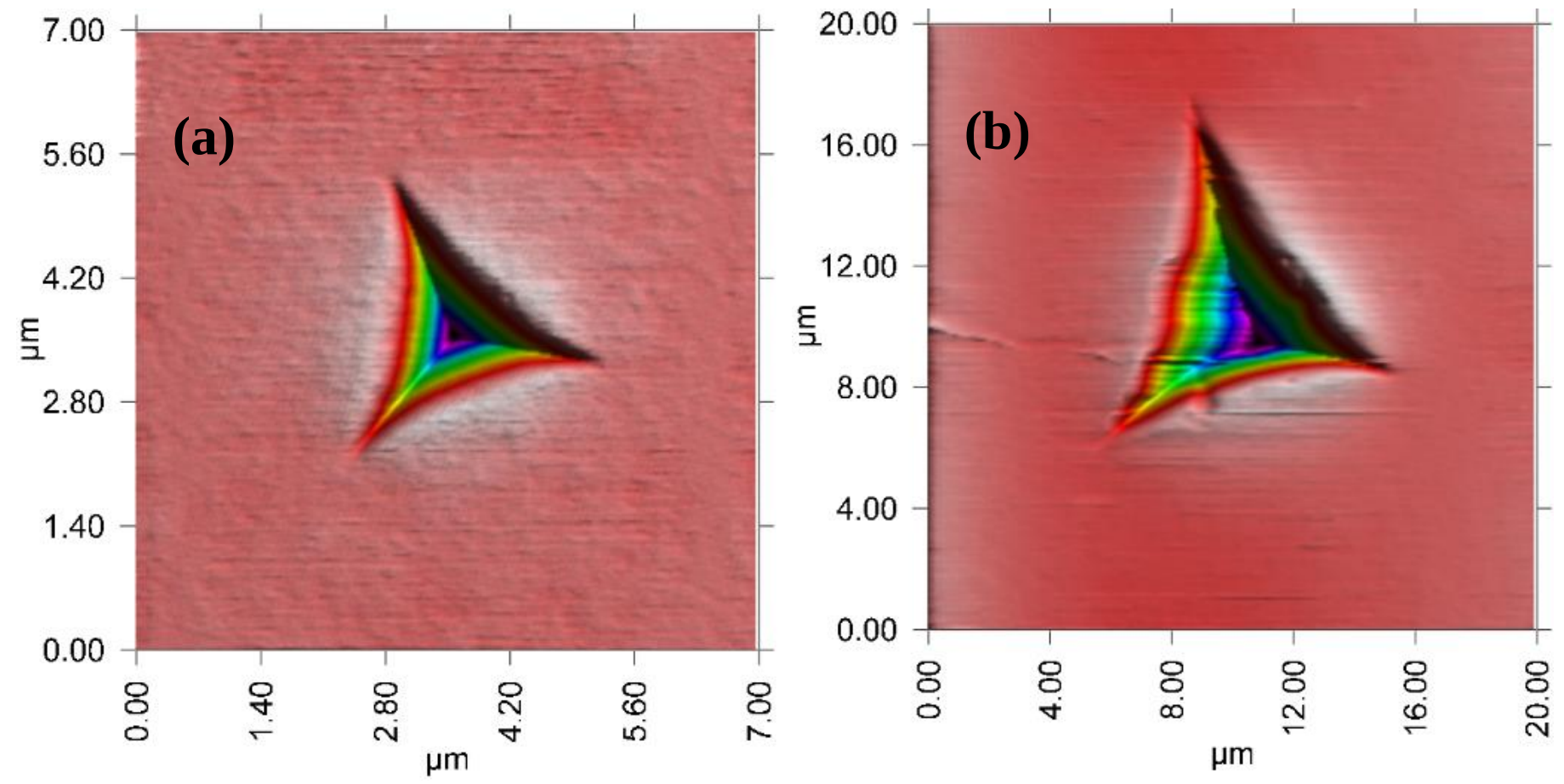

Figure S.5.1: **Top view of residual imprint for cross-linked SU-8 when $\dot{P}/P$ = 0.1 /s and $P_m$ = (a) 1000 μN, (b) 9000 μN.**

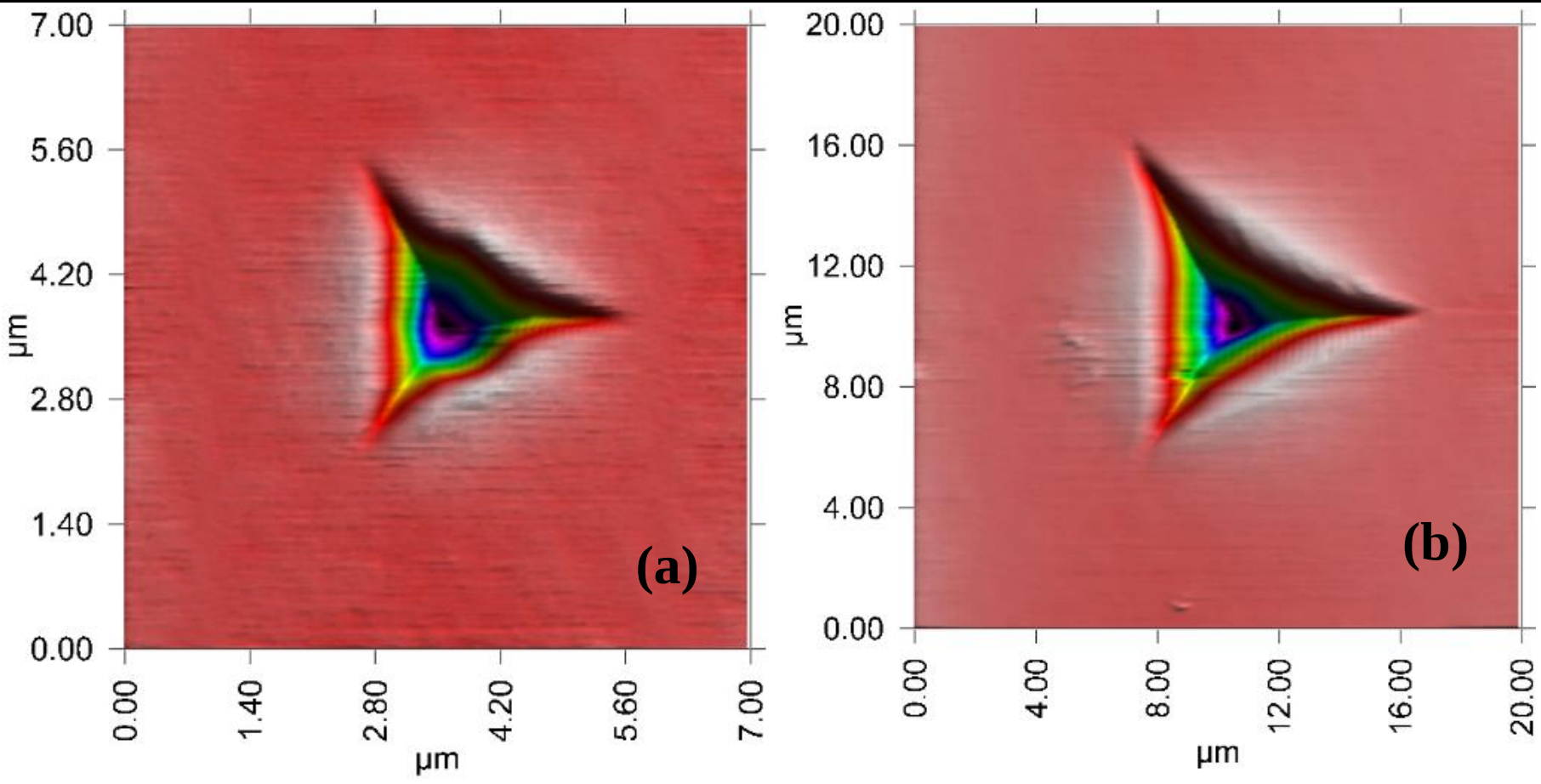

Figure S.5.2: **Top view of the residual imprints for cross-linked SU-8 for $\dot{P}/P$ = 1.0 /s and $P_m$ = (a) 1000 µN, (b) 9000 µN.**

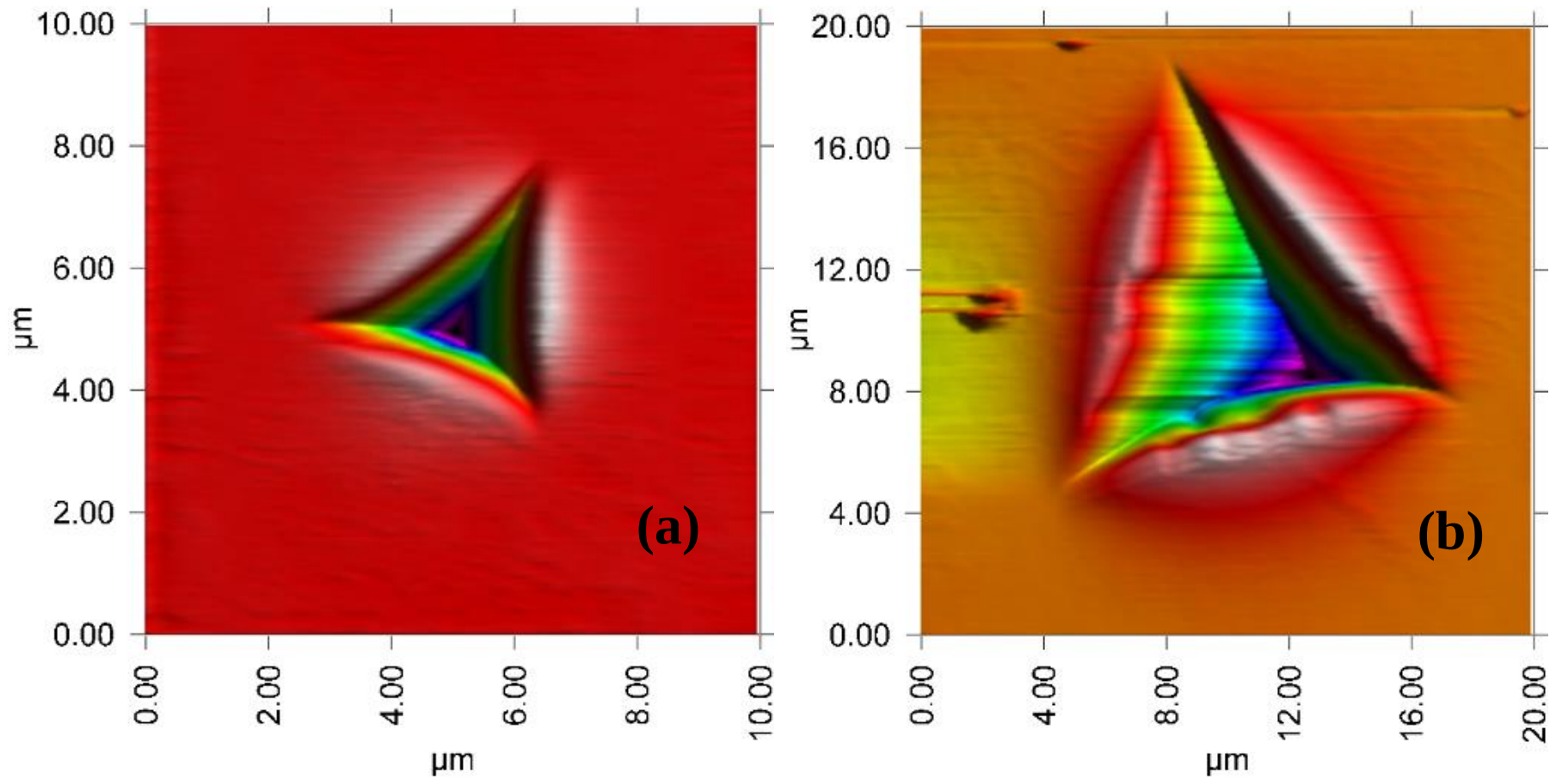

Figure S.5.3: **Top view of the residual imprints for PC for $\dot{P}/P$ = 0.1 /s and $P_m$ = (a) 1000 µN, (b) 9000 µN.**

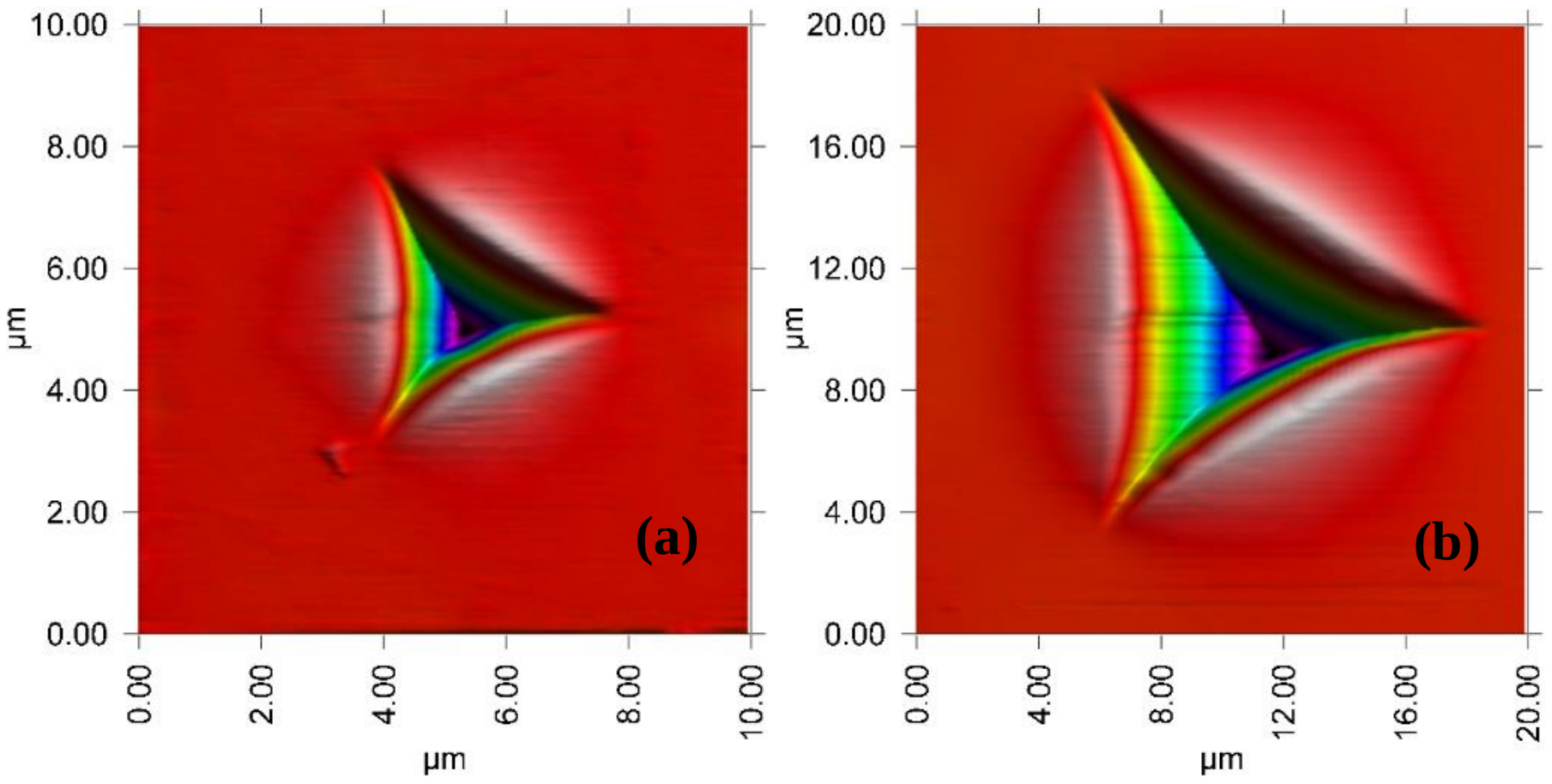

Figure S.5.4: **Top view of the residual imprints for PC for $\dot{P}/P$ = 1.0 /s and $P_m$ = (a) 1000 µN, (b) 9000 µN.**

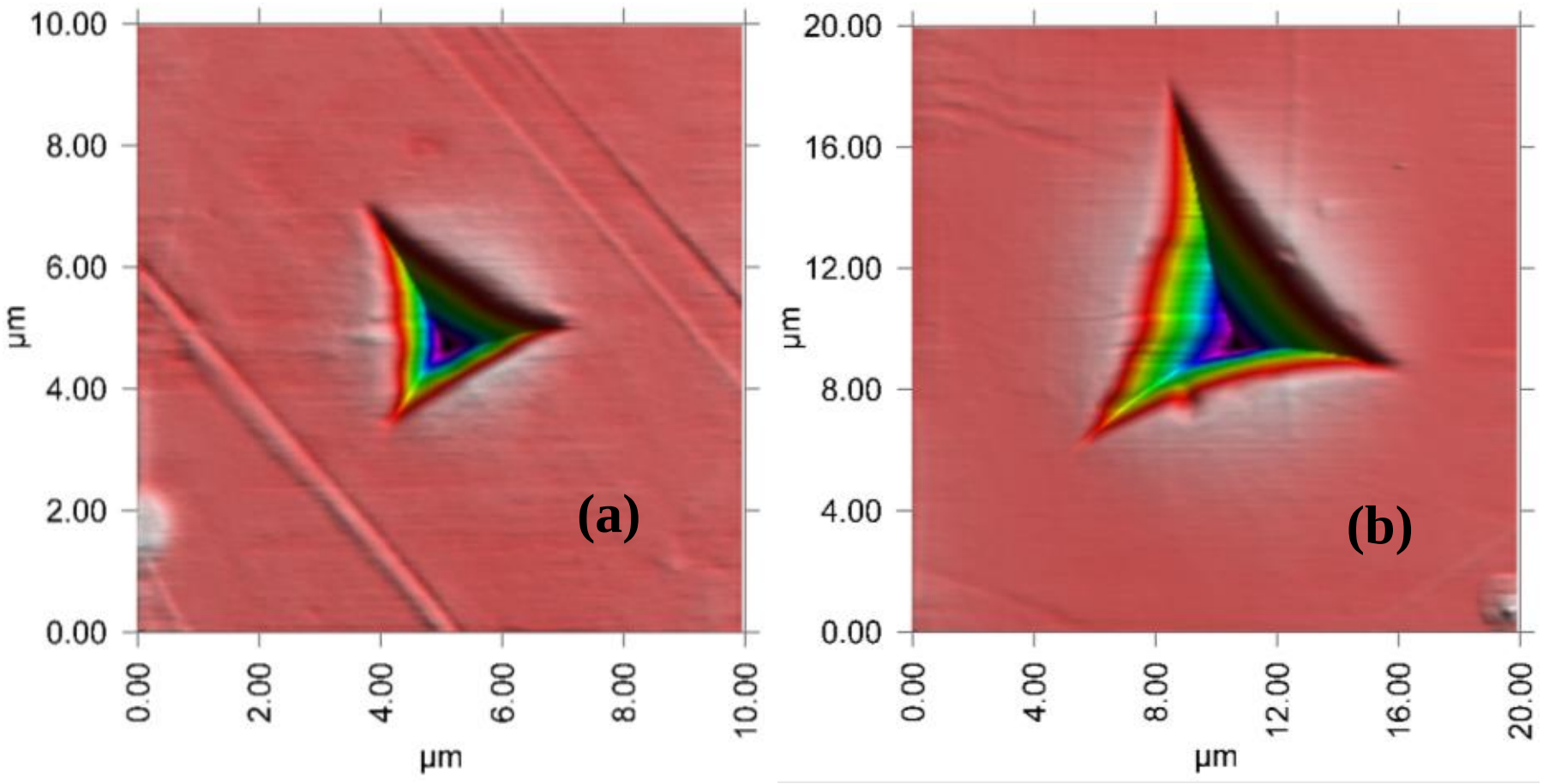

Figure S.5.5: **Top view of the residual imprints for PMMA for $\dot{P}/P$ = 0.1 /s and $P_\text{m}$ = (a) 1000 μN, (b) 9000 μN.**

# Supporting Information 6

## Indenter tip blunt height ($h_b$) estimate

The $P$-$h$ curves of the tip calibration on quartz standard, are employed to evaluate $h_b$ at different usage days, for a given tip. For each $P_{\mathrm{m}}$, the known $E_r$ = 69.6 ± 3.5 GPa, and $H$ = 9.3 ± 0.9 GPa, yield two $A_c$ values, $A_c^{Er} = \pi S^2 / 4E_r^2$ and $A_c^H = P_{\mathrm{m}} / H$, respectively. We could choose either, and we have considered the average.

The Berkovich tip contact area, $A_c(h) \approx 24.5 \times (h_c + h_b)^2$. Thus, for any calibration day, we obtain an estimate for $h_b$. For any experiment day, $h_b$ varies linearly with the age of the tip; $h_b = pt + q$, where $t$ is the duration of tip usage, measured in days (Fig. S.6.1).

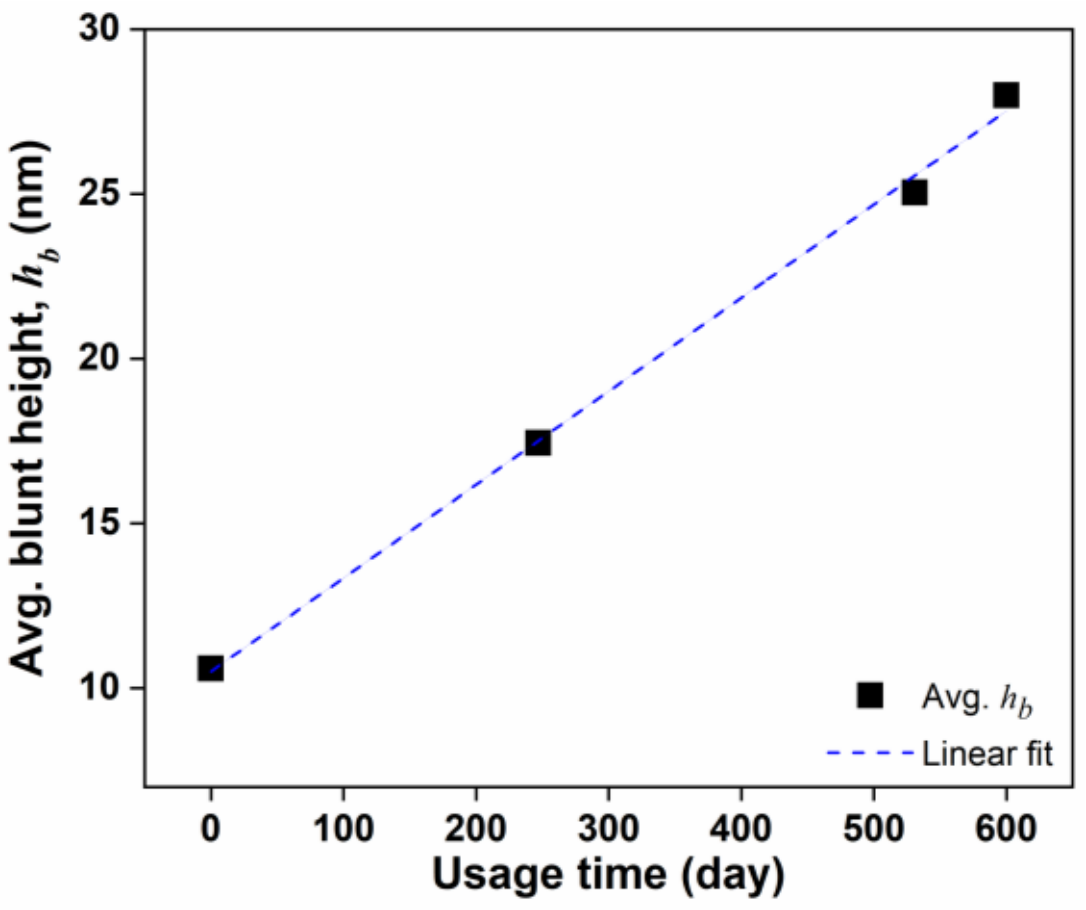

**Figure S.6.1:** Estimating tip blunt height, $h_b$, as a function of days of tip usage.

# Supporting Information 7

## True strain correlation with nanoindentation data

From Appendix B

$$\left(\frac{h_{\mathrm{ne}}}{h_{\mathrm{e}}}\right)^{c} = \frac{1+\varepsilon_y \varepsilon_{\mathrm{ne}}^{*}}{1+\varepsilon_y\left(\varepsilon^{*}-\varepsilon_{\mathrm{ne}}^{*}\right)} \qquad \text{eqn. S.7.1}$$

### 1. Cross-linked SU-8 true strain correlations

Linear fit of $\varepsilon_y$ with $\ln\dot{\varepsilon}$ yields $\varepsilon_y = 9.46\times10^{-4}\ln\dot{\varepsilon} + 5.23\times10^{-2} = p\ln\dot{\varepsilon}+q = p\left(\ln\dot{\varepsilon}_{\mathrm{t}}+\varepsilon_{\mathrm{t}}\right)+q$.

From eqn. S.1.5,

$$\varepsilon_{\mathrm{ne}} = \left(p\left(\ln\dot{\varepsilon}_{\mathrm{t}}+\varepsilon_{\mathrm{t}}\right)+q\right)\begin{bmatrix} 8.57\left(\varepsilon^{*}-0.5\right)^{6}-25.58\left(\varepsilon^{*}-0.5\right)^{5}+26.21\left(\varepsilon^{*}-0.5\right)^{4} \\ -10.67\left(\varepsilon^{*}-0.5\right)^{3}+2.32\left(\varepsilon^{*}-0.5\right)^{2} \end{bmatrix}$$

$$\varepsilon_{\mathrm{t,ne}} = \ln\left(1+\left(p\left(\ln\dot{\varepsilon}_{\mathrm{t}}+\varepsilon_{\mathrm{t}}\right)+q\right)\begin{bmatrix} 8.57\left(\varepsilon^{*}-0.5\right)^{6}-25.58\left(\varepsilon^{*}-0.5\right)^{5}+26.21\left(\varepsilon^{*}-0.5\right)^{4} \\ -10.67\left(\varepsilon^{*}-0.5\right)^{3}+2.32\left(\varepsilon^{*}-0.5\right)^{2} \end{bmatrix}\right)$$

Combining the above with eqn. S.7.1,

$$\left(\frac{h_{\text{ne}}}{h_{\text{e}}}\right)^{c} = \frac{1+\left(p\left(\ln\dot{\varepsilon}_{\text{t}}+\varepsilon_{\text{t}}\right)+q\right)\begin{bmatrix}8.57\left(\varepsilon^{*}-0.5\right)^{6}-25.58\left(\varepsilon^{*}-0.5\right)^{5}\\+26.21\left(\varepsilon^{*}-0.5\right)^{4}-10.67\left(\varepsilon^{*}-0.5\right)^{3}\\+2.32\left(\varepsilon^{*}-0.5\right)^{2}\end{bmatrix}}{1+\left(p\left(\ln\dot{\varepsilon}_{\text{t}}+\varepsilon_{\text{t}}\right)+q\right)\left[\varepsilon^{*}-\begin{pmatrix}8.57\left(\varepsilon^{*}-0.5\right)^{6}-25.58\left(\varepsilon^{*}-0.5\right)^{5}\\+26.21\left(\varepsilon^{*}-0.5\right)^{4}-10.67\left(\varepsilon^{*}-0.5\right)^{3}\\+2.32\left(\varepsilon^{*}-0.5\right)^{2}\end{pmatrix}\right]} \quad \text{eqn. S.7.2}$$

From eqn. S.1.6, $\left(\frac{h_{\text{ne}}}{h_{\text{e}}}\right)^{c} = \frac{1+\left(p\left(\ln\dot{\varepsilon}_{\text{t}}+\varepsilon_{\text{t}}\right)+q\right)\left(\varepsilon^{*}-0.65\right)}{1+\left(p\left(\ln\dot{\varepsilon}_{\text{t}}+\varepsilon_{\text{t}}\right)+q\right)\left[\varepsilon^{*}-\left(\varepsilon^{*}-0.65\right)\right]}$. This yields,

$$\left(\frac{h_{\text{ne}}}{h_{\text{e}}}\right)^{c} = \frac{1+\left(p\left(\ln\dot{\varepsilon}_{\text{t}}+\varepsilon_{\text{t}}\right)+q\right)\left(\varepsilon^{*}-0.65\right)}{1+0.65\left(p\left(\ln\dot{\varepsilon}_{\text{t}}+\varepsilon_{\text{t}}\right)+q\right)} \quad \text{eqn. S.7.3}$$

## 2. PC true strain correlations

Linear fit of $\varepsilon_y$ with $\ln\dot{\varepsilon}$ yields $\varepsilon_y = 4.81 \times 10^{-4}\ln\dot{\varepsilon} + 9.08 \times 10^{-2} = p\ln\dot{\varepsilon}+q = p\left(\ln\dot{\varepsilon}_{\text{t}}+\varepsilon_{\text{t}}\right)+q$.

From eqn. S.1.14,

$$\varepsilon_{\text{ne}} = \left(p\left(\ln\dot{\varepsilon}_{\text{t}}+\varepsilon_{\text{t}}\right)+q\right)\left[-0.49\left(\varepsilon^{*}-0.3\right)^{4}+0.72\left(\varepsilon^{*}-0.3\right)^{3}+0.44\left(\varepsilon^{*}-0.3\right)^{2}\right]$$

$$\Rightarrow \varepsilon_{\text{ne}} = \ln\left(1+\left(p\left(\ln\dot{\varepsilon}_{\text{t}}+\varepsilon_{\text{t}}\right)+q\right)\left[-0.49\left(\varepsilon^{*}-0.3\right)^{4}+0.72\left(\varepsilon^{*}-0.3\right)^{3}+0.44\left(\varepsilon^{*}-0.3\right)^{2}\right]\right)$$

By incorporating the above into eqn. S.7.3,

$$\left(\frac{h_{\text{ne}}}{h_{\text{e}}}\right)^{c}=\frac{1+\left(p\left(\ln\dot{\varepsilon}_{\text{t}}+\varepsilon_{\text{t}}\right)+q\right)\begin{bmatrix}-0.49\left(\varepsilon^{*}-0.3\right)^{4}+0.72\left(\varepsilon^{*}-0.3\right)^{3}\\+0.44\left(\varepsilon^{*}-0.3\right)^{2}\end{bmatrix}}{1+\left(p\left(\ln\dot{\varepsilon}_{\text{t}}+\varepsilon_{\text{t}}\right)+q\right)\left[\varepsilon^{*}-\begin{pmatrix}-0.49\left(\varepsilon^{*}-0.3\right)^{4}+0.72\left(\varepsilon^{*}-0.3\right)^{3}\\+0.44\left(\varepsilon^{*}-0.3\right)^{2}\end{pmatrix}\right]} \qquad \text{eqn. S.7.4}$$

From eqn. S.1.11, $\left(\frac{h_{\text{ne}}}{h_{\text{e}}}\right)^{c}=\frac{1+\left(p\left(\ln\dot{\varepsilon}_{\text{t}}+\varepsilon_{\text{t}}\right)+q\right)\left(1.10\varepsilon^{*}-0.77\right)}{1+\left(p\left(\ln\dot{\varepsilon}_{\text{t}}+\varepsilon_{\text{t}}\right)+q\right)\left[\varepsilon^{*}-\left(1.10\varepsilon^{*}-0.77\right)\right]}$. Hence,

$$\left(\frac{h_{\text{ne}}}{h_{\text{e}}}\right)^{c}=\frac{1+\left(p\left(\ln\dot{\varepsilon}_{\text{t}}+\varepsilon_{\text{t}}\right)+q\right)\left(1.10\varepsilon^{*}-0.77\right)}{1+\left(p\left(\ln\dot{\varepsilon}_{\text{t}}+\varepsilon_{\text{t}}\right)+q\right)\left(0.77-0.10\varepsilon^{*}\right)} \qquad \text{eqn. S.7.5}$$

## 3.A. PMMA [El-Qoubaa] true strain correlations

Linear fit of $\varepsilon_y$ with $\ln\dot{\varepsilon}$ yields $\varepsilon_y = 4.18\times 10^{-3}\ln\dot{\varepsilon} + 0.1116 = p\ln\dot{\varepsilon}+q = p\left(\ln\dot{\varepsilon}_{\text{t}}+\varepsilon_{\text{t}}\right)+q$.

From eqn. S.1.19,

$$\varepsilon_{\text{ne}}=\left(p\left(\ln\dot{\varepsilon}_{\text{t}}+\varepsilon_{\text{t}}\right)+q\right)\left[-0.407\left(\varepsilon^{*}-0.4\right)^{4}+0.337\left(\varepsilon^{*}-0.4\right)^{3}+0.812\left(\varepsilon^{*}-0.4\right)^{2}\right]$$

$$\Rightarrow\varepsilon_{\text{ne}}=\ln\left(1+\left(p\left(\ln\dot{\varepsilon}_{\text{t}}+\varepsilon_{\text{t}}\right)+q\right)\left[-0.407\left(\varepsilon^{*}-0.4\right)^{4}+0.337\left(\varepsilon^{*}-0.4\right)^{3}+0.812\left(\varepsilon^{*}-0.4\right)^{2}\right]\right)$$

By incorporating the above into eqn. S.7.1,

$$\left(\frac{h_{\mathrm{ne}}}{h_{\mathrm{e}}}\right)^{c}=\frac{1+\left(p\left(\ln\dot{\varepsilon}_{\mathrm{t}}+\varepsilon_{\mathrm{t}}\right)+q\right)\begin{bmatrix}-0.407\left(\varepsilon^{*}-0.4\right)^{4}+0.337\left(\varepsilon^{*}-0.4\right)^{3}\\+0.812\left(\varepsilon^{*}-0.4\right)^{2}\end{bmatrix}}{1+\left(p\left(\ln\dot{\varepsilon}_{\mathrm{t}}+\varepsilon_{\mathrm{t}}\right)+q\right)\left[\varepsilon^{*}-\begin{pmatrix}-0.407\left(\varepsilon^{*}-0.4\right)^{4}+0.337\left(\varepsilon^{*}-0.4\right)^{3}\\+0.812\left(\varepsilon^{*}-0.4\right)^{2}\end{pmatrix}\right]} \quad \text{eqn. S.7.6}$$

From eqn. S.1.20, $\left(\frac{h_{\mathrm{ne}}}{h_{\mathrm{e}}}\right)^{c}=\frac{1+\left(p\left(\ln\dot{\varepsilon}_{\mathrm{t}}+\varepsilon_{\mathrm{t}}\right)+q\right)\left(1.11\varepsilon^{*}-0.804\right)}{1+\left(p\left(\ln\dot{\varepsilon}_{\mathrm{t}}+\varepsilon_{\mathrm{t}}\right)+q\right)\left[\varepsilon^{*}-\left(1.11\varepsilon^{*}-0.804\right)\right]}$. Hence,

$$\left(\frac{h_{\mathrm{ne}}}{h_{\mathrm{e}}}\right)^{c}=\frac{1+\left(p\left(\ln\dot{\varepsilon}_{\mathrm{t}}+\varepsilon_{\mathrm{t}}\right)+q\right)\left(1.11\varepsilon^{*}-0.804\right)}{1+\left(p\left(\ln\dot{\varepsilon}_{\mathrm{t}}+\varepsilon_{\mathrm{t}}\right)+q\right)\left(0.804-0.11\varepsilon^{*}\right)} \quad \text{eqn. S.7.7}$$

## 3.B. PMMA [Hu] true strain correlations

Linear fit of $\varepsilon_y$ with $\ln\dot{\varepsilon}$ yields $\varepsilon_y$ = -2.52 × $10^{-3}\ln\dot{\varepsilon}$ + 0.074 = $p\ln\dot{\varepsilon}+q$ = $p\left(\ln\dot{\varepsilon}_{\mathrm{t}}+\varepsilon_{\mathrm{t}}\right)+q$.

From eqn. S.1.25,

$$\varepsilon_{\mathrm{ne}}=\left(p\left(\ln\dot{\varepsilon}_{\mathrm{t}}+\varepsilon_{\mathrm{t}}\right)+q\right)\left[0.996\left(\varepsilon^{*}-0.3\right)^{4}-2.50\left(\varepsilon^{*}-0.3\right)^{3}+2.33\left(\varepsilon^{*}-0.3\right)^{2}\right]$$

$$\Rightarrow\varepsilon_{\mathrm{ne}}=\ln\left(1+\left(p\left(\ln\dot{\varepsilon}_{\mathrm{t}}+\varepsilon_{\mathrm{t}}\right)+q\right)\left[0.996\left(\varepsilon^{*}-0.3\right)^{4}-2.50\left(\varepsilon^{*}-0.3\right)^{3}+2.33\left(\varepsilon^{*}-0.3\right)^{2}\right]\right)$$

By incorporating the above into eqn. S.7.1,

$$\left(\frac{h_{\text{ne}}}{h_{\text{e}}}\right)^{c}=\frac{1+\left(p\left(\ln\dot{\varepsilon}_{\text{t}}+\varepsilon_{\text{t}}\right)+q\right)\begin{bmatrix}0.996\left(\varepsilon^{*}-0.3\right)^{4}-2.50\left(\varepsilon^{*}-0.3\right)^{3}\\+2.33\left(\varepsilon^{*}-0.3\right)^{2}\end{bmatrix}}{1+\left(p\left(\ln\dot{\varepsilon}_{\text{t}}+\varepsilon_{\text{t}}\right)+q\right)\left[\varepsilon^{*}-\begin{pmatrix}0.996\left(\varepsilon^{*}-0.3\right)^{4}-2.50\left(\varepsilon^{*}-0.3\right)^{3}\\+2.33\left(\varepsilon^{*}-0.3\right)^{2}\end{pmatrix}\right]} \qquad \text{eqn. S.7.8}$$

From eqn. S.1.26, $\left(\frac{h_{\text{ne}}}{h_{\text{e}}}\right)^{c}=\frac{1+\left(p\left(\ln\dot{\varepsilon}_{\text{t}}+\varepsilon_{\text{t}}\right)+q\right)\left(1.02\varepsilon^{*}-0.504\right)}{1+\left(p\left(\ln\dot{\varepsilon}_{\text{t}}+\varepsilon_{\text{t}}\right)+q\right)\left[\varepsilon^{*}-\left(1.02\varepsilon^{*}-0.504\right)\right]}$. Hence,

$$\left(\frac{h_{\text{ne}}}{h_{\text{e}}}\right)^{c}=\frac{1+\left(p\left(\ln\dot{\varepsilon}_{\text{t}}+\varepsilon_{\text{t}}\right)+q\right)\left(1.02\varepsilon^{*}-0.504\right)}{1+\left(p\left(\ln\dot{\varepsilon}_{\text{t}}+\varepsilon_{\text{t}}\right)+q\right)\left(0.504-0.02\varepsilon^{*}\right)} \qquad \text{eqn. S.7.9}$$

## 3.C. PMMA [Ames] true strain correlations

Linear fit of $\varepsilon_y$ with $\ln\dot{\varepsilon}$ yields $\varepsilon_y = 4.86\times10^{-4}\ln\dot{\varepsilon} + 8.56\times10^{-2} = p\ln\dot{\varepsilon}+q = p\left(\ln\dot{\varepsilon}_{\text{t}}+\varepsilon_{\text{t}}\right)+q$.

From eqn. S.1.31,

$$\varepsilon_{\text{ne}}=\left(p\left(\ln\dot{\varepsilon}_{\text{t}}+\varepsilon_{\text{t}}\right)+q\right)\left[1.08\left(\varepsilon^{*}-0.3\right)^{4}-2.51\left(\varepsilon^{*}-0.3\right)^{3}+2.22\left(\varepsilon^{*}-0.3\right)^{2}\right]$$

$$\Rightarrow\varepsilon_{\text{ne,t}}=\ln\left(1+\left(p\left(\ln\dot{\varepsilon}_{\text{t}}+\varepsilon_{\text{t}}\right)+q\right)\left[1.08\left(\varepsilon^{*}-0.3\right)^{4}-2.51\left(\varepsilon^{*}-0.3\right)^{3}+2.22\left(\varepsilon^{*}-0.3\right)^{2}\right]\right)$$

By incorporating the above into eqn. S.7.1,

$$\left(\frac{h_{\mathrm{ne}}}{h_{\mathrm{e}}}\right)^{c}=\frac{1+\left(p\left(\ln\dot{\varepsilon}_{\mathrm{t}}+\varepsilon_{\mathrm{t}}\right)+q\right)\begin{bmatrix}1.08\left(\varepsilon^{*}-0.3\right)^{4}-2.51\left(\varepsilon^{*}-0.3\right)^{3}\\+2.22\left(\varepsilon^{*}-0.3\right)^{2}\end{bmatrix}}{1+\left(p\left(\ln\dot{\varepsilon}_{\mathrm{t}}+\varepsilon_{\mathrm{t}}\right)+q\right)\left[\varepsilon^{*}-\begin{pmatrix}1.08\left(\varepsilon^{*}-0.3\right)^{4}-2.51\left(\varepsilon^{*}-0.3\right)^{3}\\+2.22\left(\varepsilon^{*}-0.3\right)^{2}\end{pmatrix}\right]}$$ eqn. S.7.10

From eqn. S.1.32, $\left(\frac{h_{\mathrm{ne}}}{h_{\mathrm{e}}}\right)^{c}=\frac{1+\left(p\left(\ln\dot{\varepsilon}_{\mathrm{t}}+\varepsilon_{\mathrm{t}}\right)+q\right)\left(1.06\varepsilon^{*}-0.589\right)}{1+\left(p\left(\ln\dot{\varepsilon}_{\mathrm{t}}+\varepsilon_{\mathrm{t}}\right)+q\right)\left[\varepsilon^{*}-\left(1.06\varepsilon^{*}-0.589\right)\right]}$. Hence,

$$\left(\frac{h_{\mathrm{ne}}}{h_{\mathrm{e}}}\right)^{c}=\frac{1+\left(p\left(\ln\dot{\varepsilon}_{\mathrm{t}}+\varepsilon_{\mathrm{t}}\right)+q\right)\left(1.06\varepsilon^{*}-0.589\right)}{1+\left(p\left(\ln\dot{\varepsilon}_{\mathrm{t}}+\varepsilon_{\mathrm{t}}\right)+q\right)\left(0.589-0.06\varepsilon^{*}\right)}$$ eqn. S.7.11

# Supporting Information 8

We find that the uncertainty in the linear relationship between $\varepsilon_y$ vs $\ln\dot{\varepsilon}$, correspond to an uncertainty of ~ 2.5% in the $c$ values of cross-linked SU-8. The $\dot{\varepsilon}$, $\varepsilon$ and $c$ values, are listed in Table S.8.1 and in Table S.8.2 for minimum and maximum values respectively, of the $\varepsilon_y$ vs $\ln\dot{\varepsilon}$ slope.

**Table S.8.1:** $\dot{\varepsilon}$, $\varepsilon$ and $c$ values when $p_{\mathrm{min,SU-8}}$ = 2.39 × $10^{-4}$, $q_{\mathrm{min,SU-8}}$ = 4.73 × $10^{-2}$ for cross-linked SU-8.

| $\dot{P}/P$ (/s) | $P_{\mathrm{m}}$ (μN) | $\dot{\varepsilon}$ × $10^2$ (/s) | $\varepsilon$ × $10^2$ | $c$ |
|---|---|---|---|---|
| 0.1 | 1000 | 0.23 ± 0.00 | 6.24 ± 0.09 | 0.045 ± 0.000 |
| | 9000 | 0.17 ± 0.00 | 6.74 ± 0.08 | 0.047 ± 0.001 |
| 1.0 | 1000 | 2.70 ± 0.25 | 6.04 ± 0.00 | 0.045 ± 0.000 |
| | 9000 | 1.90 ± 0.00 | 5.78 ± 0.03 | 0.043 ± 0.000 |

**Table S.8.2:** $\dot{\varepsilon}$, $\varepsilon$ and $c$ values when $p_{\mathrm{max,SU-8}}$ = 2.30 × $10^{-3}$, $q_{\mathrm{max,SU-8}}$ = 6.05 × $10^{-2}$ for cross-linked SU-8.

| $\dot{P}/P$ (/s) | $P_{\mathrm{m}}$ (μN) | $\dot{\varepsilon}$ × $10^2$ (/s) | $\varepsilon$ × $10^2$ | $c$ |
|---|---|---|---|---|
| 0.1 | 1000 | 0.21 ± 0.00 | 6.03 ± 0.09 | 0.042 ± 0.000 |
| | 9000 | 0.16 ± 0.00 | 6.37 ± 0.09 | 0.042 ± 0.001 |
| 1.0 | 1000 | 2.79 ± 0.27 | 6.01 ± 0.00 | 0.046 ± 0.000 |
| | 9000 | 1.94 ± 0.03 | 6.18 ± 0.03 | 0.043 ± 0.000 |

# Supporting Information 9

Table S.9.1, Table S.9.2 and Table S.9.3 list the $\dot{\varepsilon}$, $\varepsilon$ and $c$ values for 0.25, 0.50 and 0.75 $r/r_{\mathrm{m}}$ positions, respectively, for the all polymers.

**Table S.9.1**: $\dot{\varepsilon}$, $\varepsilon$ and $c$ values for $r/r_{\mathrm{m}}$ = 0.25 position for cross-linked SU-8, PC and PMMA.

| | $\dot{P}/P$ (/s) | $P_{\mathrm{m}}$ (μN) | $\dot{\varepsilon} \times 10^2$ (/s) | $\varepsilon \times 10^2$ | $c$ |
|---|---|---|---|---|---|
| Cross-linked SU-8 | 0.1 | 1000 | 0.43 ± 0.00 | 6.23 ± 0.07 | 0.045 ± 0.000 |
| | | 9000 | 0.41 ± 0.00 | 6.49 ± 0.09 | 0.043 ± 0.000 |
| | 1.0 | 1000 | 5.00 ± 0.18 | 5.98 ± 0.00 | 0.041 ± 0.000 |
| | | 9000 | 3.75 ± 0.25 | 5.72 ± 0.00 | 0.035 ± 0.000 |
| PC | 0.1 | 1000 | 0.97 ± 0.02 | 12.17 ± 0.29 | 0.077 ± 0.002 |
| | | 9000 | 0.65 ± 0.01 | 12.38 ± 0.34 | 0.073 ± 0.002 |
| | 1.0 | 1000 | 6.81 ± 0.26 | 11.60 ± 0.37 | 0.068 ± 0.003 |
| | | 9000 | 6.13 ± 0.28 | 11.09 ± 0.59 | 0.064 ± 0.003 |
| PMMA [30] | 0.1 | 1000 | 1.08 ± 0.01 | 12.47 ± 0.22 | 0.084 ± 0.001 |
| | | 9000 | 0.39 ± 0.00 | 12.58 ± 0.28 | 0.079 ± 0.001 |

**Table S.9.2:** $\dot{\varepsilon}$, $\varepsilon$ and $c$ values for $r/r_{\mathrm{m}}$ = 0.50 position for cross-linked SU-8, PC and PMMA.

| | $\dot{P}/P$ (/s) | $P_{\mathrm{m}}$ (μN) | $\dot{\varepsilon} \times 10^2$ (/s) | $\varepsilon \times 10^2$ | $c$ |
|---|---|---|---|---|---|
| Cross-linked SU-8 | 0.1 | 1000 | 0.72 ± 0.00 | 6.21 ± 0.07 | 0.042 ± 0.000 |
| | | 9000 | 0.62 ± 0.00 | 6.26 ± 0.09 | 0.040 ± 0.000 |
| | 1.0 | 1000 | 8.42 ± 0.18 | 5.47 ± 0.00 | 0.037 ± 0.000 |
| | | 9000 | 4.67 ± 0.25 | 5.44 ± 0.00 | 0.028 ± 0.000 |
| PC | 0.1 | 1000 | 1.45 ± 0.01 | 10.59 ± 0.29 | 0.067 ± 0.002 |
| | | 9000 | 1.01 ± 0.01 | 11.84 ± 0.34 | 0.068 ± 0.002 |
| | 1.0 | 1000 | 11.68 ± 0.26 | 9.95 ± 0.37 | 0.063 ± 0.003 |
| | | 9000 | 8.65 ± 0.28 | 10.43 ± 0.59 | 0.056 ± 0.003 |
| PMMA [30] | 0.1 | 1000 | 2.60 ± 0.01 | 11.11 ± 0.22 | 0.077 ± 0.001 |
| | | 9000 | 0.62 ± 0.00 | 11.87 ± 0.28 | 0.072 ± 0.001 |

**Table S.9.3:** $\dot{\varepsilon}$, $\varepsilon$ and $c$ values for $r/r_{\mathrm{m}}$ = 0.75 position for cross-linked SU-8, PC and PMMA.

| | $\dot{P}/P$ (/s) | $P_{\mathrm{m}}$ (μN) | $\dot{\varepsilon}$ × $10^2$ (/s) | $\varepsilon$ × $10^2$ | $c$ |
|---|---|---|---|---|---|
| Cross-linked SU-8 | 0.1 | 1000 | 2.42 ± 0.00 | 5.54 ± 0.07 | 0.038 ± 0.000 |
| | | 9000 | 1.11 ± 0.00 | 5.87 ± 0.09 | 0.033 ± 0.000 |
| | 1.0 | 1000 | 10.37 ± 0.17 | 5.10 ± 0.00 | 0.032 ± 0.000 |
| | | 9000 | 7.33 ± 0.25 | 5.22 ± 0.00 | 0.022 ± 0.000 |
| PC | 0.1 | 1000 | 4.33 ± 0.01 | 10.29 ± 0.34 | 0.063 ± 0.002 |
| | | 9000 | 1.63 ± 0.01 | 10.29 ± 0.34 | 0.056 ± 0.002 |
| | 1.0 | 1000 | 16.48 ± 0.26 | 9.45 ± 0.37 | 0.060 ± 0.003 |
| | | 9000 | 12.18 ± 0.28 | 9.10 ± 0.59 | 0.042 ± 0.003 |
| PMMA [30] | 0.1 | 1000 | 3.13 ± 0.01 | 9.98 ± 0.22 | 0.071 ± 0.001 |
| | | 9000 | 0.93 ± 0.00 | 11.12 ± 0.28 | 0.065 ± 0.001 |

# Supporting Information 10

## S.10.1. Stress state vs corresponding strain state

We have defined a stress state, $\Sigma = \sigma^* - \sigma_0^*$. We plot $\Sigma$ vs $\Phi$ in Fig. S.10.1.

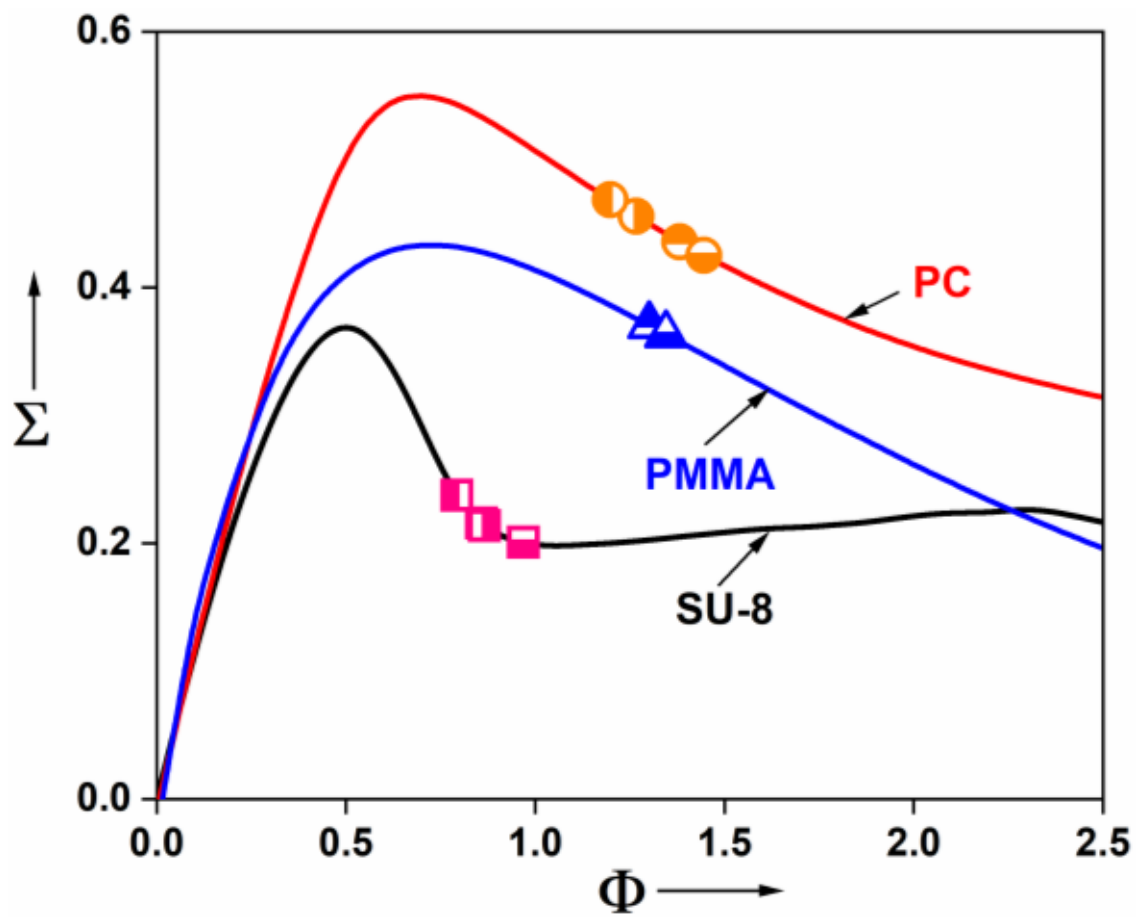

Figure S.10.1: $\Sigma$ **vs** $\Phi$ for **cross-linked SU-8, PC and PMMA**. The equivalent strains of the nanoindentation experiments are marked. ■, ●, ▲ **represent** $P_{\mathrm{m}}$ **= 1000 μN and** $\dot{P}/P$ **= 0.1 /s,** ■, ●, ▲ **represent** $P_{\mathrm{m}}$ **= 9000 μN and** $\dot{P}/P$ **= 0.1 /s,** ■, ● **represent** $P_{\mathrm{m}}$ **= 1000 μN and** $\dot{P}/P$ **= 1.0 /s and** ■, ● **represent** $P_{\mathrm{m}}$ **= 9000 μN and** $\dot{P}/P$ **= 1.0 /s.**